\documentclass[10pt,a4paper]{article}
\usepackage{graphics,epsfig}
\usepackage{palatino}
\usepackage{amstext,amsfonts,amssymb,amsthm,amsmath,mathrsfs,epic,eepic}
\usepackage{subfigure}  
\usepackage{color}
\usepackage[normalem]{ulem}
\usepackage[left]{lineno}
\usepackage{setspace}
\usepackage{graphics,epsfig}
\usepackage{color}
\usepackage{caption}
\usepackage{capt-of}
\usepackage{array}
\usepackage{multirow}

\DeclareMathOperator*{\argmin}{arg\,min}

\begin{document}

\newcommand{\figref}[1]{Fig~\ref{#1}}
\newcommand{\figsref}[1]{Figs~\ref{#1}}
\newcommand{\figssref}[2]{Figs~\ref{#1}-\ref{#2}}
\newcommand{\etal}{\emph{et al}}
\newcommand{\be}{\begin{equation}}
\newcommand{\ee}{\end{equation}}
\definecolor{darkgreen}{rgb}{0.0,0.33,0.0}
\definecolor{midgreen}{rgb}{0.0,0.5,0.0}
\newcommand{\TODO}[1]{{\bf \color{midgreen}{($\bigstar$ #1)}}}
\newcommand{\TF}[1]{{\color{midgreen}{#1}}}
\newcommand{\E}{\mbox{erf}}
\renewcommand{\eqref}[1]{Eq.~(\ref{#1})}
\newcommand{\eqsref}[1]{Eqs.~(\ref{#1})}
\newcommand{\eqssref}[2]{Eqs.~(\ref{#1})-(\ref{#2})}
\newcommand{\eqnumref}[1]{(\ref{#1})}

\title{Intrinsic limits to gene regulation by global crosstalk}

\author{Tamar Friedlander, Roshan Prizak, C\u{a}lin C. Guet, Nicholas H. Barton, \and Ga\v{s}per Tka\v{c}ik\\
Institute of Science and Technology Austria, Am Campus 1, A-3400 \\ Klosterneuburg, Austria}

\maketitle

\begin{abstract}
Gene regulation relies on the specificity of transcription factor (TF)-DNA interactions. Limited specificity may lead to crosstalk: a regulatory state in which a gene is either incorrectly activated due to noncognate TF-DNA interactions or remains erroneously inactive. Since each TF can have numerous interactions with noncognate cis-regulatory elements, crosstalk is inherently a global problem, yet has previously not been studied as such. We construct a theoretical framework to analyze the effects of global crosstalk on gene regulation. We find that crosstalk presents a significant challenge for organisms with low-specificity TFs, such as metazoans. Crosstalk is not easily mitigated by known regulatory schemes acting at equilibrium, including variants of cooperativity and combinatorial regulation. Our results suggest that crosstalk imposes a previously unexplored global constraint on the functioning and evolution of regulatory networks, which is qualitatively distinct from the known constraints that act at the level of individual gene regulatory elements.
\end{abstract}

\clearpage

\section*{Introduction}
Life depends on the specificity of molecular recognition to ensure that essential reactions only occur between cognate substrates even when similar noncognate substrates are present, sometimes in large excess. A paradigmatic example is that of the aminoacyl tRNA synthetase~\cite{yamane_experimental_1977}
, which uses kinetic proofreading~\cite{hopfield_kinetic_1974} 
to load  appropriate amino acids onto matching  tRNAs. This and other examples---including DNA replication, ligand sensing \cite{mora_physical_2015}, protein-protein interactions~\cite{swain_role_2002,skerker_rewiring_2008,johnson_nonspecific_2011,zhang_constraints_2008,ouldridge_robustness_2014,rowland_crosstalk_2014}, 
 recognition events in the immune system~\cite{mckeithan_kinetic_1995,lalanne_principles_2013} and 
molecular self-assembly~\cite{murugan_undesired_2015}
---indicate
that biology places a large premium on the reduction of unintended ``crosstalk'', a generic term that encompasses all potentially disruptive processes due to reactions between noncognate substrates.

Molecular recognition is fundamental also to transcriptional regulation, the primary mechanism by which cells control gene expression. The specificity of this regulation ultimately originates in the binding interactions between special regulatory proteins, called transcription factors (TFs), and short regulatory sequences on the DNA, called binding sites.
Although each type of TF preferentially binds certain regulatory DNA sequences, a large body of evidence shows that this binding specificity is limited, and that TFs bind other noncognate targets as well~\cite{von_hippel_non-specific_1974,wunderlich_different_2009,johnson_dark_2005,maerkl_systems_2007,rockel_islim:_2012}.
These additional binding targets were previously discussed in the context of their effect on the TF concentration \cite{burger_abduction_2010,Sheinman_how_2012}. However, if these sites happen to also be  regulatory elements of other genes, non-cognate binding not only depletes  TF molecules, but could also actively interfere with gene regulation. This suggests that the crosstalk problem is global: in a pool of TF molecules of different chemical species co-expressed at any one time, each molecule has a small probability of erroneously regulating some subset of all genes. As the regulatory system grows in complexity, the number of potential noncognate interactions will grow faster than the number of cognate interactions. While this makes the problem biologically relevant and theoretically interesting, existing work has mostly considered a reduced setting, by computing binding probabilities for a single TF to cognate vs noncognate sites~\cite{gerland_physical_2002,sengupta_specificity_2002,bintu_transcriptional_2005,lynch_evolutionary_2015}. Such a reduced description thus overlooked the effect of this TF on the (mis)regulation of genes that were not its cognate regulatory targets. Motivated by this observation, our primary goal here is to develop a new framework for crosstalk that captures its global nature, by simultaneously treating multiple TFs and multiple regulatory binding sites.  Moreover, the focus of prior work has been on how to achieve reliable gene regulation by cognate TFs~\cite{todeschini_transcription_2014}, while the complementary question of how to prevent erroneous regulation
by noncognate TFs has remained largely unexplored (but see~\cite{bird_gene_1995}). As a result, it remains unclear whether
crosstalk places strong constraints on the ability of cells to orchestrate their gene expression programs, and to what extent different molecular mechanisms could relax any such constraints.

To address these questions quantitatively, we construct a model of crosstalk in transcriptional regulation that satisfies three key requirements for biophysical plausibility.
First, the model should
be global. Global models, where many targets are simultaneously regulated by different TFs, will properly capture the faster-than-linear growth in the number of possible noncognate interactions as the number of TFs increases, and the difficulty in ensuring that recognition sequences for all TFs remain sufficiently distinct. Second, the model should
explicitly account for differential activation of genes depending on  regulatory conditions. Consequently---and in contrast to previously studied cases of molecular recognition~\cite{hopfield_kinetic_1974}---the distinction between ``erroneous'' and ``correct'' outcomes of regulation will depend on the presence / absence of the regulatory signals. In particular, the ability of the regulatory system to  keep genes reliably inactive 
when appropriate, despite crosstalk interference, will emerge as an important consideration. Third, textbook models of transcriptional regulation assume that TF-DNA interactions happen in equilibrium~\cite{bintu_transcriptional_2005,phillips_napoleon_2015}.
This assumption, which is supported experimentally for prokaryotic regulation~\cite{ackers_quantitative_1982,kinney_using_2010}
and which underlies the majority of modeling and bioinformatic applications, puts strong constraints on models of crosstalk. In this work, we explore its consequences in depth; we report on out-of-equilibrium schemes elsewhere~\cite{cepeda-humerez_stochastic_2015}.

Using our biophysical model we identify the parameters that have a major influence on 
crosstalk severity. While some of these parameters, such as the free concentration of TFs, are difficult to estimate, we show that there exists a lower bound to crosstalk with respect to these parameters. This implies the existence of a ``crosstalk floor,'' which cannot be overcome even if TF concentrations were optimally adjusted by the cell, by various feedback mechanisms or otherwise, and compensated for sequestration at noncognate sites.

Our model allows us to ask a number of fundamental questions:
How does the severity of crosstalk depend on the number of (co-expressed) genes or the biophysical properties of TF-DNA interactions, such as binding site length and binding energy, for which we have reliable estimates? How do the regulatory strategies of prokaryotes compare to those of eukaryotes? Do complex regulatory schemes, such as combinatorial regulation by activators and repressors, or cooperative activation, lower 
crosstalk, as is often implied~\cite{todeschini_transcription_2014}?

Many biophysical constraints have been shown to shape the properties of genetic regulatory networks, e.g., programmability~\cite{gerland_physical_2002}, response speed~\cite{mangan_structure_2003},
noise in gene expression and dynamic range of regulation~\cite{tkacik_information_2011,dubuis_positional_2013,friedlander_adaptive_2011,friedlander_cellular_2008},
robustness~\cite{von_dassow_segment_2000}
and evolvability of the regulatory sequences~\cite{payne_robustness_2014,stern_is_2009}.
Most of these constraints, however, could be understood at the level of individual genetic regulatory elements. Crosstalk, as analyzed here, is special: while it originates locally due to biophysical limits to molecular recognition, its cumulative effect only emerges globally. At the level of a single genetic regulatory element, crosstalk can always be avoided by  
increasing the concentration of cognate TFs or introducing multiple binding sites in the promoter.
It is only when we self-consistently consider that these same cognate TFs act as noncognate TFs for 
other genes, or that new binding sites in the promoter drastically increase the number of noncognate binding configurations, that crosstalk constraints become clear.

\begin{figure}
\centerline{\includegraphics[width=\textwidth]{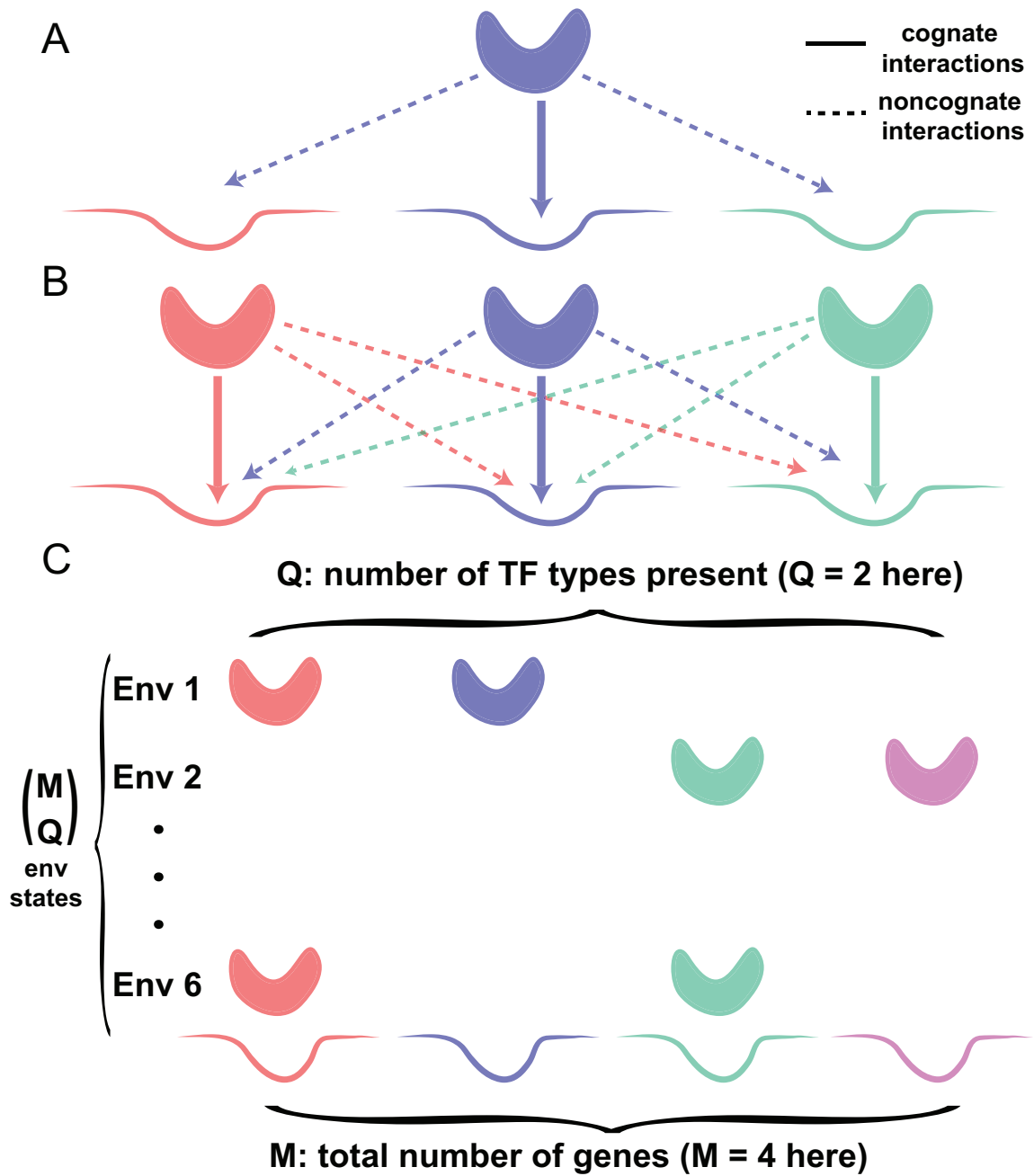}}
\caption{\textbf{Crosstalk in gene regulation.} {\bf (A)} A TF preferentially binds to its cognate binding site, but can also bind noncognate sites, potentially causing crosstalk---an erroneous activation or repression of a gene. {\bf (B)} In a global setting where many TFs regulate many genes, the number of possible noncognate interactions grows quickly with the number of TFs; additionally, it may become difficult to keep TF recognition sequences sufficiently distinct from each other.
{\bf (C)} Cells respond to changing environments by attempting to activate subsets of their genes. In this example, the total number of genes is $M=4$, and different environments (here, 6 in total) call for activation of different subsets with $Q=2$ genes. To control the expression in every environment, TFs for $Q$ required genes are present, while the TFs for the remaining $M-Q$ genes are absent. Because of crosstalk, TFs can bind noncognate
sites, generating a pattern of gene expression that can differ from the one required.
}\label{Fig:crosstalk}
\end{figure}

\section*{Results}

\subsection*{A thermodynamic model of global crosstalk}
We start by introducing a \emph{basic model} of regulation, in which each gene will be regulated in the simplest possible manner by a dedicated TF type, and the mechanism of regulation will be identical for every gene. For this basic model, where the limits to crosstalk are analytically computable, we will outline the reasoning, sketch the derivation, and interpret the results in the main text. We will then relax our simplifying assumptions in a variety of ways, and extend the analysis to more elaborate regulatory schemes, such as different flavors of cooperative or combinatorial regulation. We will summarize the corresponding results later in this section and present  detailed computations in the Supplementary Information.

We consider a cell that contains $M$ genes, which need to be transcriptionally regulated. In the basic model, each gene is associated with a single binding site of length $L$ basepairs, and a unique kind of TF, which---in environments where the TF is expressed---preferably binds to that binding site to activate transcription.
We assume that the genes are inactive, unless a TF binds to their binding site.
We later relax this simplification to cases where each TF regulates several genes.
Every TF can also bind other (noncognate) binding sites, albeit with lower probability, as schematized in Figs~\ref{Fig:crosstalk}A,\,B. These noncognate interactions will contribute to crosstalk in our model.

We employ a thermodynamic model of regulation~\cite{ackers_quantitative_1982,von_hippel_specificity_1986,lynch_evolutionary_2015}, which postulates that the gene expression level depends on the equilibrium occupancy of TFs at the regulatory sites  on the DNA. This model has been widely used to predict gene expression and has been experimentally validated in various systems~\cite{he_thermodynamics-based_2010,sherman_thermodynamic_2012,Fakhouri_deciphering_2010}. In this framework, the binding probability of a TF to any binding site, cognate or noncognate, is determined by two factors: the effective concentration of TFs, and the binding energy.

We assume that the binding energy only depends on the number of mismatches  between a
particular binding site and the consensus sequence unique to the
given TF. Each binding site can thus exist in either of the three possible states \cite{von_hippel_specificity_1986}: {(i)} bound by a cognate TF; {(ii)} bound by a noncognate TF; or {(iii)} unbound. Binding of the cognate factor {(i)} is energetically the most favorable state and is assigned the energy $E=0$. The unbound state {(iii)} is usually energetically least favorable with energy $E_a>0$. Between these two extremes there exist
noncognate-bound configurations {(ii)} with intermediate energies that depend only on
the number of nucleotide mismatches $d$ between the consensus sequence of the TF and the sequence of a given binding site, i.e., $E(d) = \epsilon\,d$, where $\epsilon$ is the energy per mismatch. This mismatch energy model provides a tractable approximation to more detailed models~\cite{kinney_using_2010}, and has been extensively used in the literature~\cite{berg_selection_1987,gerland_physical_2002}.

Gene regulation gives cells the ability to differentially activate subsets of their genes in a manner appropriate to the environmental conditions, signals, cell type, or time. In our basic model, we imagine a cell that responds to different environments by activating different subsets of $Q$ genes (out of a total of $M$ genes),
while keeping the remaining $M-Q$ genes inactive (see Fig~\ref{Fig:crosstalk}C). As regulation unfolds, the regulatory network thus switches between equilibrium states where any choice of $Q$ out of $M$ genes could be activated; to make the problem tractable, we assume that all these choices are equally probable. In a given environment, activating a particular subset of $Q$ out of $M$ genes is achieved by expressing the corresponding $Q$ TF types. The remaining $M-Q$ TF types, corresponding to the genes that should remain inactive, are absent in the cell.

How does the cell express the correct set of TFs for any particular environment, and at what concentrations are these TFs expressed? The issue is made seemingly even more complicated by the fact that the TF concentration reflects the total number of TF molecules in the cell, as well as any possible effects due to nonspecific TF localization or sequestration on the DNA and elsewhere~\cite{burger_abduction_2010, Sheinman_how_2012,weinert_scaling_2014}. What we will show below is that even if the TF presence and concentrations were perfectly adjusted to the environment, a residual level of crosstalk---representing a lower bound or intrinsic limit---is inevitable. Since we are interested precisely in this limit, we will not need to specify the mechanisms by which cells control their TF concentrations, which likely involve complex regulatory network dynamics with feedback loops; instead, we will mathematically look for the lowest achievable crosstalk and show that even in an optimal scenario crosstalk can present a serious regulatory problem.

In our model, the crosstalk error can be separated into two contributions that can be computed using basic statistical mechanics:
\begin{enumerate}
\item For a gene $i$ that should be {\bf active} and whose {\bf cognate TF is therefore present}, error occurs if its binding site is bound by a noncognate regulator (activation out of context due to crosstalk), or if the binding site is mistakenly unbound (gene is inactive). This happens with probability
\begin{equation}
x_1(i) = \frac{e^{-E_a} + \sum_{j\neq i} C_j e^{-\epsilon d_{ij} }}
{C_i + e^{-E_a} + \sum_{j\neq i} C_j e^{-\epsilon d_{ij}}},
\label{x1}
\end{equation}
where $C_j$ is the concentration of the $j$th TF, $d_{ij}$ is the number of mismatches between the $j$th TF consensus sequence and the binding site of gene $i$, and $\epsilon$ the energy per mismatch; all energies are measured in units of $k_BT$. Here we consider activation by a non-cognate TF as crosstalk; reasons for this choice, as well as an alternative model where such cross-activation is not considered an error state, are presented in SI Section~4.
\item For a gene $i$ that should be {\bf inactive} and whose {\bf cognate TF is therefore absent}, crosstalk error only happens if its binding site is bound by a noncognate regulator (erroneous activation) rather than remaining unbound. This happens with probability
\begin{equation}
x_2(i) = \frac{\sum_{j\neq i} C_j e^{-\epsilon d_{ij}}} {e^{-E_a} + \sum_{j\neq i} C_j e^{-\epsilon d_{ij}}}.
\label{x2}
\end{equation}
\end{enumerate}
%


We define the global crosstalk $X$ as the expected fraction of erroneously regulated genes.
In our basic model where all genes are identically regulated and TFs for genes that need to be activated are present at equal concentrations (i.e., $C_j=C/Q$, where $C$ is the total concentration of all TFs and $Q$ is the number of distinct TF species present simultaneously), we show in the SI that the crosstalk is
\begin{equation}
X=\frac{Q}{M}x_1 + \frac{M-Q}{M}x_2. \label{rerror}
\end{equation}
%
Global crosstalk $X$ ranges between zero (no erroneous regulation) and one (every gene is mis-regulated).
%

\begin{figure}
\centerline{\includegraphics[width=\textwidth]{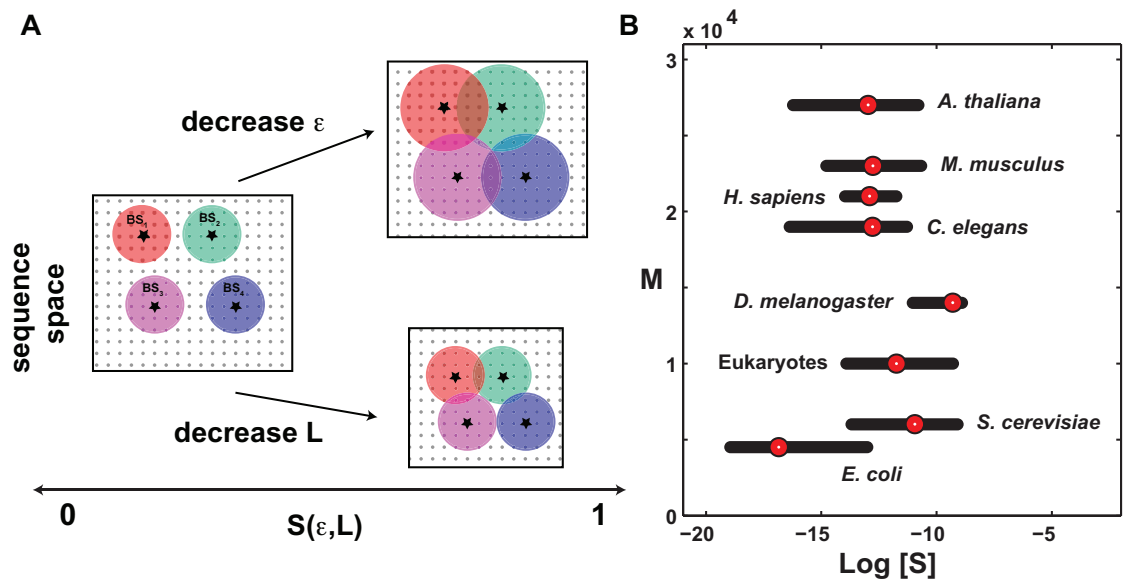}}
\caption{\textbf{Binding site similarity $S$ and number of genes $M$  are basic determinants of crosstalk.}
{\bf (A)} Binding site similarity, $S(\epsilon,L)$, determines the likelihood that a TF will bind noncognate sites, if recognition sequences are of length $L$ and the energy per mismatch  is $\epsilon$. A schematic diagram of sequence space packing by different TFs: sequences (dots) in a colored circle are likely to be bound by the TF whose consensus is the circle's center star. Smaller $L$ contracts the sequence space and makes crosstalk (circle overlap) more likely (larger $S$); crosstalk is increased (larger $S$) also by smaller $\epsilon$, which expands the circle radius.
 {\bf (B)} Typical values for the number of genes, $M$, and binding site similarity, $S(\epsilon,L)$, across different taxa, estimated from genomic databases. For each organism, we find a distribution of $S$ over its reported TFs (dots = median of the distribution, black bars = $\pm 1$-quartile range; see SI Section~5.4 for details). 
}\label{Fig:S}
\end{figure}

The major determinant of crosstalk is the likelihood of TFs to bind noncognate sites, which is determined by the similarity between cognate and noncognate sites. In the global setting, making a particular site less similar to all the remaining sites can only happen at the cost of making the remaining sites more  similar amongst themselves. For a large number of sites we describe this effect by introducing an \emph{average binding site similarity measure}  $S_i$ between the binding site of gene $i$ and all others, defined as:

\begin{equation}
\sum_{j\neq i} C_j e^{-\epsilon d_{ij} }\equiv C S_i(\epsilon,L) \approx C \sum_d P(d) e^{-\epsilon d},
\label{S}
\end{equation}
where $P(d)$ is the distribution of mismatches between all pairs of binding sites in our model and $C$ is the total concentration of all TFs. In the following we assume full symmetry between the genes, so that for every $i$, $S_i=S$. $S$ depends solely on the binding sites, but it carries no functional meaning in the absence of any TF, namely when $C=0$. We emphasize that this quantity, $S$, is not arbitrary, but rather emerges from our calculations in Eqs~(\ref{x1},\ref{x2}); a related measure of the likelihood of olfactory or immune receptors to bind an arbitrary ligand from a large repertoire has been previously introduced and measured \cite{lancet_probability_1993}.
$S_i$ is  proportional to the probability of the $i$-th TF to bind any noncognate binding site. The highest level of similarity, $S=1$, occurs if all sites are identical ($d=0$). Similarity is very low, $S\approx 0$, if the sites are all significantly different from each other.
The shorter the binding site length $L$ is and the weaker the binding energy $\epsilon$, the larger $S$ gets and the less distinguishable the sites are (Fig.~\ref{Fig:S}A); simultaneously, we expect the crosstalk to increase, an intuition we will make precise in the following section.

Binding site similarity $S(\epsilon,L)$ of Eq~(\ref{S}) could be directly measured, by experimentally probing the average TF binding affinity to a large repertoire of known binding sites. Alternatively, $S$  can be estimated from bioinformatic data.
In Fig~\ref{Fig:S}B we used databases of known TF binding sites to extract organism-specific estimates for $S$.
Under certain assumptions about how binding sites are organized in sequence space, $S$ can be also computed theoretically. If the binding sites were random sequences of length $L$,
one can derive a simple analytical expression for $S$ (see SI): $S(\epsilon,L)=\left(\frac{1}{4} + \frac{3}{4}e^{-\epsilon}\right)^L $.
We also studied more realistic  models for how TF binding sites are organized, e.g., taking into account the possibility of TFs to bind reverse-complemented sites (SI Section~5.2); an improved biophysical model for mismatch energy that saturates with the number of mismatches (SI Section~5.3); and a model of binding sites that have evolved to be maximally distinct~(SI Section~5.1).
All these variations ultimately only affect the value of $S$ while leaving the crosstalk formalism unchanged. We therefore carried out our main computations as a function of $S$ directly. To estimate typical crosstalk for values of $S$ that are biophysically realistic, we assumed that binding sites are distributed as randomly as possible in the sequence space while avoiding excessive similarity (i.e., we used the results of Fig~S14 with $d_{\rm min}=2$).

\begin{figure}
\centerline{\includegraphics[width=\textwidth]{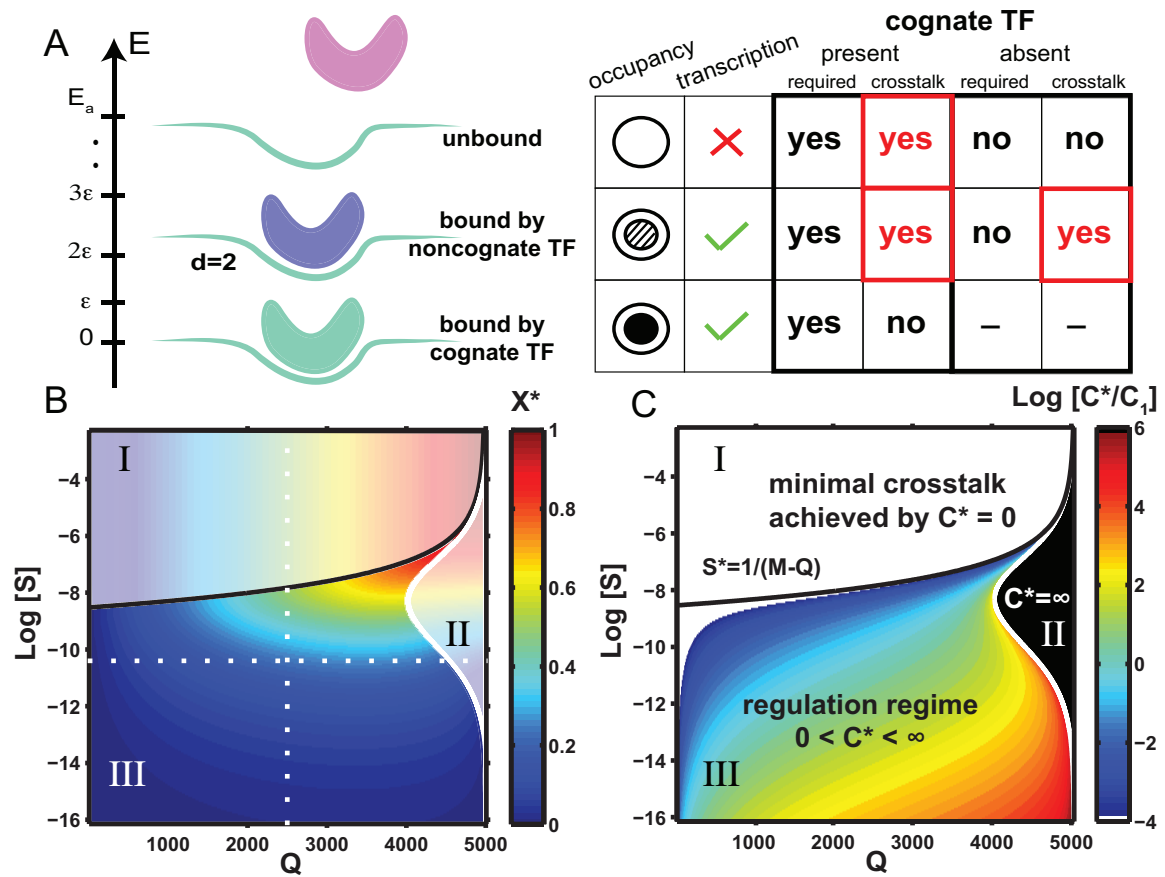}}
\caption{\textbf{Basic model with one activator binding site per gene exhibits three distinct regulatory regimes.}
{\bf (A)} Each binding site can be in either of the three possible states with different corresponding energies: bound by a cognate factor ($E=0$, green molecule), bound by a noncognate factor with $d$-mismatches ($E=\epsilon\, d$, here a blue molecule with $d=2$), or unbound ($E=E_a$, pink molecule). The table shows which of these states lead to transcription and which of these outcomes is considered as crosstalk
when the cognate TF is present and the gene is required to be active (left), 
or if it is absent and the gene is required to be inactive (right). 
{\bf (B)} Minimal crosstalk 
$X^*$, shown in color, as a function of the number of coactivated genes $Q$ and binding site similarity, $S$.  Three different regulatory regimes are separated by black and white boundary lines (analytical expressions in SI), identical between panels (B) and (C). Dotted lines refer to the ``baseline parameters'' ($Q=2500, M=5000, \log(S)=-10.5$ - represents $L=10$, $\epsilon=2$ with $d_{\rm min}=2$.) that we use in all subsequent figures if not specified differently. {\bf (C)} Optimal TF concentration, $C^*$, that minimizes the crosstalk, relative to $C_1$, the optimal concentration at baseline parameters. For high binding site similarity (large $S$), the crosstalk
is minimized at $C^*=0$ (white region, I: ``no regulation regime''). For $Q\rightarrow M$ and intermediate $S$, the crosstalk 
is minimized at $C^*\rightarrow \infty$ (black region, II: ``constitutive regime''). In a large, biologically plausible intermediate regime, crosstalk
is minimized at a finite nonzero TF concentration (color, III: ``regulation regime'').}
\label{Fig:BasicModel}
\end{figure}

\subsection*{Basic crosstalk model exhibits three distinct regulatory regimes}
While we can reasonably estimate the major determinants of crosstalk in our model---the number of genes typically coactivated, $Q$, the total number of genes, $M$, and the binding site similarity $S$---it is harder to determine the appropriate value for the total concentration of available TFs, $C$. This is not only due to the lack of quantitative data, but also because the relation between the total copy number of TFs in a cell and the concentration of TFs that are available for binding may be complicated~\cite{burger_abduction_2010}. We thus opted for an alternative approach: we look for a concentration $C^*$
that minimizes the crosstalk error $X$. An optimal $C^*$ emerges as a trade-off between activating the $Q$ genes that should be active (for which a higher concentration is beneficial) and avoiding the activation of the remaining $M-Q$ genes (for which a lower concentration is beneficial). Such a minimum, $X^*=X(C^*)$, is
a lower bound on crosstalk, which can be analytically computed in the mean field approximation (SI Section~1), as well as validated numerically by simulation (SI Section~2). This level of crosstalk cannot be decreased even if a cell could perfectly adjust its TF concentrations to the environment and optimally compensate the concentrations for nonspecific binding and sequestration.

First, we consider a fixed number of total genes, $M=5\,000$, and ask how crosstalk depends on the number of co-activated genes, $Q$, and the binding site similarity, $S$, in our basic model, summarized in~Fig~\ref{Fig:BasicModel}A.
The optimization yields three distinct regulatory regimes, illustrated in Figs~\ref{Fig:BasicModel}B,\,C. For larger values of $S$ where binding sites are very similar, regulation is so non-specific that crosstalk
is minimized by having no TF at all, i.e., at $C^*=0$ (region I).
This regime, which happens whenever $S>1/(M-Q)$,  is dysfunctional and thus biologically implausible. Interestingly, the resulting fundamental limit to $S$, or to how similar binding sites can ever get while still permitting functional regulation, is set by $M-Q$, the typical number of genes that must remain inactive in each environment. This highlights the strong constraint on the regulatory system of keeping undesired gene activation levels low despite crosstalk interference.

As the organism tries to activate increasingly large subsets of genes in each environment and $Q$ increases, the optimal concentration $C^*$ climbs until we reach a regime where $C^*$ formally diverges (region II), shown in Fig~\ref{Fig:BasicModel}C and Fig~S2. In this limit, however, a biologically plausible solution would simply be to constitutively express the majority of the genes rather than using transcriptional regulation to do so, thus avoiding any possible crosstalk interference. This strategy might be applicable for organisms living in nearly constant environments, such as obligatory parasites.

Finally, there is a broad region (region III) in the $(S,Q)$ plane where crosstalk is minimized by a finite positive value for the optimal TF concentration. In this regime, which we call the ``regulation regime'' since it corresponds to the biological notion of regulation, crosstalk is given to a very good approximation by
\be
X^* = \frac{Q}{M}\left(-S (M - Q) + 2 \sqrt{S(M-Q)}\right).
\label{eq:X}
\ee
This simple expression for $X^*$ is one of our key results.  It is independent of the energy gap between cognate and unbound states, $E_a$; increasing this gap only lowers the optimal concentration, $C^*$, while leaving the crosstalk unchanged.
The crosstalk depends both on the fraction of genes that need to be activated,
$Q/M$, as well as on the total number of genes that need to be inactive, $M-Q$, in a typical environment. This dependence also suggests that it is costly to maintain genes that are never expressed, arguing against unlimited accumulation of obsolete genes in organisms. Crosstalk $X^*$ in the regulation regime is dominated by the second term of Eq~(\ref{eq:X}), and thus increases as $\sim \sqrt{S}$ and as $Q\sqrt{M-Q}/M$ 
for sufficiently small $S$.
At the boundary between regions I and III, where regulation breaks down, $S(M-Q)=1$, hence $X^*=Q/M$ and is independent of $S$ throughout region I, because all genes that need to be active are in a crosstalk state due to absence of TFs.  Alternatively, we can view Eq~(\ref{eq:X}) as a function of $M$, the total number of genes, at a fixed fraction of genes typically activated, $Q/M$. In that case we can see that the average binding site similarity $S$ sets the limit to the maximum number of  genes in the organism, $M \lesssim 1/S$, if the system is to stay in region III where regulation is effective. This is confirmed in Fig.~S1 by a detailed analysis of crosstalk for an organism with $M=20\,000$ genes, a typical number for a metazoan.

A quick inspection of Fig~\ref{Fig:BasicModel}B shows that crosstalk in the basic model is surprisingly high for an organism of $M=5\,000$ genes of which typically a half ($Q=M/2$) would be activated in each environment, and with TF specificity typical of metazoans ($\log(S)=-10.5$). At these ``baseline'' parameters, the crosstalk limit is $X^*\approx 0.23$, implying that almost a quarter of the genes at any time would be in an erroneous regulatory state. This suggests that global crosstalk is a serious constraint and that more complex regulatory mechanisms have evolved, in part, to permit reliable regulation despite noncognate TF binding. In what follows, we examine variants of the basic model to assess the robustness of our theoretical conclusions and compare, quantitatively, the crosstalk limit for different regulatory scenarios at our ``baseline'' parameter set. These results are detailed below as well as in the SI, and are summarized in Table~\ref{Tab:Models}.

\begin{table}[h]
\caption{
\label{Tab:Models}
Comparison of crosstalk levels between the different variants of the model. Baseline parameters are: $Q=2500$, $M=5000$, $\log(S)=-10.5$ - equivalent to an optimal packing model where distinct binding sites are different from each other in at least 2 bp ($d_{\min}=2$) with binding sites of $L=10$ bp and binding energy $\epsilon=2\;k_BT$ per mismatch. }
\small{
\begin{tabular}{|l||c|l|}
  \hline
  {\bf Model} & {\bf Crosstalk}   & {\bf Remarks} \\
              & (at baseline  & \\
              & parameters) & \\
 \hline \hline
  Basic model (activators-only) & 0.23 &  \\
  Basic model (repressors-only) & 0.23 &  \\
  Mixed model (activators + repressors) & 0.14 & 2000 genes expressed in $20\%$ of the env., \\
  & & 3000 genes in $70\%$. \\
      Genes of unequal importance & 0.31 & 10\% of the genes are important and penalized\\
                                  &      & $10\times$ the ``normal'' rate. \\
                                &      &The resulting error per important gene\\
                                &      & decreases to 0.1, \\
                                &      & but for the other genes increases to 0.33.  \\

  Unequal weights for the two error types & 0.17 & $b=0.5$, weight of erroneously-active genes is \\
  & & half that of genes that are erroneously inactive. \\
  Each TF regulates exactly $\Theta=10$ genes & 0.08 &  Also holds for $P(\Theta)\sim $ Poisson($\bar{\Theta}=10$).\\
  Activators +  global non-specific repressor & 0.23 &  cannot reduce crosstalk.\\
  Activators + specific repressors (non-overlapping) & 0.2 &  \\
  Activators + specific repressors (overlapping) & 0.15 &  \\
  Perfect AND-gate combinatorial regulation & 0.07	& Uses only $\sim\sqrt{M}$ TF species. \\
  Generic cooperativity  & 0.064 & e.g., dimerization, direct TF-TF contacts,\\
                          &      & TF/nucleosome competition, etc. \\
                                                    & & 2 bindings sites, each of length $L=10$. \\
  Cooperativity exclusive to cognate binding & 0.006 &  currently unknown molecular mechanisms\\
& &   2 bindings sites, each of length $L=10$. \\
   \hline
   \end{tabular}
   }
\end{table}

\subsection*{Crosstalk constraints exist also in variations of the basic model}
We first ask whether  the existence of regimes where regulation in the basic model is ineffective (region I and II) could be an artefact of penalizing expression of unnecessary proteins equally to the incorrect expression of the necessary proteins. To study this, we vary the relative contribution of the two components of crosstalk error, $x_1$ and $x_2$ from Eqs~(\ref{x1},\ref{x2}), to the total crosstalk, $X$. In Fig.~S5 we show that all three regimes reported for the basic model exist generically, although their boundaries may shift (see also Table~\ref{Tab:Models} and SI Section~1.3).

Next, we ask how our results change if all genes do not contribute equally to the total crosstalk, $X$. We thus split genes into two groups: ``important'' genes contribute to the crosstalk error more than ``normal'' genes, but---in order to compute the lower bound on crosstalk---we allow TFs of the basic model to redistribute optimally between the two groups. Fig.~S6 shows that in this scenario the crosstalk for important genes can be reduced, but only at a cost of increasing the crosstalk error for normal genes. Our theoretical framework can  be extended easily to treat multiple heterogenous groups of genes.

Next, we examine the situation where each TF can regulate more than one target gene. Specifically, the cell still contains $M$ genes in total out of which $Q$ need to be activated in each environment; in contrast to the basic model, each TF now activates groups of $\Theta$ genes, which are assumed to have identical binding sites (if the sites are not identical, one can show that the crosstalk only worsens). In this case the achievable crosstalk is lower than in the basic model, as expected: the regulatory network is trading off detailed control over individual genes for crosstalk improvement. Surprisingly, however, the crosstalk $X^*$ decreases only by a factor of $\sqrt{\Theta}$ (Table~\ref{Tab:Models}; see SI Section 1.5), making it unlikely 
that crosstalk constraints could be made negligible solely by implementing gene regulation at a very coarse level.

Finally, we modify our basic model to use repression instead of activation to regulate target genes. In the basic model, the default state for each gene is to be inactive, with transcription proceeding only when an activator is bound; in the modified model, the default state for each gene is to be expressed, which can be prevented by binding of a repressor. We find a simple mathematical relation between the crosstalk equations for the basic model and its repressor-only version (SI Section~1.2), showing that the repressor-only case exhibits the same three regulatory regimes and the same range of crosstalk values. One can also consider mixed models, where activation is used for some genes and repression for the others. Unless symmetry between genes is broken such that some genes need to be activated in more environments than other genes, crosstalk is minimized by pure strategies (using either only activators or only repressors); mixed strategies can become optimal when the symmetry is broken (see SI Section~3).

\subsection*{Crosstalk is not easily mitigated by complex regulatory schemes}
So far we considered the simplest cis-regulatory element architecture with a single TF binding site. Most genes, especially in eukaryotes, employ complex regulatory elements with multiple TF binding sites, some of which have been suggested in the literature to increase the effective binding specificity of TFs or protect the binding sites from spurious binding~\cite{bird_gene_1995,todeschini_transcription_2014}. By implication, such effects are expected to also reduce crosstalk. We next use our theoretical framework to study quantitatively under what conditions that may be the case.

\begin{figure}
\centerline{\includegraphics[width=\textwidth]{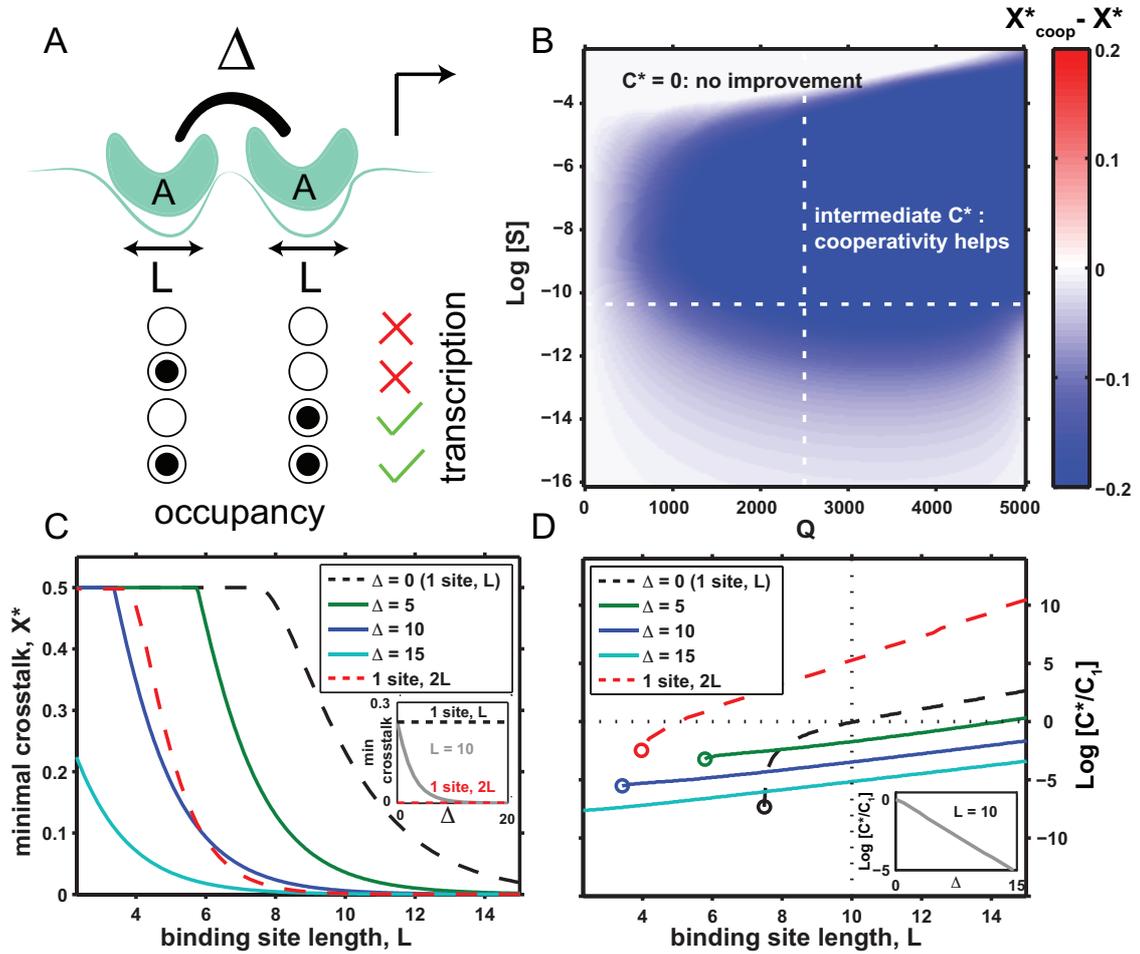}} 
\caption{\textbf{Cooperative regulation reduces crosstalk and the required optimal TF concentration.}
{\bf (A)} Cognate binding configurations (noncognate not shown) for two sites of length $L$ leading to transcription (green check) or not (red cross); doubly occupied promoter gains a cooperative energy $\Delta$. Transcription proceeds only when the proximal (rightmost) site is occupied.
{\bf (B)} Difference in minimal crosstalk, shown in color, 
between the cooperative model and the basic model of Fig~\ref{Fig:BasicModel}, $X^*_{\rm coop}-X^*$, for cooperative interaction strength $\Delta=10$. Cooperativity significantly reduces crosstalk (blue; at baseline parameters shown with white dashed lines, $X^*_{\rm coop}=0.006$ here vs. $X^*=0.23$ in the basic model) and shrinks the ``no regulation'' ($C^*=0$) regime.
{\bf (C)} Minimal crosstalk error, $X^*$, vs. binding site length $L$ for different values of cooperative energy $\Delta$ shows that strong cooperativity can decrease the crosstalk beyond the basic model with binding site of length $2L$ (red).
{\bf (D)} Optimal TF concentration, $C^*$, required to minimize crosstalk, decreases with increasing cooperativity $\Delta$ for all $L$, and is consistently below the single-site basic model with site length of either $L$ (black) or even $2L$ (red). Circles denote transition to the ``no regulation'' ($C^*=0$) regime at low $L$ (large $S$), showing that cooperativity extends the ``regulation regime.'' In {\bf (C)}-{\bf (D)} we convert $S$ values to the equivalent binding site lengths $L$ utilizing the random sequence model.
}\label{Fig:cooperativity}
\end{figure}

\textbf{Cooperativity.}
We extend our basic model such that each gene is influenced by two nearby binding sites of length $L$ to which cognate TFs can bind cooperatively. For simplicity we assume that cooperativity occurs between TFs of the same type, although the framework can be extended to more general cases. This molecular configuration of two cognate DNA-bound proteins is favored by an additional energy contribution $\Delta$. We assume that only one of the two sites controls transcriptional activity directly (here, the site proximal to the gene start, e.g., by polymerase recruitment~\cite{ackers_quantitative_1982}), while the other -- here, the distal site -- helps stabilize the binding to the proximal site, as schematized in Fig~\ref{Fig:cooperativity}A. In this model, as $\Delta$ goes to zero, the distal binding site has no effect on regulation, and we recover the basic model of regulation by a single binding site (Fig~\ref{Fig:BasicModel}).

To assess whether cooperative regulation can reduce crosstalk, we compute the minimal achievable crosstalk, $X^*_{\rm coop}$, and compare this in Fig~\ref{Fig:cooperativity}B with the minimal crosstalk of the basic model, $X^*$. We find that cooperativity can significantly reduce crosstalk in a large part of the ``regulation regime,'' which itself extends towards larger $S$. Examining in detail how the crosstalk behaves in Fig~\ref{Fig:cooperativity}C, we see that at a fixed binding site length $L$, minimization of crosstalk prefers strong cooperativity $\Delta$; nevertheless, the improvement in crosstalk is bounded and as $\Delta$ grows, saturates at a limiting value. In this limit, crosstalk
can approach and even drop below the crosstalk of the basic model with a binding site which is twice as long.
This is a relevant comparison because cooperative regulation does, in fact, have access to a total of $2L$ base pairs of recognition sequence. Furthermore, the optimal TF concentration $C^*$ required in the cooperative case is \emph{lower} than in the single site case (Fig~\ref{Fig:cooperativity}D), making cooperativity a realistic crosstalk reduction mechanism.

The crucial assumption of the cooperative model presented above is that cooperative interaction between two TF molecules can only occur when they bind their cognate binding sites and never otherwise. This is a very restrictive assumption that is unlikely to hold in many documented models of cooperativity. For example, if cooperative interaction energy $\Delta$ originates in protein-protein interactions between the two TF molecules of the same species, this energy will plausibly be gained even when these same TF molecules come into contact while binding two nearby noncognate sites. Similarly, synergistic activation~\cite{todeschini_transcription_2014} or nucleosome-mediated cooperativity~\cite{mirny_nucleosome-mediated_2010} models also imply that noncognately-bound factors could contribute towards cooperativity, violating our assumption that cooperativity is exclusive to cognate binding.

To relax this assumption and study the effects of the resulting ``noncognate cooperativity,'' we recompute accordingly the crosstalk improvement relative to the basic model, as shown in Fig~S19.  Not surprisingly, we find that allowing cooperative interactions between TFs of the same type when bound noncognately leads to much smaller reductions in crosstalk compared to cooperativity that is exclusive to cognate binding, as shown in Table~\ref{Tab:Models}.
When noncognate cooperativity is allowed, we can also look at the strong cooperativity (large $\Delta$) limit and compare crosstalk improvement due to two TFs cooperatively binding two sites of length $L$, to the basic model of a single TF binding a site of length $2L$. Now, cooperative regulation by two TFs is always inferior to the regulation by a single factor with a longer binding site (see SI Section~6).

Dimerization of TFs is very common among prokaryotes, where TF monomers often dimerize in solution before binding to DNA. If the two binding sites in our model predominantly act as half-sites for the binding of a single dimer, the relevant equations for crosstalk are identical to noncognate cooperativity in the large $\Delta$ limit, with $C$ being the concentration of monomers. Our theory is thus also applicable to this case, although dimerization in solution is often not considered a canonical example of cooperative regulation. Cooperative interactions conditional on DNA binding have been less frequently reported but are also known to occur in prokaryotes (e.g., on proximal binding of two dimers); in experimentally documented cases, the interaction energies are weaker, $\Delta\sim 3\;k_BT$ \cite{ackers_quantitative_1982}, which still facilitates crosstalk reduction although it is accordingly smaller  (Fig~S19).


The two cases of cooperativity we considered here represent two extremes of a spectrum: cooperative interaction is either possible exclusively at the cognate site, or at all sites equally. There likely exist intermediate situations which help limit the occurrence of spurious cooperative interactions. A simple example of such a mechanism could utilize the positioning of the binding sites on the DNA: TF cooperative binding is limited only to pairs of sites which are appropriately spaced. If different TF types use different spacing, the harmful effects of cooperativity at a particular noncognate site pair will be restricted to a subset of TFs. More complex geometrical arrangements, e.g., cooperative interactions involving DNA looping or allosteric effects between the two TFs and the DNA \cite{merino_cooperative_2015}, could provide similar benefits. While possible in principle, these benefits should be considered as hypothetical, since direct experimental support for cooperativity that is exclusive to cognate binding is still lacking.

\begin{center}
\includegraphics[width=\textwidth]{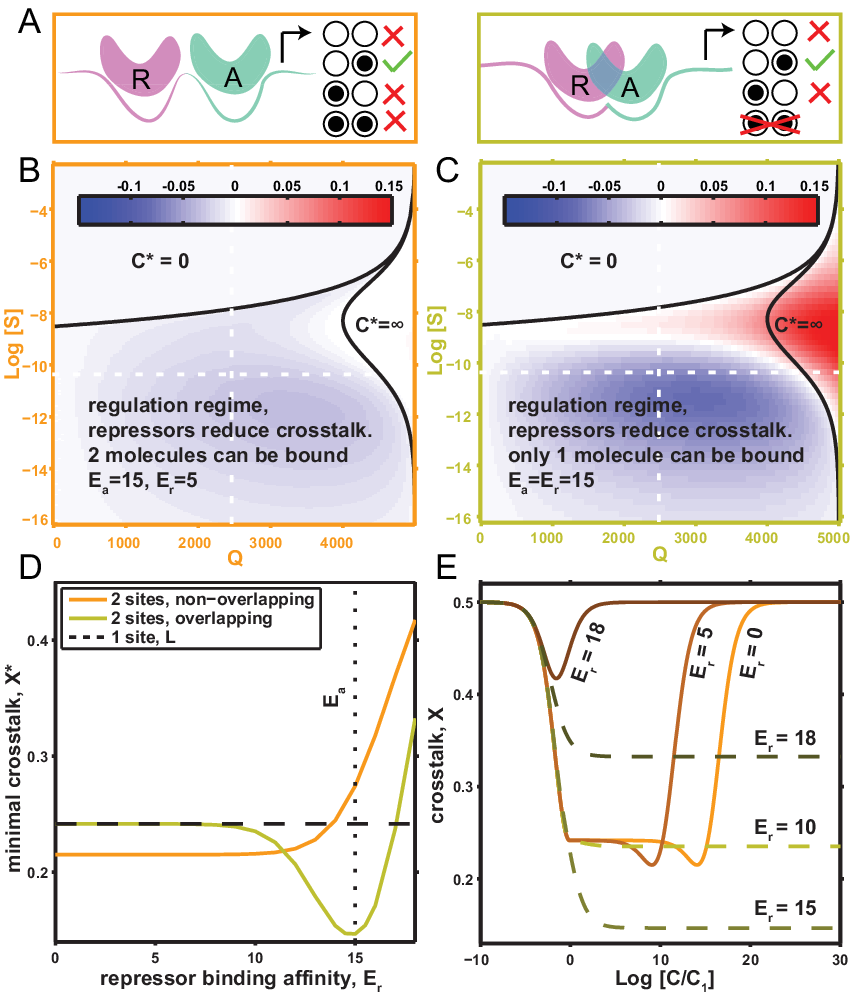}
\captionof{figure}{
\textbf{Combinatorial regulation by activators and repressors yields marginal improvements in crosstalk error.}
{\bf (A)} Separate (left) or overlapping (right) binding sites for activators A and repressors R. A subset of binding configurations for cognate regulators is shown; transcription proceeds (green) only when the A site is bound by the cognate activator 
and the R site is unbound.
{\bf (B,C)} Difference (shown  in color) between minimal crosstalk achievable with activator-repressor regulation, and the basic model of Fig~\ref{Fig:BasicModel}.
With optimal value for the affinity of repressor sites ($E_r$) selected in both cases, a small overall improvement in crosstalk error is seen in (B), and a larger improvement, but localized to $\log S\lesssim -10$, in (C).
At baseline parameters (white dashed lines), $X^*=0.2$ for the non-overlapping case, $X^*=0.15$ for the overlapping case and $X^*=0.23$ in the basic model.
{\bf (D)} Dependence of the crosstalk 
on the repressor binding affinity $E_r$ (activator affinity fixed at $E_a=15$). When $E_r>E_a$, the crosstalk quickly increases: instead of helping prevent erroneous activation, repressors themselves bind too frequently in noncognate configurations, aggravating the crosstalk. For non-overlapping sites scenario, $E_r\ll E_a$ is optimal, whereas in the overlapping sites case, $E_r=E_a$ is optimal.
{\bf (E)} Dependence of crosstalk 
on the total concentration, $C$, of transcription factors, for non-overlapping sites case (orange-brown curves representing different $E_r$, as indicated) and overlapping sites case (green curves representing different $E_r$, as indicated). The total concentration is optimally split between activators and repressors for each $C$, and is reported relative to the optimal concentration $C_1$ of the basic model.
\label{Fig:repressor}}
\end{center}

\noindent\hrulefill

\textbf{Combinatorial regulation by activators and repressors. }
An important contribution to crosstalk is the erroneous activation of genes that should remain inactive. One might argue that any kind of global repression could alleviate this problem by preventing spurious transcription. We explored this scenario by extending our basic model to include an additional nonspecific repressor (SI Section~8). Perhaps not surprisingly, we find that the minimal achievable crosstalk error in this extended scheme is exactly the same as in the basic setup, regardless of the concentration and the affinity of the sites.


We next turned our attention to a sequence-specific repression mechanism. In an extension to our basic model, we equipped each gene with both an activator and a repressor site, such that each of these sites has its own cognate regulator (activator or repressor). For the $Q$ genes that should be active, only their $Q$ cognate  
activators (but not repressors) were present. For the remaining $M-Q$ genes that should be inactive only their cognate  
repressors (but not activators) were present. Repressor sites could have a different affinity ($E_r$) than the activator sites ($E_a$). To look for the minimal achievable crosstalk, we optimized over the concentration of activators, repressors, and the affinity $E_r$. Importantly, we considered two possible molecular arrangements on the promoter: in the \emph{non-overlapping sites} scenario (Fig~\ref{Fig:repressor}A, left) the two binding sites could be occupied by regulatory molecules simultaneously, whereas in the \emph{overlapping sites} scenario (Fig~\ref{Fig:repressor}A, right), either the activator or repressor site, but not both, could simultaneously be occupied. Whether this exclusion happens because the two binding sites literally overlap or due to more complex mechanisms is not crucial for our results. We assumed that a bound repressor inactivates transcription, regardless of the activator state; for a detailed list of molecular configurations on the promoter, see SI Section~9.
In the non-overlapping case, small ($\sim 10\%$ at baseline parameters) decreases in crosstalk error are nominally possible, as shown in Fig~\ref{Fig:repressor}B. A detailed examination, however, argues against this mechanism for crosstalk reduction. Optimization in Fig~\ref{Fig:repressor}D assigns the repressor sites a very weak, or even vanishing, affinity for the TFs, $E_r\ll E_a$: in essence, the repressor sites energetically favor staying empty to the same amount as binding a cognate repressor, to fight off noncognate binding. As a costly consequence, the optimal concentration of the required TFs needs to be larger by an unreasonable factor, $\sim 20\,000$-fold, relative to the basic model, to achieve this small crosstalk reduction gain.

The overlapping case provides a greater crosstalk reduction ($\sim 35\%$ at baseline parameters), as shown in Fig~\ref{Fig:repressor}C. The optimal repressor sites have similar affinity to their cognate TFs as do the optimal activator sites, $E_r\sim E_a$; the benefit of the repressors quickly vanishes if this condition is not met. The total required regulator concentration now no longer has a clearly defined optimum, but does exhibit a  plateau where the crosstalk is minimized. Importantly, as shown in Fig~\ref{Fig:repressor}E, this plateau is reached for concentrations only somewhat higher than in the baseline case, making this solution biologically plausible.

In sum, the case for combinatorial regulation by activators and repressors is complicated.
Combinatorial regulation provides a smaller absolute improvement than cooperativity, but this improvement is also centered around smaller values for binding site similarity, $\log(S)\lesssim -10$, where the crosstalk of the basic model is itself already lower. In contrast to our initial expectation, this small gain is realistically achievable only with one of the two regulatory schemes considered, and only when its parameters are properly tuned.

\textbf{AND-gate combinatorial regulation.}
Lastly, we considered the simplest AND-gate regulation scenario. The expression state of each gene is determined by the occupancy of two binding sites; in particular, activation is achieved by binding of a precisely specified, unique pair of cognate activating TFs. Crucially, in the ``perfect combinatorial regulation'' scenario, $\sqrt{2 M}$ TF species (instead of $M$, as in the basic model) are sufficient to specifically regulate any subset of the $M$ genes.
As we show in SI Section~7 and summarize in Table~\ref{Tab:Models}, this leads to a sizeable crosstalk reduction. Using $\sqrt{2 M}$ TF species means, on average, $\Theta = M/\sqrt{2M}$ regulated genes per TF. If sets of $\Theta$ genes were regulated jointly by a common TF, crosstalk should decrease as $\sim \sqrt{\Theta}$, as we argued above.  Figure~S21 shows that for the AND-gate the decrease is somewhat smaller, but unlike in the simple scenario where each TF regulates groups of $\Theta$ genes with no possibility of control over individual genes, the AND-gate allows each gene to be regulated individually. While this combinatorial strategy allows crosstalk reduction and has been documented at specific promoters, we point out that the predicted, square-root scaling of the number of TF species with the total number of genes, $M$, is inconsistent with published reports~\cite{nimwegen_scaling_2003,maslov_toolbox_2009}, making it unlikely that crosstalk reduction is achieved through genome-scale combinatorial control as analyzed here.

\section*{Discussion}
Finite specificity of recognition reactions is a fact of life at the molecular scale.
In transcriptional regulation, which takes place in a mix of cognate and noncognate transcription factor species, the consequences of this fact could be severe---but have surprisingly not been taken to their logical conclusion so far. Here, we constructed a theoretical framework for crosstalk that accounts for all possible cross interactions between regulators and their binding sites. This global model enabled us to compute the lower bound on crosstalk and assess the effectiveness of various regulatory schemes. We derived limits to reliable gene regulation that depend only on the total number of  genes $M$, the typical number of co-activated genes, $Q$, and the average level of similarity between pairs of binding sites, $S$.

We find that these parameters robustly define three possible regulatory regimes. A nonzero TF concentration that minimizes crosstalk exists only when binding sites are sufficiently distinguishable from each other and the typical number of co-activated genes is not extreme. We call this the ``regulation regime.'' The other two regimes are anomalous cases where regulation is dysfunctional. 
Looking closely at the boundaries between the three regulatory regimes, we find that  the average similarity between binding sites, $S$, puts an upper bound to the total number of genes that an organism can effectively regulate~\cite{itzkovitz_coding_2006}.

An analogous problem exists in protein-protein interaction networks, where protein function requires strong binding to a few partner-proteins but avoidance of binding to all the others~\cite{zhang_constraints_2008,johnson_nonspecific_2011}. Previous works have studied the evolution of such networks by applying a  combination of positive and negative design using computer simulations, concluding that ``negative design'' seriously constrains the possible architectures~\cite{sear_highly_2004,sear_specific_2004,myers_satisfiability_2008,johnson_nonspecific_2011}. As a quantitative measure for the likelihood of specific vs. nonspecific interactions, Johnson \etal~used the minimal energy gap between specific and nonspecific interactions, in analogy to our measure of binding sites similarity $S$. They found a power-law scaling of the energy gap with the total number of proteins in the network and also found that it  depends inversely on the size of binding surface, $L$ -- both results are in qualitative agreement with ours for the total number of genes $M$ and length of the binding sites $L$. Similarly, a larger binding domain was found to enable a larger number of specific interactions in a protein mixture when other nonspecific interactions are excluded~\cite{sear_specific_2004}. Johnson \etal~also found that network designs in which some proteins have multiple specific partners (``hubs'') have higher crosstalk compared to networks with only pairwise interactions. At this point protein-protein interaction networks significantly differ from TF-DNA interactions: if multiple binding sites share a common TF, these binding sites cannot bind each other, as would be the case for different protein species interacting with a common hub. Zhang \etal~ identified a trade-off between proteome diversity and concentration due to crosstalk considerations, concluding that the numbers found experimentally are close to the possible limit~\cite{zhang_constraints_2008}. Protein concentrations face trade-off: they should be high enough to form specific interactions, but not so high as to form many non-specific ones. The optimal TF concentration in our model is determined by a similar trade-off. Analogous problems due to explosion of non-cognate configurations were studied in the context of prebiotic metabolism~\cite{schuster_taming_2000} and the immune system, where receptors are selected to recognize foreign peptides, but avoid binding self-peptides~\cite{kosmrlj_how_2008}. In the context of TF-DNA interactions Sengupta \etal~\cite{sengupta_specificity_2002} studied how the evolutionary mutation-selection balance tunes TF specificities to its DNA targets and how this depends on the number of targets. They identified a trade-off between avoiding the loss of current targets (for which a lower specificity is favored) and avoiding the spurious recruitment of new ones (for which a higher specificity is favored); they also report an inverse relation between the number of different targets and the TF specificity for each. An intriguing direction for future research is to explore how crosstalk might limit the complexity of regulatory networks in an evolutionary setting.

\begin{table}[h]
\caption{
\label{Tab:ProkEuk}
Comparison of relevant parameters and crosstalk values between prokaryotes and eukaryotes. }
\small{
\begin{tabular}{|l|c|c|}
  \hline
                                            & prokaryotes                    & eukaryotes              \\
                                            &&                                                          \\
  \hline
   binding site length                      & 10-20 bp                       & 6-10 bp                   \\&&\\
  binding site similarity, $S$              & $-20\lesssim\log(S)\lesssim-13$ & $-15\lesssim\log(S)\lesssim-9$                  \\&&\\
  number of genes, $M$                     & a few thousands      & $5,000-20,000$                   \\&&\\
  crosstalk in the basic model             & $1\%-10\%$           & 20\%-50\%  (depending on $M$)      \\&&\\
  crosstalk with cooperative regulation & $<1\%$                  & $\sim 10\%$                      \\
  \hline
\end{tabular}
}
\end{table}

Where do real organisms find themselves in this parameter space?
Prokaryotes tend to have longer binding sites and fewer genes than eukaryotes. In Table~\ref{Tab:ProkEuk} we present typical biophysical parameters for each and the resulting crosstalk estimates. While for prokaryotes we expect crosstalk to easily be between $1\%$ and $10\%$ even if each gene is regulated by a single site, and below $1\%$ for biophysically realistic  cooperative regulation, for eukaryotes the situation is significantly different. Even for a short genome of $M=5\,000$ genes, such as yeast, or for longer genomes of metazoans where most of the genes have been non-transcriptionally silenced, we expect minimal crosstalk of $X^*=0.23$. In an organism with $M=20\,000$ regulated genes crosstalk would increase substantially according to the basic model: more than 40\% of all genes would be erroneously regulated. Incorporating known constraints on the biophysics of TF-DNA interaction (Figs.~S16, S17) increases crosstalk even further and pushes metazoan regulation towards the anomalous regime.


Complex regulatory schemes increase the specificity of gene regulation by cognate factors, and high specificity was tacitly assumed to provide automatic resilience against crosstalk. In contrast, our  analysis of several complex regulatory mechanisms  reveals a more intricate picture. We focused on two broad classes of regulatory mechanisms. The first class comprises various schemes of cooperative regulation.  Cooperativity can lower crosstalk because it effectively increases the binding site length and energy and thus reduces binding site similarity. We found that the effectiveness of cooperativity for reducing crosstalk crucially depends on the strength of the cooperative interaction and on whether cooperative interactions are restricted exclusively to cognate sites. With respect to cooperative interaction strength, the optimal crosstalk reduction happens at very strong cooperativity, but this might be hard to realize biophysically. Commonly reported values are indeed small ($3-5\;k_BT$), comparable to the energetic contribution of only $1-2$ bp in the TF-DNA interaction~\cite{ackers_quantitative_1982,kinney_using_2010}. With respect to cooperative interactions being exclusive to cognate binding, such regulatory schemes, while optimal for crosstalk reduction, would require additional sequence recognition mechanisms, and it is unclear to what extent they exist or how effective they are. If cooperative interactions can occur at non-cognate sites as well, as is the case for most documented mechanisms of cooperativity, its effectiveness in mitigating crosstalk is significantly diminished. The second class of mechanisms we considered relies on combinatorial regulation by multiple TFs. As a representative example we studied combinatorial regulation by  activators and repressors. Contrary to the common expectation that repression should eliminate spurious gene activation~\cite{bird_gene_1995,todeschini_transcription_2014}, we found various mechanisms to be either ineffective (global repression) or providing marginal global improvement at best (activator-repressor regulation with overlapping binding sites).  While crosstalk can indeed be mitigated for particular gene(s) by employing a complex promoter architecture, this inevitably comes at a cost for the regulation of other genes.
The intuitive explanation for the limited benefit of combinatorial schemes is that adding new regulatory components---in this case, repressors and their respective binding sites---drastically increases the number of possible noncognate interactions, thereby potentially aggravating, instead of mitigating, the crosstalk problem.  A similar detrimental effect due to growth in the number of undesired configurations with the number of molecular species has been reported in the study of molecular self-assembly~\cite{murugan_undesired_2015}. A potentially powerful set of mechanisms are therefore schemes in which combinatorial regulation is used primarily to decrease the required number of molecular species, as in the simple AND-gate example we explored in SI Section~7. Further work is needed to fully elucidate crosstalk limits in more general models of combinatorial control and cooperativity,  with interesting parallels to precision in biochemical sensing, in equilibrium as well as out-of-equilibrium scenarios~\cite{govern_energy_2014,mora_physical_2015,skoge_chemical_2013,cepeda-humerez_stochastic_2015}.

An interesting result of our study is that various schemes of molecular control logic at promoters and enhancers~\cite{buchler_schemes_2003}, while nearly equivalent in the absence of crosstalk, can behave very differently in the presence of noncognate regulators~\cite{shinar_rules_2006}. For example, the issue of cooperative interactions during noncognate binding is a striking demonstration of how a seemingly microscopic detail may influence global crosstalk, while it has no bearing on the aspects for which cooperativity has been studied traditionally: its ability to sharply activate the cognate gene in response to small increases in TF concentration. 
A similar remark applies for the case of overlapping vs nonoverlapping binding sites in the combinatorial regulation scenario. By going beyond mean-field approximations, this could be extended to biologically relevant situations where pairs of binding sites overlap so as to share large sequence fragments~\cite{hermsen_transcriptional_2006}.  Clearly, there is a need to further understand signal processing
at complex promoters~\cite{rieckh_noise_2014}, and calls for experimental measurements of crosstalk in various regulatory architectures.

Direct measurements of crosstalk are challenging precisely because crosstalk is a global effect and experimentally influencing noncognate binding in a controlled manner is difficult. An alternative approach would be to search for indirect signatures of crosstalk~\cite{sasson_mode_2012}. A promising line of research supported by a large body of recent experimental evidence would be to examine ``pervasive transcription'' in eukaryotes~\cite{johnson_dark_2005,clark_reality_2011} as a proxy for erroneous initiation, perhaps due to crosstalk interference.

Taken together, our findings suggest that global crosstalk represents a strong constraint in eukaryotic regulation that can be mitigated, but not easily removed. Initially, this conclusion was based on a greatly simplified model of gene regulation. We succeeded in relaxing many of our assumptions only to find that crosstalk constraints remain significant. This is because the major determinant of crosstalk is the  binding site similarity $S$, which primarily depends on the typical mismatch energy $\epsilon$ and the length of the binding sites, $L$. While crosstalk could be reduced by extending binding site length and/or augmenting the binding energy, both parameters are severely constrained by a combination of biophysical and evolutionary factors. The scale of the mismatch energy is set by the energetics of hydrogen bonds to $\sim 2-4\;k_BT$, while the length of individual binding sites in eukaryotes appears strongly constrained by evolutionary considerations to $\sim 10$ bp~\cite{sengupta_specificity_2002,stewart_why_2012,tugrul_dynamics_2015}. Moreover, the performance of complex regulatory schemes, which appear beneficial at first glance, is also limited by the explosion of possible noncognate configurations that may lead to erroneous regulation. These constraints should apply universally, beyond the specific mechanisms we analyzed in detail: any regulatory scheme operating at equilibrium, no matter how complex, faces a fundamental limit to its achievable error, for reasons that led Hopfield to propose kinetic proofreading~\cite{hopfield_kinetic_1974}.

The main conclusion of our work is that crosstalk in gene regulation is far from being a solved problem. We find several commonly studied regulatory mechanisms to be insufficient for eliminating  
crosstalk in metazoans, at least when acting alone.  While it is theoretically possible that a combination of equilibrium mechanisms acting in unison could achieve low crosstalk levels, this possibility is by no means obvious and indeed appears unlikely. Alternatively, cells might have evolved out-of-equilibrium solutions where energy is deliberately spent to counteract the detrimental effects of crosstalk; example mechanisms  could include permanent gene silencing, localization of transcriptional activity to specific cellular compartments, or molecular reaction schemes for gene regulation that implement variants of kinetic proofreading~\cite{cepeda-humerez_stochastic_2015}.

\textbf{Acknowledgments}
The research leading to these results has received funding from the People Programme (Marie Curie Actions) of the European Union's Seventh Framework Programme (FP7/2007-2013) under REA grant agreement Nr. 291734 (T.F.), ERC grant Nr. 250152 (N.B.), and Austrian Science Fund grant FWF P28844 (G.T.). We thank Rok Grah, Tiago Paix\~{a}o, Georg Rieckh, Thomas Sokolowski and Marcin Zagorski for critical reading of the manuscript, and Dominik Schr\"oder for valuable insights to our mean-field approximation.

\renewcommand{\thefigure}{S\arabic{figure}}   
\renewcommand{\theequation}{S\arabic{equation}}   
%
%

\clearpage

\baselineskip = 14 pt \leftline{\large \bf Intrinsic limits to gene regulation by global crosstalk -- Supporting Information}
\bigskip
\baselineskip = 14 pt \leftline{Tamar Friedlander, Roshan Prizak, C\u{a}lin C. Guet, Nicholas H. Barton and Ga\v{s}per Tka\v{c}ik}
\bigskip
\leftline{\today}
\hrule\bigskip\bigskip



\section{Basic model -- analytical solution}
\label{sec:BasicModel}
We assume that the genome of a cell contains $M$ ``target'' genes, each of which is regulated by a single unique transcription factor binding site (BS). In the basic formulation, there exist also $M$ distinct TF types, such that each TF can preferentially activate its corresponding target gene by binding to its binding site. At any point in time, however, not all $M$ TF types are present: we assume that only subsets of size $Q \leq M$ are present at some nonzero concentration, and that the optimal gene regulatory state for the cell would be to express exactly and only those genes for which the $Q$ corresponding TFs are present.

Let regulation be determined by the (mis)match between the binding site sequence and the recognition sequence of any transcription factor.
 Each binding site is  associated with a single TF type with which it forms a perfect match -- this is the cognate TF for the given binding site. However, each site could also occasionally be bound by other (noncognate) TFs, at an energetic cost of a certain number of mismatches.
Following earlier works \cite{von_hippel_specificity_1986, gerland_physical_2002}, we assume that the contribution of mismatches at individual positions in a binding site to the binding energy is equal, additive, and independent. We define the energy scale such that binding with cognate TF has zero energy and all other binding configurations have positive energies, proportional to the number of mismatches $d$, $E=\epsilon d$, where $\epsilon$ is the per-nucleotide binding energy. The unbound state has energy $E_a$ with respect to the cognate bound state. The different states and their energies are illustrated in Fig. 3A in the main text. We employ a thermodynamic model to calculate the equilibrium binding probabilities of cognate and noncognate factors to each binding sequence.

TFs can also be non-specifically bound to the DNA. These configurations only sequester TFs from free solution, but do not directly interfere with gene expression. As explained later, we will lump together the TFs freely diffusing in the solution, as well as nonspecifically bound TFs and any other TF ``reservoirs'' into one effective concentration of available TFs (equivalently, we work with the chemical potential of the available TFs using the grand-canonical ensemble).

Previous studies calculated the probability of a given transcription factor to be bound or unbound to certain DNA sequences \cite{gerland_physical_2002}.  These probabilities were calculated assuming that the site is vacant or bound by the TF under study, but not bound by TFs of other  types. This approach is cumbersome  when a large number of TF types are considered simultaneously, because the probability that the site is bound by other factors is non-negligible, and due to steric hinderance, a site cannot be bound by more than one molecule at any given time. Previous studies also proceeded by using the canonical ensemble. These two modeling choices together make the problem of many TFs binding to multiple binding sites coupled and not easily tractable, because one would need to enumerate all possible combinations of TF-BS states. However, an alternative and much simpler approach is to employ the grand-canonical ensemble, and calculate the binding probabilities for the binding sites, rather than for the TFs. The necessary assumption is that binding sites behave independently (e.g., they are sufficiently separated on the DNA so that binding at one site does not overlap the binding at another, or if it does, this is treated explicitly).
Underlying the grand-canonical ensemble is the assumption that TFs are present at sufficient copy numbers, so that  the binding of a single site under consideration does not appreciably affect the chemical potential of the remaining TFs. Experimental support for such decoupling and  the applicability of the grand-canonical approach has been  demonstrated recently~\cite{weinert_scaling_2014}.
In the following we assume equal concentrations of all TF types.

We distinguish two contributions to crosstalk:
\begin{enumerate}
\item
  For a gene $i$ that should be active and whose cognate TF
is therefore present,  error occurs if its binding
site is bound by a noncognate regulator (activation out of
context due to crosstalk), or if the binding site is unbound (gene is inactive).
This happens with probability
  \begin{equation}
x_1^i(\{C_j\}) = \frac{e^{-E_a} + \sum_{j\neq i} C_j e^{-\epsilon d_{ij} }}
{C_i + e^{-E_a} + \sum_{j\neq i} C_j e^{-\epsilon d_{ij}}},
\label{eq:x1}
\end{equation}

where $C_j$ is the concentration of the $j$th TF, $d_{ij}$ is the number of mismatches between the $j$th TF consensus sequence and the binding site of gene $i$,  $\epsilon$ the energy per mismatch and $E_a$ the energy difference between unbound and cognate bound states; all energies are measured in units of $k_BT$.


\item
For a gene $i$ that should be inactive and whose cognate
TF is therefore absent, crosstalk error only happens if
its binding site is bound by a noncognate regulator (erroneous
activation) rather than remaining unbound. This
happens with probability
\begin{equation}
x_2^i(\{C_j\}) = \frac{\sum_{j\neq i} C_j e^{-\epsilon d_{ij}}} {e^{-E_a} + \sum_{j\neq i} C_j e^{-\epsilon d_{ij}}}.
\label{eq:x2}
\end{equation}
\end{enumerate}

In general $x_{1,2}$ depend on the specific set of pair-wise distances $d_{ij}$ between the consensus sequence of each TF present and the site of gene $i$. Hence they could vary between genes, and even for each gene different sets of TFs can yield different values of crosstalk. In the following we assume a fully symmetric setup, such that all genes are equivalent in their sensitivity to crosstalk ($x_{1,2}$ is independent of $i$). We assume that for each gene the mismatches $d_{ij}$ of all the noncognate TFs are distributed according to a probability density $p(d)$ (independent of the gene). For a particular gene $i$, clearly different sets of TFs provide different pairwise distances $d_{ij}$. However, for $Q\gg 1$ the fraction of sets of same size $Q$ that yield distances which are distributed very differently from $p(d)$ is small. In the following we neglect this fraction and assume that all choices of $Q$ TFs yield exactly the same crosstalk contribution $x_{1,2}(Q,M)$; this mean-field assumption is explicitly validated by numerical simulations in SI Section~\ref{sec:meanField}. We will also consider that all possible sets of $Q$ TFs (sets of genes that need to be active) are equally likely to occur.

See SI Section~\ref{sec:Alt_xtalk} for the alternative definitions of $x_1$ and $x_2$.

Our next step is to calculate  total crosstalk as a function of the above parameters (the total number of binding sites $M$ and the number of TF types available at any given time $Q$).
 We define  total crosstalk  as the \emph{fraction} of genes found in any of the possible  erroneous states.
We assume that the particular choice of $Q$ TFs that are present is random (hence we average over all possible ways to choose $Q$ out of $M$ TFs). In reality only certain sets of TFs need to be active together in which case the genes that are co-activated could have mutually similar binding sites, especially if they were regulated by the same TF, compared to genes that are activated separately, possibly by different TFs. In SI Section~\ref{sec:thetagenes} we treat a simple extension of our model where each TF can co-regulate several target genes. We also assume equivalence between the two types of error (we relax this assumption below in SI Section~\ref{sec:errortypes}).

Clearly, if each of the $Q$ genes that should be active has probability $x_1$ to be in any of the crosstalk states, then the  \emph{expected number} of genes in that state is $Qx_1$. Similarly, of the genes that should be inactive  the \emph{expected number} that are in crosstalk state is $(M-Q)x_2$. To obtain the \emph{fraction} of genes in any of the crosstalk states we simply divide by the total number of genes $M$:

 \be
 \label{eq:X}
 X(Q,M,x_1,x_2) = x_1 \frac{Q}{M} + x_2 \frac{M-Q}{M}.
 \ee

\noindent Using the definition of $S$ introduced in the main text
 \begin{equation}
\sum_{j\neq i} C_j e^{-\epsilon d_{ij} } =
\frac{C}{Q} (Q-1) \sum_d P(d) e^{-\epsilon d}\approx
C \sum_d P(d) e^{-\epsilon d} \equiv
 C S_i(\epsilon,L), \label{S}
\end{equation}

where we approximated $Q-1 \approx Q$ which is valid for $Q\gg 1$ (an assumption we make here and throughout the paper).
 $S(\epsilon,L)$ is an average similarity measure between all pairs of binding sites. If  binding site sequences are drawn randomly from a uniform distribution, $S=(\frac{1}{4}+\frac{3}{4}e^{-\epsilon})^L$. This is easy to derive: since  individual base pairs are assumed to be statistically independent, at each position the probability of a random sequence to be identical to a given TF consensus sequence is $1/4$, whereas with probability $3/4$ it is different, implying a decrease of $e^{-\epsilon}$ in binding energy. Since the complete binding site consists of $L$ independent base pairs, this expression for a single base pair is now raised to the power of $L$.

 The expressions for $x_{1,2}$ read: 



\begin{subequations}
\label{eq:x_12}
\begin{align}
x_1&=\frac{e^{-E_a} + CS}{\frac{C}{Q} + e^{-E_a} + CS}\\
x_2&=\frac{CS}{e^{-E_a} + CS}.
\end{align}
\end{subequations}

The two extreme cases occur when TF concentrations are either zero or very large (Table~\ref{Tab:basicModelErr}).
If $C=0$, $x_1=1$ and $x_2=0$, i.e.,  $x_1$ is maximal due to binding sites that should be bound, while zero error for $x_2$ occurs due to binding sites that should be unbound. The total error then amounts to the fraction of genes that need to be activated $X(C=0)=Q/M$. At the other extreme, if $C\rightarrow \infty$, $x_1=SQ/(1+SQ))$ and $x_2\approx 1$, i.e., no site is left unbound. The magnitude of $x_1$ error due to noncognate binding   is determined by the binding site similarity $S$. If $QS\ll 1$, $x_1\approx QS - (QS)^2$. The total crosstalk then amounts to $X(C\rightarrow \infty) = 1 - \frac{Q/M}{1+SQ}$. If $SQ\ll 1$, $X \approx 1 - \frac{Q}{M}(1 - SQ)$.



\begin{table}[h]
\begin{tabular}{|c|c|c|c|}
  \hline
                                            & $x_1$                    & $x_2$             & crosstalk, X \\
                                            &&&                                                          \\
  \hline
                                            & $\frac{e^{-E_a} + CS}{\frac{C}{Q} + e^{-E_a} + CS}$
                                                                      & $\frac{CS}{e^{-E_a} + CS}$
                                                                                 & $\frac{Q}{M}x_1 + \frac{M-Q}{M}x_2$ \\&&&\\
  $C = 0$                           & 1                       & 0                 & $Q/M$ \\&&&\\
  $C = \infty$                      & $\frac{SQ}{1+SQ}$       & 1                 & $1-\frac{Q/M}{1+SQ}$ \\&&&\\
  optimal $C$; only activators       & $\frac{1+QZ}{1+Z/S+QZ}$ & $\frac{QZ}{1+QZ}$ & $\frac{Q}{M}\frac{1+QZ}{1+Z/S+QZ}+\frac{M-Q}{M}\frac{QZ}{1+QZ}$ \\&&&\\
  optimal $C$; activators and global repressor & $\frac{1+QZ}{1+Z/S+QZ}$ & $\frac{QZ}{1+QZ}$ & $\frac{Q}{M}\frac{1+QZ}{1+Z/S+QZ}+\frac{M-Q}{M}\frac{QZ}{1+QZ}$ \\&&&\\
  \hline
\end{tabular}
\caption{
\label{Tab:basicModelErr}
Crosstalk  errors in the basic model. Per-gene errors of the two types: $x_1$ is the error of a site whose cognate TF exists and the site should therefore be bound, but is either unbound or bound by a noncognate factor. $x_2$ is the error of a site whose cognate factor does not exist, and the site should therefore be unbound, but is bound by a noncognate factor. The last column shows the total crosstalk, averaged over all $M$ sites.}
\end{table}

Next, we analyze the dependence of crosstalk on various parameters. One unknown in these expressions is the TF concentration $C$. Because we are searching for a lower bound on  crosstalk, we can find the concentration that minimizes $X$.
Taking the derivative of $X$ and solving for its zeros,
\[\frac{\partial}{\partial C} X(Q,M,x_1,x_2)=0,\] we find two potential extrema

\[C^\ast_{1,2}= \frac{Q e^{-E_a} \left(S (S M Q-Q (S Q+2)+M)\pm\sqrt{S (M-Q)}\right)}{S \left(-M (S Q+1)^2+S Q^2 (S Q+3)+Q\right)},\]

but only one of them can yield non-negative concentration values (and is consistently a minimum):
\be C^\ast= \frac{Q e^{-E_a} \left(S (S M Q-Q (S Q+2)+M)-\sqrt{S (M-Q)}\right)}{S \left(-M (S Q+1)^2+S Q^2 (S Q+3)+Q\right)}.
\label{eq:Copt}
\ee

For small $S$ the leading terms in the optimal concentration are
\be
C^*= \frac{e^{-E_a}Q}{\sqrt{S(M-Q)}} - \frac{e^{-E_a}Q(M-2Q)}{M-Q} - \frac{e^{-E_a}Q^2(2M-3Q)\sqrt{S}}{M-Q}^{3/2} + O[S].
\label{eq:C_leadingterm}
\ee

Substituting \eqref{eq:Copt} back into \eqref{eq:X} yields the minimal achievable crosstalk:
\be
X^\ast= \frac{Q}{M}\left(-S (M - Q) + 2 \sqrt{S(M-Q)}\right).
\label{eq:minX}
\ee

For a constant number of co-activated genes $Q$, $X^\ast$ increases to leading order like the square root of $S$,
\be
X^\ast = \frac{2Q\sqrt{M-Q}}{M}\sqrt{S}+O[S].
\ee

Substituting $C^\ast$ into the single gene crosstalk expressions \eqssref{eq:x1}{eq:x2}, we obtain the minimal per-gene crosstalk

\begin{subequations}
\begin{align}
x_1^\ast &= \sqrt{S(M-Q)}\\
x_2^\ast &= SQ\left(\frac{1}{\sqrt{S(M-Q)}} - 1\right).
\label{eq:x12_min}
\end{align}
\end{subequations}

Since crosstalk must be in the range [0,1] and $M \geq Q$, this solution is only valid under the condition that $S(M-Q)<1$.
Thus, minimal crosstalk has 3 regimes:
\begin{enumerate}
\item
For $S>1/(M-Q)$, crosstalk is minimized by taking $C=0$. This is the ``no regulation'' regime. In this case,  crosstalk amounts to $Q/M$, which is simply the fraction of genes that were supposed to be activated (but are not due to lack of their TFs).
\item
For $Q>Q_{\mbox{max}}(S,M)$, crosstalk is minimized by taking $C\rightarrow \infty$; this is the ``constitutive regime.''
$Q_{\mbox{max}}(S,M)$ is given by two of the roots of the 4th order equation, $S(M + SMQ - 2Q - SQ^2) - \sqrt{S(M-Q)}=0$, solved for $Q$.
We find the boundaries between the 3 different regulatory regimes by solving for $C^\ast(S,M,Q) = 0$.
\item
Otherwise, there is an optimal concentration $0<C^\ast<\infty$, given by \eqref{eq:Copt}, that minimizes crosstalk; this is the ``regulation regime.''
\end{enumerate}

\begin{figure}[h!]
\centering
\subfigure[]{
\includegraphics[width=0.45\textwidth]{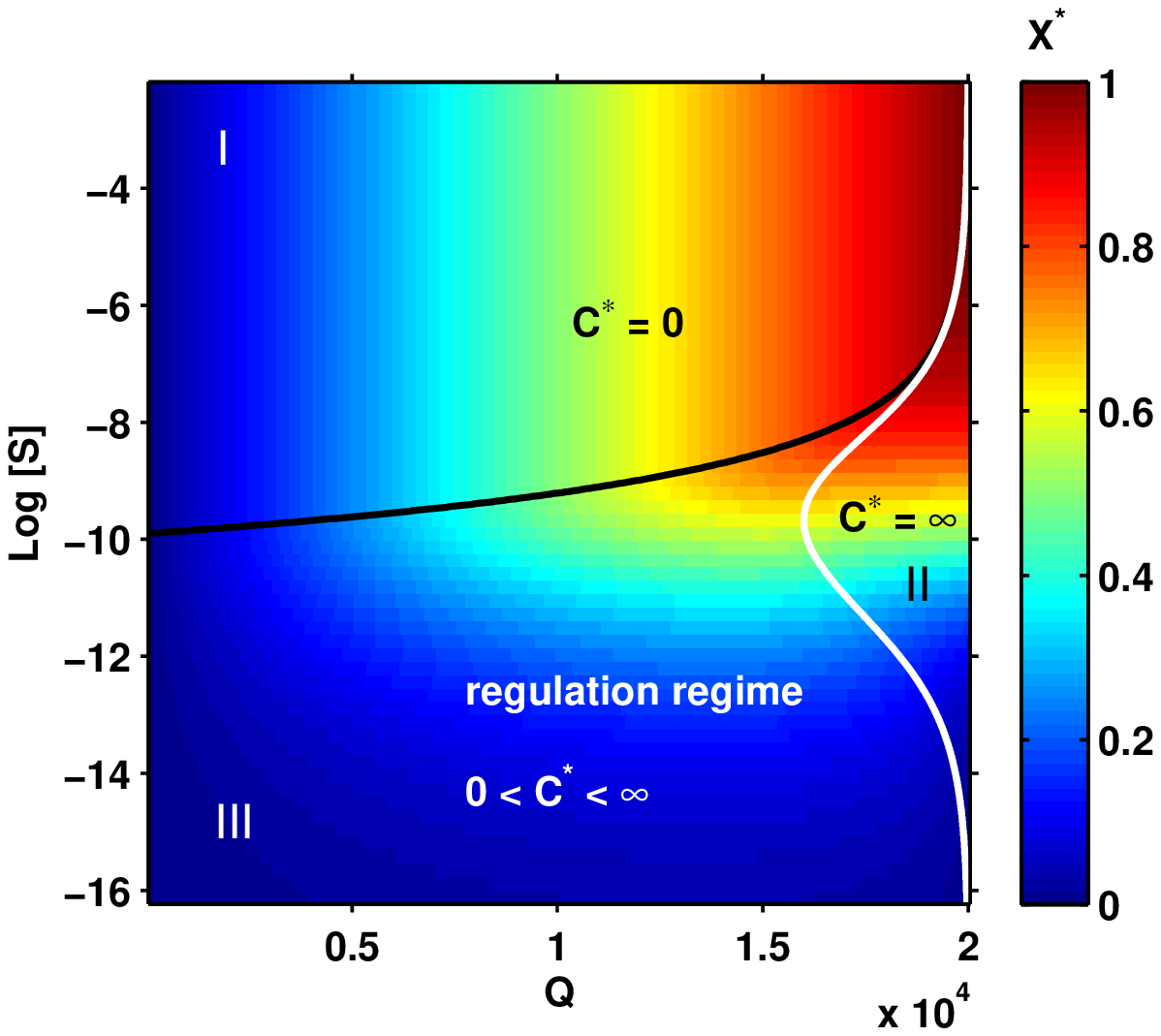}
\label{fig:3B_M_20000} }
\subfigure[]{
\includegraphics[width=0.45\textwidth]{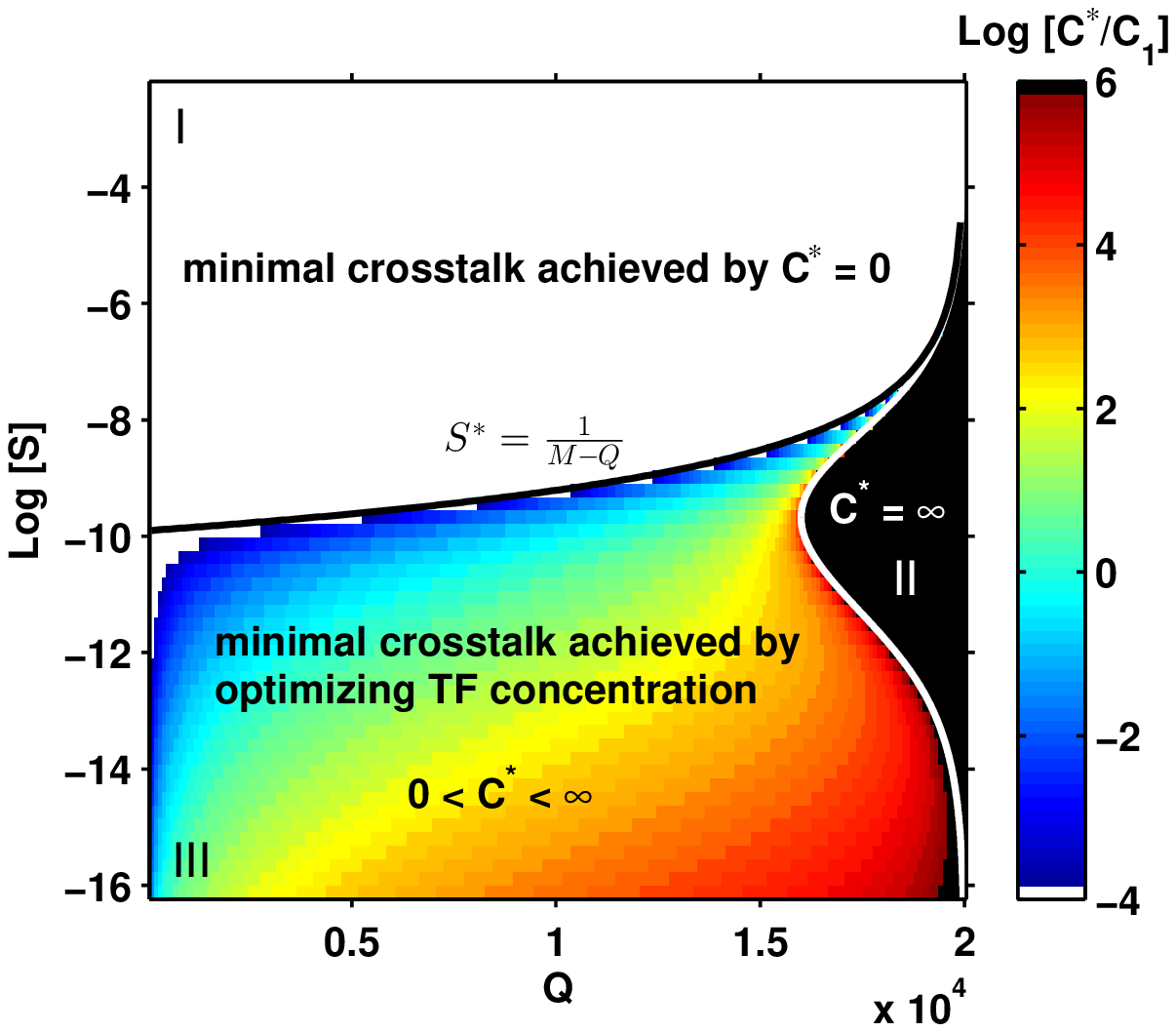}
\label{fig:3C_M_20000} }
  \caption[]
 { \label{fig:basic_M_20000} \textbf{Crosstalk in the basic model for $M=20,000$}. Panel (a) shows the minimal crosstalk, $X^*$; panel (b) shows the optimal TF concentration, $C^*$. These results are analogous to Fig.~3 of the main paper, which is computed for $M=5000$. The results for two different $M$ are qualitatively similar and show 3 different regimes of regulation. We make the following observations: {\bf (i)} for larger $M$, the $C^*=0$ regime expands to include lower $S$ values, as expected from the analytical solution for the regime boundaries; {\bf (ii)} if the fraction of co-activated genes, $Q/M$, remains constant, the crosstalk \emph{increases} with $M$, as it also depends on the absolute number of inactive genes $M-Q$ (see \eqref{eq:minX}). The discrepancies at small $Q$ between the black solid curve separating the ``no regulation'' and ``regulation'' regimes, and the numerically computed $C^*$ values are due to the approximation $Q-1\approx Q$. }
  \end{figure}

\begin{figure}[!h]
\centering
\subfigure[]{
\includegraphics[width=0.4\textwidth]{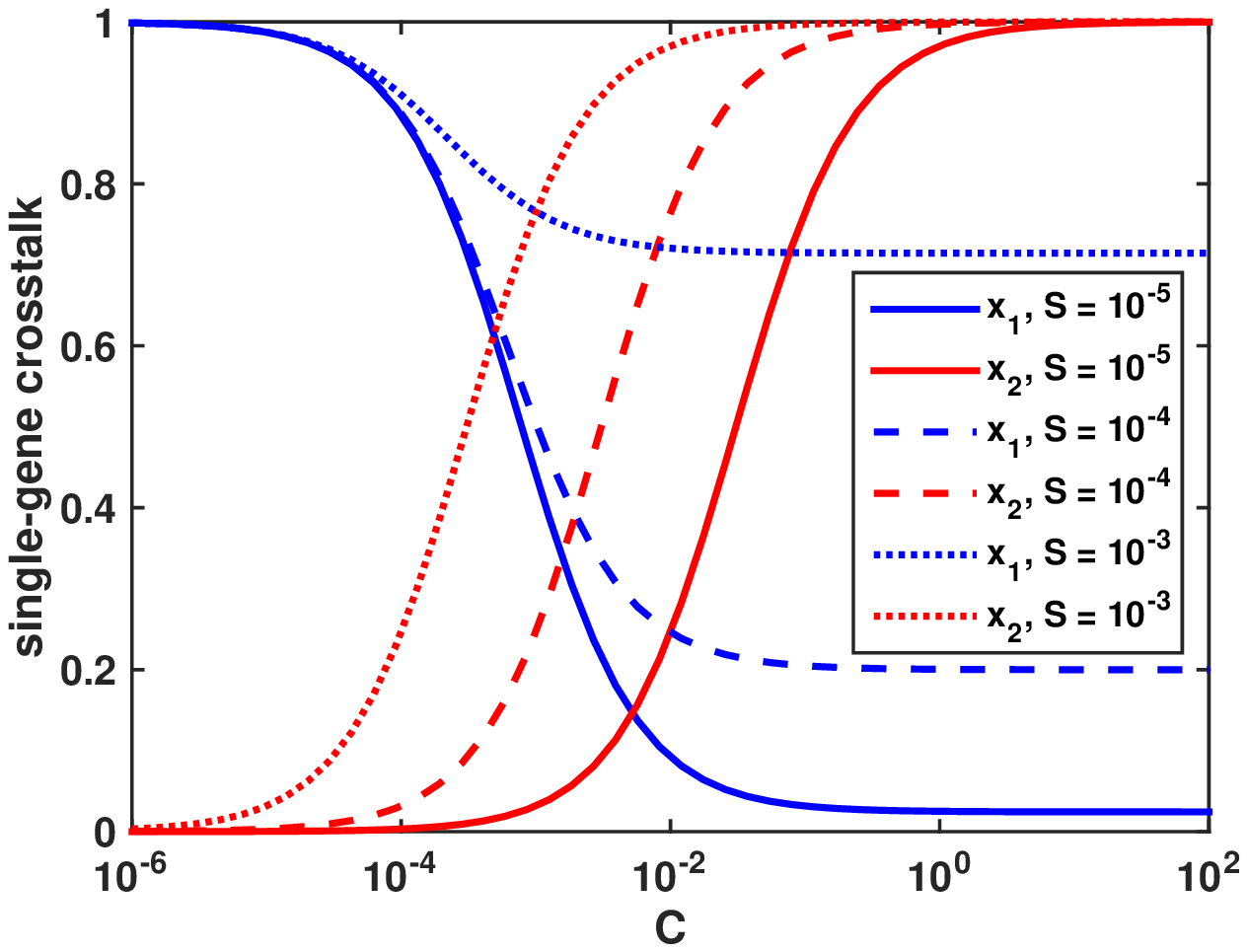}
\label{fig:x1x2_vs_C_S}
}
\subfigure[]{
\includegraphics[width=0.4\textwidth]{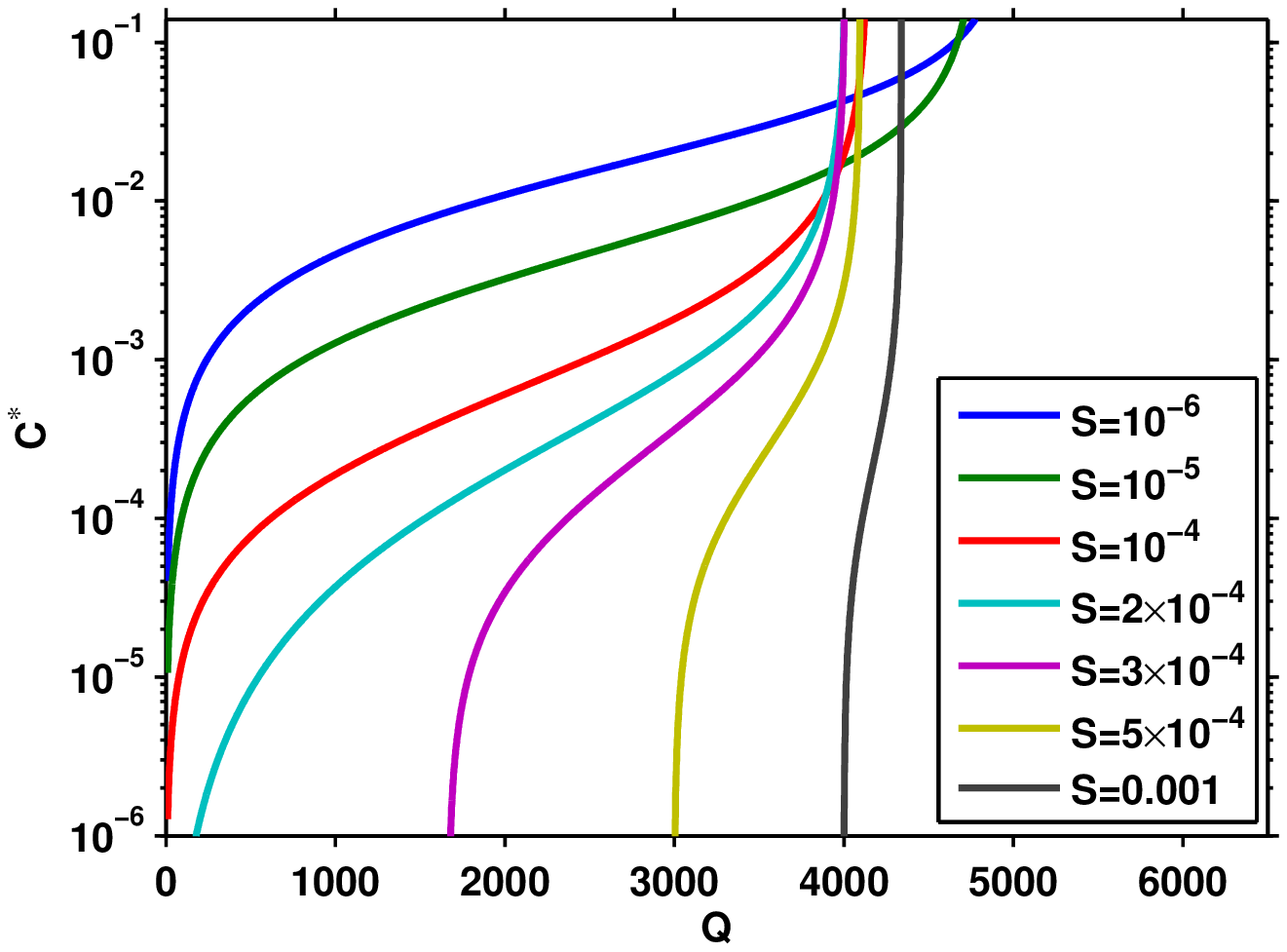}
\label{fig:optimal_C_vs_Q} }
  \captionof{figure}
 {\label{fig:C_dependence}
 \textbf{How is optimal TF concentration $C^*$ determined?}
 (a) $x_1$ crosstalk component (genes that should be active) decreases with TF concentration $C$, whereas $x_2$ crosstalk component (genes that should remain inactive) shows the opposite trend. Curves of $x_1$ and $x_2$ (crosstalk of a single gene) vs. $C$ are illustrated for various values of $S$. While $x_2$ can be fully eliminated if $C=0$, $x_1$ has a residual component which depends on $S$ even for infinite $C$. Both crosstalk types increase with the similarity between the binding sites $S$ (compare curves with various $S$ values).
 (b) The optimal concentration $C^*$ is a decreasing function of the similarity $S$ for all $Q$ values.  At fixed $M$, the optimal TF concentration, $C^\ast$, diverges with the number of co-activated genes, $Q$. This leads to the ``constitutive regime,''  where crosstalk is mathematically minimized by taking $C=\infty$.  Shown is the optimal concentration $C^\ast$ as a function of the number of co-activated genes $Q$, for various $S$ values; $M$ is fixed at $5000$.  The value of $Q$ at which $C$ diverges depends on $S$. For small $Q$, we require $M-1/S<Q$, otherwise the optimal concentration is in the $C^\ast=0$ regime. For the lower $S$ values crosstalk can be minimized for $0<Q<Q_{\mbox{max}}<M$, whereas for higher $S$ values there exists also a value for $Q_{\mbox{min}}$, such that $0<Q_{\mbox{min}}<Q<Q_{\mbox{max}}<M$. In other words, higher $S$ leads to a narrower range of $Q$ where the crosstalk can be effectively minimized. }
  \end{figure}

The boundary between the first and third region is at $S^*=\frac{1}{M-Q}$ and the boundary between the second and the third is at $S^*=\frac{-2M+3Q\pm\sqrt{Q(5Q-4M)}}{2Q(M-Q)}$.
Hence, the second region (where $C^*=\infty$) only applies for $Q>\frac{4M}{5}$.
\figref{fig:optimal_C_vs_Q} illustrates the dependence of the TF concentration $C^\ast$, which minimizes crosstalk, on the number of co-activated genes $Q$. It demonstrates how the range in which $0<C^\ast<\infty$ gets narrower when $S$ increases.
\figref{fig:basic_M_20000} demonstrates crosstalk and $C^\ast$ values for $M=20,000$ (compare to Fig. 3 in the main text with $M=5000$).

\begin{figure}[h!]
\centering
\includegraphics[width=0.45\textwidth]{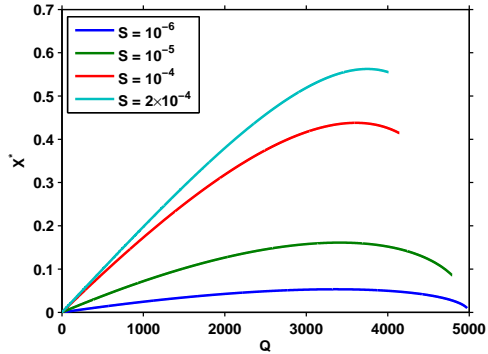}
  \caption[]
 { \label{fig:Xopt_vs_Q} \textbf{Minimal crosstalk $X^*$ is an increasing function of the similarity $S$ and has a non-monotonous dependence on the number of active genes $Q$.}
 The balance between genes that need to be active ($x_1$ crosstalk type) and genes that need to remain inactive ($x_2$ crosstalk type) causes a non-monotonous dependence of the \emph{total crosstalk} on the number of active genes $Q$, which has a maximum at an intermediate $Q$ value. Curves are shown only in the regulation regime, where crosstalk is minimized by a finite TF concentration. The curves are truncated at the point of transition to regime II where TF concentration formally diverges to infinity.
}
  \end{figure}


\subsection{Basic model: Dependence on variables}
\subsubsection{Dependence on TF concentration}
The optimal TF concentration $C^*$ in our model arises as a trade-off between the $Q$ genes that need to be active (for which a higher $C$ is favored) and the $M-Q$ genes that need to be inactive (for which a lower $C$ is favored). Note, however, the asymmetry between the two crosstalk types: while the $x_2$ component (genes that should remain inactive) can be completely suppressed by having no TF ($C=0$), the opposite does not hold. The $x_1$ component (genes that should be active) cannot be fully eliminated even for infinitely high $C$, because of the cross-activation between the distinct genes that should be active; see \figref{fig:C_dependence}(a).
This trade-off varies with the relative weights of $x_1$ and $x_2$, which depend on both $Q$ and $S$. We find that a concentration $C^*$ that minimizes crosstalk exists only in the third regime (``regulation regime''). In the first regime where $S<1/(M-Q)$, binding sites are so similar that crosstalk due to the inactive $M-Q$ genes dominates the total crosstalk. Hence the choice  of $C^*=0$ completely eliminates $x_2$ crosstalk, and minimizes the total crosstalk. In the second regime, where a large number of genes $Q$ need to be active, crosstalk due to the $Q$ active genes dominates ($x_1$ type),  hence $C^*$ diverges to infinity. \figref{fig:C_dependence}(b) illustrates curves of the optimal concentration $C^*$ as a function of the number of active genes $Q$ for constant values of $S$. As $Q$ increases, the relative weight of the genes that need to be active increases, hence $C^*$ is always a  monotonously increasing function of $Q$.

\subsubsection{Dependence on the similarity $S$}
Both crosstalk types $x_1$ and $x_2$ increase with the similarity $S$ (see \figref{fig:C_dependence}(a)). For a fixed $Q$, $C^*$ \emph{decreases} as a function of $S$. Again, this is because for larger $S$ the weight of the genes that should remain inactive is more significant, hence the trade-off shifts towards lower TF concentrations (but the minimal crosstalk $X^*$ still increases!). This behavior applies only in the regulation regime, hence for $ M - \frac{1}{S} < Q < Q_{\mbox{max}}$. For larger values of $Q$ ($Q > Q_{\mbox{max}}$), a more complex behavior is found because by changing $S$ we pass through all three regimes: $C^*$ then first decreases, then diverges (because it enters the second regime), but then decreases back again.

\subsubsection{Dependence on the number of active genes $Q$}
The two crosstalk types show opposite dependence on the number of active genes $Q$: crosstalk per gene that needs to be active ($x_1$) decreases with $Q$, whereas crosstalk per gene that needs to remain inactive increases with $Q$. The total crosstalk is a weighted sum of both with varying weights, hence it is not surprising that the \emph{total crosstalk} has a non-monotonous dependence on the number of active genes $Q$ with a maximum at an intermediate value; see \figref{fig:Xopt_vs_Q}.
The optimal TF concentration $C^*$ increases with the number of active genes $Q$; see \figref{fig:C_dependence}(b).

\subsection{Basic model with regulation by repressors only}
Our basic model assumed that all gene regulation is achieved by using specific activators to drive the expression of genes that would otherwise remain inactive. An alternative formulation of the problem postulates that genes are strongly expressed without TFs bound to their regulatory sites, but need to be repressed by the binding of specific regulators to stop their expression. Indeed, many bacterial genes seem to be regulated in this way.
We thus studied this complementary model, in which all regulators are repressors instead of activators. We assume, as before, that $Q$ out of $M$ genes should be active, but now this implies that $M-Q$ types of cognate repressors are present for all the genes that should remain inactive.

The expressions for crosstalk per gene that should be active ($x_1$) or inactive ($x_2$) read:
\begin{subequations}
\label{eq:x_12_rep}
\begin{align}
 x_1 &= \frac{CS}{e^{-E_a} + CS} \\
x_2 &= \frac{e^{-E_a} + CS}{\frac{C}{M-Q} + e^{-E_a} + CS}.
\end{align}
\end{subequations}

The total crosstalk is still

\be
X= \frac{Q}{M}x_1 + \frac{M-Q}{M}x_2.
\ee

\eqsref{eq:x_12_rep} are mathematically identical to \eqsref{eq:x_12}, where the roles of $Q$ and $M-Q$ are simply swapped. Not surprisingly, the minimal crosstalk in this case is:

\begin{subequations}
\begin{align}
x_1^\ast &= \frac{(M-Q)S(1 - QS)}{QS + \sqrt{QS}}\\
x_2^\ast &= \sqrt{QS} \\
X^* &= \frac{M-Q}{M}(2\sqrt{QS}- QS),
\label{eq:x_repressor_min}
\end{align}
\end{subequations}
 which is valid for $S<1/Q$.

 The optimal TF concentration that minimizes crosstalk is now

 \be
 \label{eq:Copt_rep}
 C^* = \frac{e^{-E_a}(M-Q)(1-QS)}{\sqrt{QS} + QS(2-QS) + MS(QS - 1)}.
 \ee

\begin{figure}[h!]
\centering
\subfigure[]{
\includegraphics[width=0.45\textwidth]{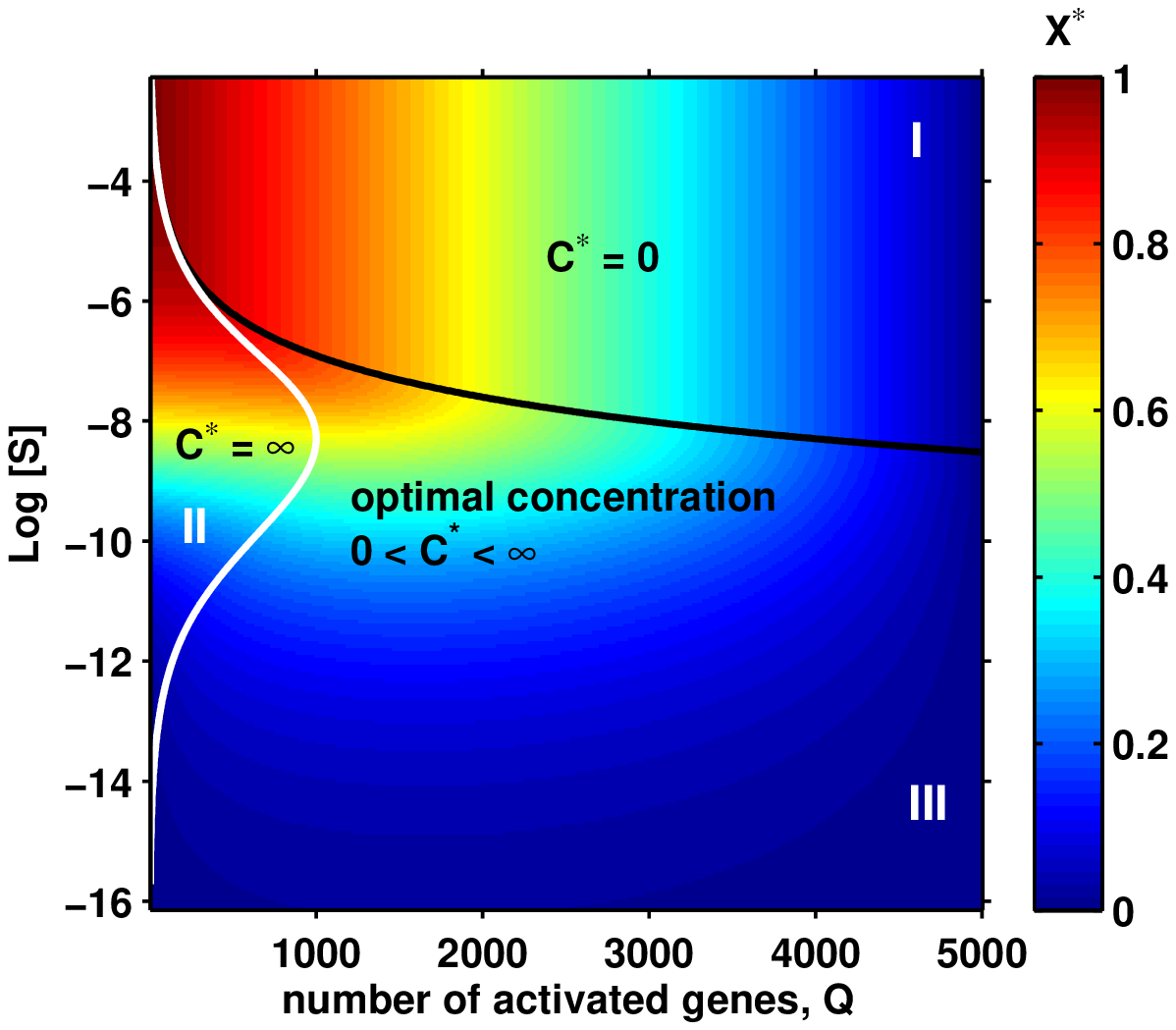}
}
\subfigure[]{
\includegraphics[width=0.45\textwidth]{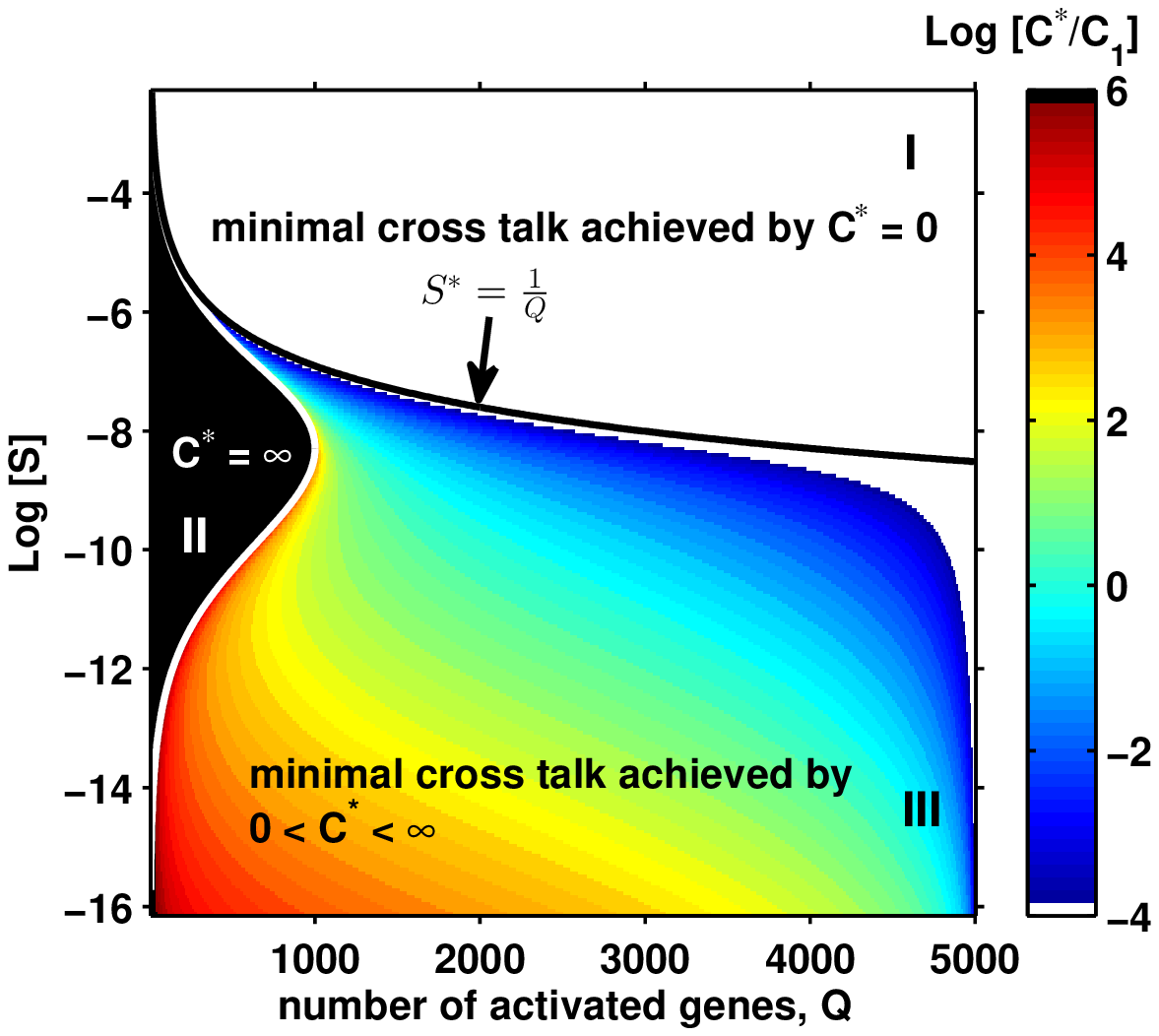} }
  \caption[]
 { \label{fig:BasicModel_rep} \textbf{Crosstalk in the basic model with regulation by repressors alone is a mirror image of regulation with activators only}. Panel (a) shows the minimal crosstalk, $X^*$; panel (b) shows the optimal TF concentration, $C^*$. These results are analogous to Fig.~3 of the main paper, which is computed for regulation with activators only.
 The observed picture is an exact mirror image of Fig. 3 of the main text, namely $Q$ maps to $M-Q$, where we keep the convention that $Q$ denotes the number of genes that should active. The difference is that in the activator-model activating $Q$ genes requires $Q$ types of activators, whereas in the repressor model this requires $M-Q$ types of repressors.
}
  \end{figure}

 The minimal crosstalk and optimal concentration are illustrated in \figref{fig:BasicModel_rep}.
 It retains the 3 regulatory regimes observed with activators only:
 \begin{enumerate}
\item
For $S>1/Q$ we obtain the ``no regulation'' regime where crosstalk is minimized by taking $C=0$.
\item
For $Q<Q_{\mbox{min}}(S,M)$ we obtain the ``constitutive regime'' where crosstalk is minimized by taking $C\rightarrow \infty$. $Q_{\mbox{min}}$ is obtained when $C^*$ of \eqref{eq:Copt_rep} diverges (the denominator equals to zero).
\item
Otherwise, there is an optimal concentration $0<C^\ast<\infty$, given by \eqref{eq:Copt_rep}, that minimizes crosstalk; this is the ``regulation regime.''
\end{enumerate}
 The three regions are marked with Roman numerals, in accordance with Fig.~3 of the main text. The boundaries between the three regimes are now:
 $S^*=1/Q$ (between regimes I and III) and $S^*=\frac{M-3Q\pm \sqrt{(M-Q)(M-5Q)}}{2Q(M-Q)}$ (between regime II to both I and III).

 The results are clearly a mirror image of the results shown in Fig.~3 of the main text for the activator-only basic model. They can be obtained simply by mapping $Q \rightarrow M-Q$.
 Since we keep the convention that $Q$ is the number of genes that are active, the difference in regulation strategies amounts to having either $Q$ activator types and keeping $M-Q$ binding sites unbound (activator-only) or having $M-Q$ repressor types and keeping $Q$ binding sites unbound.
 Comparing the expressions for minimal crosstalk, \eqref{eq:x_repressor_min} to \eqref{eq:minX}, we conclude that crosstalk depends on the \emph{fraction} of TFs that are expressed and on the \emph{absolute number} of binding sites that need to remain unbound.

\subsection{Breaking the symmetry between the two crosstalk types}
\label{sec:errortypes}
In our basic model we made a simplifying assumption that the two crosstalk types, $x_1$ and $x_2$, have equal weights: not activating a gene that should be active or erroneously activating a gene that should be silenced are assumed to be equally disadvantageous. We now relax this symmetry by allowing different weights, $a$ and $b$, for the two crosstalk types, to model possible differences in their biological significance. \eqref{eq:X} for the total crosstalk now takes the form:

\be
X = a\frac{Q}{M}x_1 + b\frac{M-Q}{M}x_2.
\ee
The expression for the optimal TF concentration then reads:
\be
C^\ast(a,b) = \frac{e^{-E_a}Q(\pm \sqrt{abS(M-Q)} - S(aQ - b(M-Q)(1+SQ)))}{S(aSQ^2 - b(M-Q)(1 + SQ)^2)},
\ee
where again only one of the two solutions yields non-negative concentration values. The resulting minimal crosstalk is:
\be
X^\ast(a,b)= \frac{Q}{M}(-Sb (M - Q) + 2 \sqrt{abS(M-Q)}).
\ee

\begin{figure}[h!]
\centering
\includegraphics[width=0.45\textwidth]{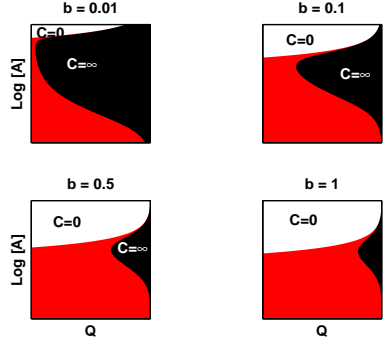}
  \caption[]
 { \label{fig:x-talk_types} \textbf{The three different regulatory regimes robustly exist even if the relative weight of the two crosstalk types vary.} To break the symmetry between the two error types we consider a redefined crosstalk, $X(b)=\frac{Q}{M}x_1 + b \frac{M-Q}{M}x_2$ (in the basic model $b=1$). For different values of $b$ (the cost of mis-activating genes that should remain inactive), all three regulatory regimes are preserved, although their boundaries shift. The weight of the first crosstalk type (mis-regulating genes that should be active) is equal in all cases. Red shows the ''regulation regime,'' ($0<C^*<\infty$). As erroneous activation is penalized less (decreasing $b$), the ``no regulation'' ($C^*=0$, white) regime shrinks, whereas the constitutive expression regime ($C^*=\infty$, black) expands, as expected.
}
  \end{figure}

 Setting $a=b=1$ reduces the above formula to the previous solution, \eqssref{eq:Copt}{eq:minX}. Note the asymmetry between the two crosstalk types: if $b=0$, i.e., when crosstalk in genes that should remain inactive is insignificant, the minimal achievable crosstalk equals zero. This is not true in the other extreme case, when $a=0$.
In \figref{fig:x-talk_types} we show that the three different regulatory regimes still exist under this generalized definition of crosstalk, but their boundaries may shift.

\subsection{Breaking the symmetry between the co-activated genes}
In our basic model we imposed full symmetry between the $Q$ co-activated genes: they contributed equally to crosstalk and  all $Q$ types of TFs were assumed to exist in equal concentrations.
We now relax these assumptions. We examine the situation in which a fraction $h$ of these $Q$ genes is more important to the functioning of the cell. Mathematically, we postulate that the per-gene crosstalk error for the important genes contributes with a $\gamma$-times higher weight to the total crosstalk relative to the non-important genes. We introduce an additional degree of freedom to the model, by allowing the concentration of the TFs to split unevenly between important and other genes: each important gene has TFs present at concentration $C_0$, while a TF of a non-important gene is present at concentration $C_0=\eta C_1$.

\noindent As $h Q C_0 + (1-h) Q C_1 = C$ we obtain:
\begin{subequations}
\begin{align}
C_1 &= \frac{C}{Q}\frac{1}{(1-h+h\eta)} \\
C_0 &= \eta C_1 = \frac{C}{Q}\frac{\eta}{(1-h+h\eta)}
\end{align}
\end{subequations}
If either $h=0$ or $\eta=1$ this reduces back to the basic model with $C_0=C_1=C/Q$.
 The total crosstalk now takes the form:

\begin{subequations}
\begin{align}
X &= \gamma h \frac{Q}{M} x_0 + (1-h) \frac{Q}{M} x_1 + \frac{M-Q}{M} x_2  \\
x_0 &=  \frac{e^{-E_a} + CS\left(1-\frac{\eta}{Q(1 + h(\eta - 1))} \right)}
{e^{-E_a} + \frac{\eta C/Q}{1 + h(\eta - 1)} + CS\left(1-\frac{\eta}{Q(1 + h(\eta - 1))} \right) }\\
x_1 &=  \frac{e^{-E_a} + CS\left(1-\frac{1}{Q(1 + h(\eta - 1))} \right)}
{e^{-E_a} + \frac{C/Q}{1 + h(\eta - 1)} + CS\left(1-\frac{1}{Q(1 + h(\eta - 1))} \right) }\\
x_2 &=  \frac{CS}{e^{-E_a} + CS},
\end{align}
\end{subequations}
where $x_0$ is the per-gene error of the important genes, $x_1$ is the error of other genes that need to be activated, and $x_2$, as before, denotes crosstalk at genes that need to be kept inactive.

\begin{figure}[h!]
\centering
\includegraphics[width=0.45\textwidth]{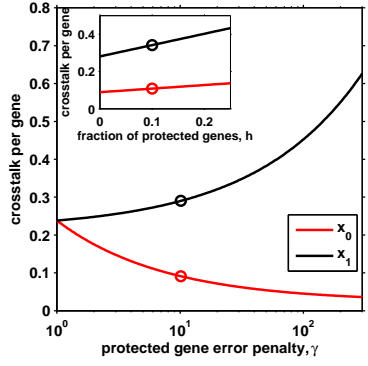}
  \caption[]
 { \label{fig:important_genes} \textbf{Crosstalk can be reduced for a subset of important genes at the cost of increasing the total crosstalk.} To break the symmetry between genes, we define
a fraction $h$ (out of $Q$) genes as important,
 having $\gamma$-times higher contribution to the total crosstalk.
TF concentration for these genes is optimized separately, subject to the total TF concentration $C$ remaining fixed  to its optimal value in the symmetric, $\gamma=1$, case. We show the crosstalk per important
gene, $x_0$ (red), and per a normal gene, $x_1$ (black), as a function of $\gamma$ (for $h=0.1$). The inset
shows the same as a function of $h$ (for $\gamma=10$).
Per-gene crosstalk
increases approximately linearly with $h$ and important
genes achieve $\sim\sqrt{\gamma}$ smaller crosstalk relative to normal genes.
}
  \end{figure}

We can optimize numerically for both the total TF concentration $C$ and the factor $\eta$ by which the TF concentration of the important genes is amplified. Alternatively, we can assume that $C$ remains fixed at the optimal value for the case where all genes are equally important, and only optimize for $\eta$. We display the latter option in \figref{fig:important_genes}, to explore crosstalk at varying $h$ under equal resource constraints.

The special case when only a single gene is important is analytically solvable assuming $Q\gg 1$, yielding:
\be
X^\ast_{\mbox{1 important gene}} \approx \frac{-SQ(M-Q)+2\sqrt{S(M-Q)}(Q - 1 + \sqrt{\gamma})}{M} .
\label{eq:Xopt_1_special_gene}
\ee

In particular the per-gene errors read:
\begin{subequations}
\begin{align}
x_0^\ast &= \frac{\sqrt{S(M-Q)}}{\sqrt{\gamma}} \\
x_1^\ast &= \sqrt{S(M-Q)}  \\
x_2^\ast &= \frac{-SQ(M-Q) + \sqrt{S(M-Q)}(Q-1+\sqrt{\gamma})}{M-Q}.
\end{align}
\end{subequations}

The error of the single important gene can be reduced at most by a factor of $\sqrt{\gamma}$ relative to the other co-activated genes. The $x_1^\ast$ error for the other $Q-1$ genes remains the same, because we assumed that $Q\gg 1$. Interestingly, the $M-Q$ genes that need to be kept inactive suffer an increase in crosstalk as a consequence of protecting the important gene.

\subsection{Every transcription factor regulates $\Theta$ genes}
\label{sec:thetagenes}
In the basic model we considered a regulatory scheme in which every gene has its own unique TF type. This allows for maximal flexibility in regulating each gene individually. Real gene regulatory networks typically have fewer TFs than the number of target genes, so that at least some transcription factors regulate several genes.
Here we consider a simple extension of the basic model, in which each TF regulates $\Theta$ genes (with identical binding sites) rather than one. We assume no overlap between the sets of genes regulated by various TFs, so that the total number of TFs species is now $\Theta$ times smaller than before. If $Q$ genes should be active, then $Q/\Theta$ TF species should be present in a given condition.
Assuming that $Q/\Theta \gg 1$, we can approximate $Q/\Theta - 1 \approx Q/\Theta$ as before. The only change from the basic crosstalk formulation is in $x_1$, because the concentration of  cognate factors is now $\Theta$ times larger than before:
\begin{subequations}
\begin{align}
x_1^{\Theta} &= \frac{e^{-E_a} + CS}
{\frac{C}{Q/\Theta} + e^{-E_a} + CS} \\
x_2^{\Theta} &= \frac{CS}
{e^{-E_a} + CS}.
\end{align}
\end{subequations}

This formulation is analytically solvable, yielding
\begin{subequations} 
\begin{align}
X_{\Theta}^* &= \frac{Q}{M}\left(-\frac{S}{\Theta} (M - Q) + 2 \sqrt{\frac{S}{\Theta}(M-Q)}\right)\\
x_1^{\Theta*} &=\frac{\sqrt{S(M-Q)}}{\sqrt{\Theta}}\\
x_2^{\Theta*} &= \frac{SQ}{\Theta}\left(\frac{\sqrt{\Theta}}{\sqrt{S(M-Q)}} - 1\right)\\
C_{\Theta}^* & = \frac{e^{-E_a}Q (\Theta - S(M-Q))} {S^2(M-Q)Q + S(M-2Q)\Theta + \sqrt{S(M-Q)}\Theta^{3/2}}.
\end{align}
\end{subequations}

The equations for minimal crosstalk are equivalent to the basic model if we map $S\rightarrow S/\Theta$. Since crosstalk depends on $\sqrt{S}$ to first order, this amounts to crosstalk reduction by a factor of $\sqrt{\Theta}$.

For small $S$ the leading term in the optimal concentration is
\be C_{\Theta}^*= \frac{1}{\sqrt{\Theta}}\frac{e^{-E_a}Q}{\sqrt{ S(M-Q)}} + O(1). \ee
These gains in crosstalk have, however, been achieved by sacrificing the ability to regulate each gene individually: now, the smallest set of genes that can be co-activated is of size $\Theta$. Typically, TFs might constitute $\gtrsim 10\%$ of the genes \cite{nimwegen_scaling_2003}; with $\Theta\sim 10$, the crosstalk could be reduced by a factor of $\sim 3$ at best.

\subsubsection{Non-constant $\Theta$}
Until now, we assumed that each TF regulates exactly $\Theta$ genes. This assumption can be relaxed using numerical simulations; in particular, we considered the case where the number of genes that each TF regulates is a random variable drawn from a specified distribution. We started by defining which TF controls which sets of genes through explicit enumeration of binding site sequences. We assumed that the number of genes that a given TF regulates is approximately Poisson distributed (with mean $\Theta$) and that all these regulated genes use the same sequence for their binding site, equal to the consensus sequence of the cognate TF. We then sample the environments in which $Q$ out of the total of $M$ genes are active; given the regulatory network structure, not all $Q$ picks out of $M$ can be realized, as is also the case with constant $\Theta$ model. The crosstalk is evaluated in each environment exactly, by computing all thermodynamic states of all binding sites, and is subsequently averaged by Monte Carlo sampling through the possible environments. This extension to the model introduces no new parameters, so its crosstalk and regime boundaries can be straightforwardly compared to the model where $\Theta$ is constant. We find that Poisson-distributed $\Theta$ changes crosstalk at a below-percent level, and produces no notable shifts in regime boundaries, showing that our results are robust with respect to this particular distributional assumption.

\section{Validity of the mean-field assumption}
\label{sec:meanField}

In computing crosstalk at given $M$ and $Q$, we have made a mean-field assumption on the similarity measure $S$. For a given set of binding site sequences in the sequence space (total $M$ in number), this
amounts to assuming that the distribution of neighbours for each binding
site comes from the same underlying distribution. For a particular selection of $Q$ genes, for each binding site $i$ from the $M$ binding sites, similarity $S_i$ can be defined using $d_{ij}$ where $j \ne i$ indexes over the binding sites of the $Q$ selected genes.

\begin{eqnarray}
S_i = \displaystyle \sum_{j \ne i} e^{-\epsilon d_{ij}}
\end{eqnarray}

From this, we have for crosstalk for a particular selection of $Q$ genes,

\be
\begin{split}
X(\{S_i\}) &= \dfrac{1}{M}\Big[\displaystyle \sum_{i \in Q} x_1(S_i) + \displaystyle \sum_{i \in M-Q} x_2(S_i)\Big] \\
&= \dfrac{1}{M}\Big[\displaystyle \sum_{i \in Q}\dfrac{e^{-E_a} + CS_i }{C/Q + e^{-E_a} + CS_i} + \displaystyle \sum_{i \in M-Q}\dfrac{CS_i }{e^{-E_a} + CS_i}\Big]
\end{split}
\ee

where $x_1(S_i)$ and $x_2(S_i)$ depend on $S_i$ as shown. We are interested in the mean crosstalk $X = \langle X(\{S_i\}) \rangle$ over all selections of $Q$ out of $M$ genes, which requires us to know the full distribution of $S_i$. The crosstalk is then

\begin{eqnarray}
X = \langle X(\{S_i\}) \rangle = \dfrac{1}{M}\Big[\displaystyle \sum_{i \in Q} \langle x_1(S_i) \rangle + \displaystyle \sum_{i \in M-Q} \langle x_2(S_i) \rangle \Big].
\end{eqnarray}

In the mean-field assumption, we have $\langle x_1(S_i) \rangle \approx x_1(\langle S_i \rangle) = x_1(S)$ and $\langle x_2(S_i) \rangle \approx x_2(\langle S_i \rangle) = x_2(S)$, which gives us

\begin{eqnarray}
X = \dfrac{Q}{M}x_1(S) + \dfrac{M-Q}{M} x_2(S).
\end{eqnarray}

\begin{figure}[h!]
\centering{\includegraphics[scale=0.4]{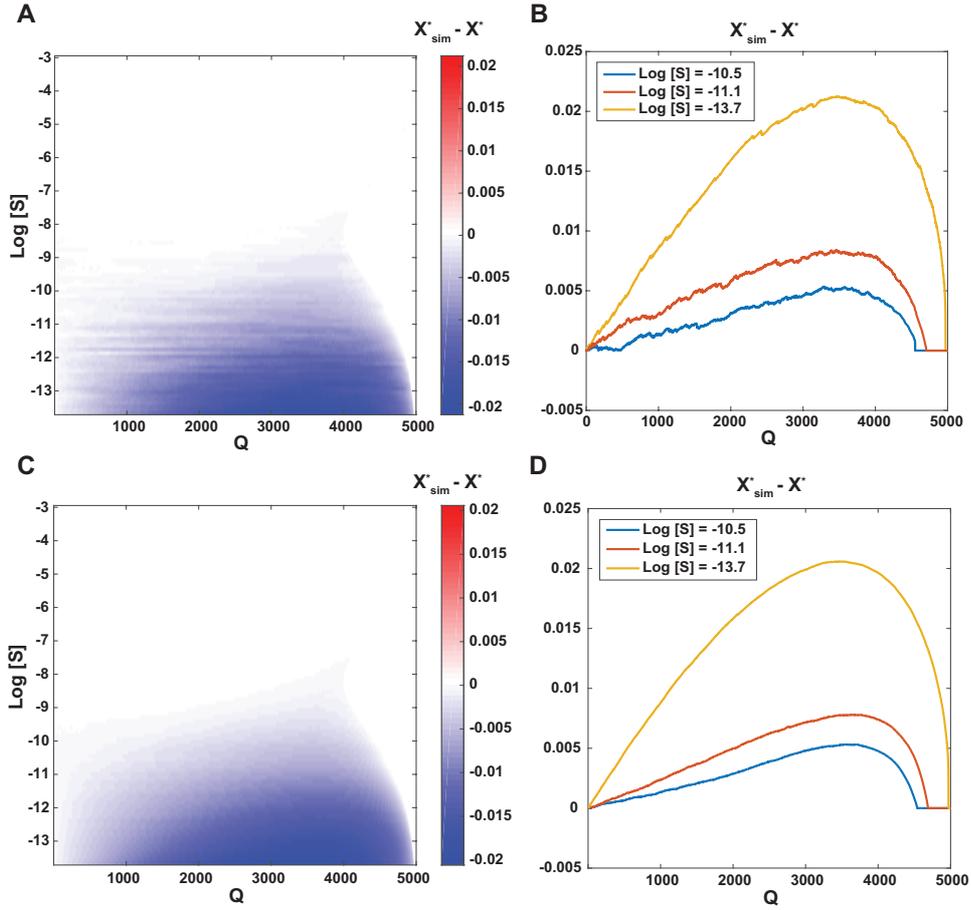}}
\caption{\textbf{Comparison of mean-field results and numerical simulations}. On the left, we
plot the difference in optimal crosstalk between simulations and the
mean-field approach, $X_{\rm sim}^{*}-X^{*}$, for different $Q$ and $S$.
On the right, we plot $X_{\rm sim}^{*}-X^{*}$ against $Q$ for three
different $S$. Here, $M=5000$, $L=10$, and $S$ has been varied by tuning $\epsilon$. $X_{\rm sim}^{*}$ is a Monte Carlo estimate of the mean crosstalk, obtained over $n_{\rm sel}$ different selections of $Q$ out of $M$ genes. $n_{\rm sel}=1$ in the top row, and $n_{\rm sel}=30$ in the bottom row. The mean-field approach is in general a very good
approximation of the simulations. The maximal crosstalk difference
is less than $0.02$, and decreases with increasing $S$.}
\label{mf1}
\end{figure}

\begin{figure}[h!]
\centering{\includegraphics[scale=0.4]{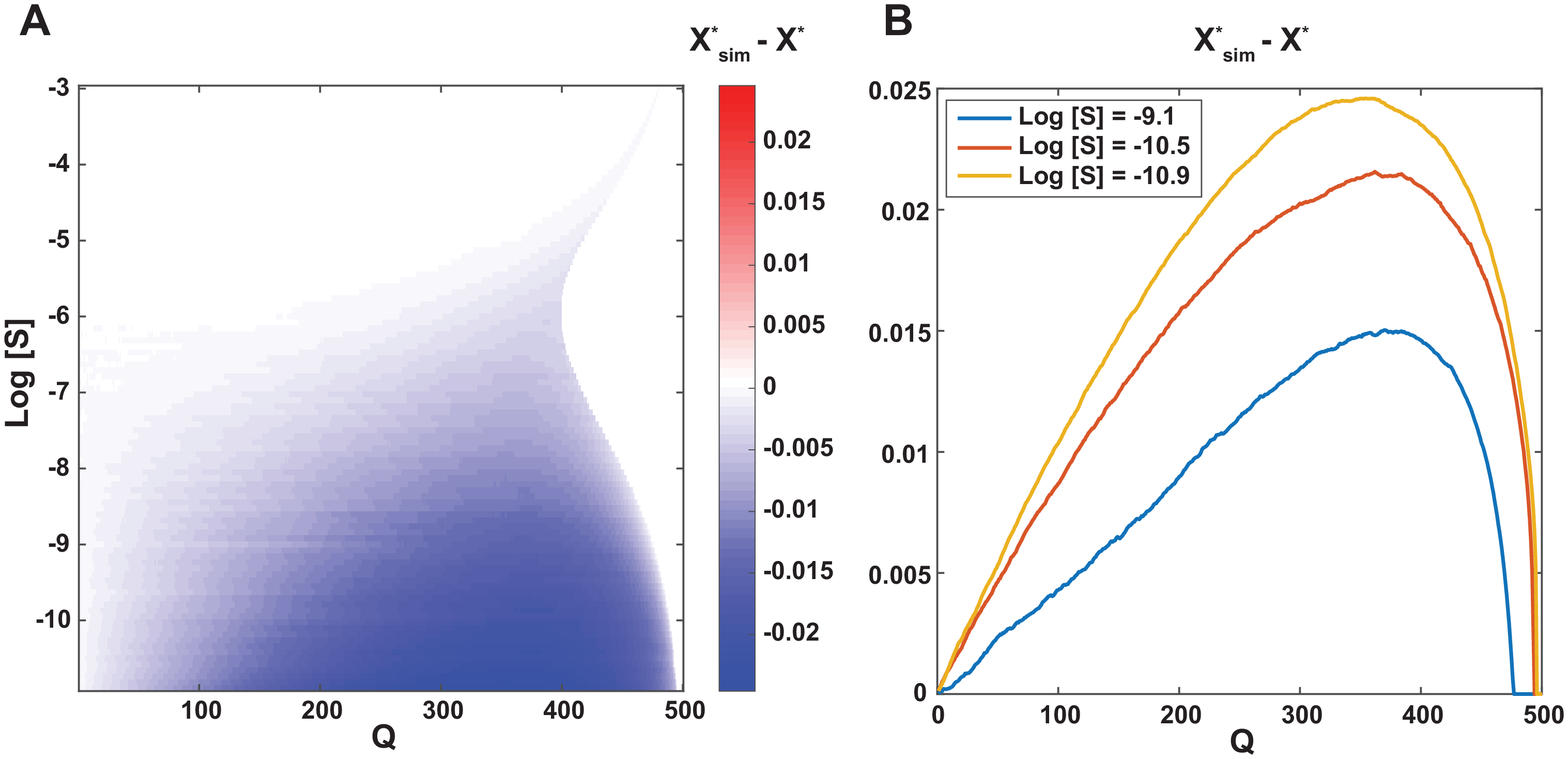}}
\caption{\textbf{Comparison of mean-field results and numerical simulations}. On the left, we
plot the difference in optimal crosstalk between simulations and the
mean-field approach, $X_{\rm sim}^{*}-X^{*}$, for different $Q$ and $S$.
On the right, we plot $X_{\rm sim}^{*}-X^{*}$ against $Q$ for three
different $S$. Here, $M=500$, $L=8$, and $S$ has been varied by tuning $\epsilon$. $X_{\rm sim}^{*}$ is a Monte Carlo estimate of the mean crosstalk, obtained over $n_{\rm sel}=100$ different selections of $Q$ out of $M$ genes. Again, as with $M=5000$, the mean-field approach is a very good approximation of the simulations. The maximal crosstalk difference
is only slightly larger than $0.02$.}
\label{mf2}
\end{figure}

From this, one can obtain the optimal crosstalk $X^{*}$. To check the validity of such a mean-field assumption, we performed numerical simulations by drawing lists of $M$ binding sites from the sequence space, computing optimal crosstalk $X_{\rm sim}^{*}$ by explicit enumeration of all thermodynamic states, and comparing this with the mean-field crosstalk $X^{*}$. In detail, we first picked $M$ binding sites (to regulate $M$ genes) randomly from the sequence space and held this choice fixed. Now, for each $Q$, we performed $n_{\rm sel}$ different selections of $Q$ out of $M$ genes. For each such selection, after computing the binding site mismatches and occupancies, we compute the crosstalk. To get the mean crosstalk for $Q$, we perform a Monte Carlo estimate of the mean crosstalk over these $n_{\rm sel}$ different selections of $Q$ out of $M$ genes. Figures~\ref{mf1} and \ref{mf2} show that the mean-field crosstalk systematically over-estimates
the actual crosstalk, but nevertheless remains a very good approximation to the true crosstalk.

\section{Mixed models}
\label{sec:mixedModels}

In the baseline model we consider $M$ genes, all of which are regulated
either solely by  activators or solely by repressors. Here, we consider mixed models,
 i.e., models that utilize repression to control one subset of genes and activation to control the other genes. Let's assume that $M_{A}$ genes are regulated by activators
and $M_{R}$ genes are regulated by repressors, where  $M=M_{A}+M_{R}$.
In a particular environment, let's assume that $Q$ genes need to
be ON. Out of these, let's assume that $Q_{A}$ genes are activator-regulated
 and $Q_{R}$ genes are repressor-regulated, where $Q=Q_{A}+Q_{R}$. For activating $Q$ genes, the number of TFs present now amounts to $T=Q_{A}+M_{R}-Q_{R}$: $Q_{A}$ activators and $M_{R}-Q_{R}$ repressors. As before, $S$ is the similarity of the binding sites
and $C$ the total concentration of TFs (activators+repressors). The
concentration of a particular TF type, when present, will now be $C/T$.
We assume that any non-cognate interaction (``activation out-of-context''
or ``repression out-of context'') counts as a crosstalk
error. We distinguish 4 types of per-gene crosstalk errors:

An activator-regulated gene that needs to be ON, should be bound
by the cognate activator. The unbound state and any non-cognate binding
(non-cognate activator or repressor) are crosstalk states:

\begin{align}
x_{1}^{A}= & \dfrac{e^{-E_{a}}+CS}{\frac{C}{T}+e^{-E_{a}}+CS}\qquad(Q_{A}\text{ out of }M\text{ genes}).
\end{align}

An activator-regulated gene that needs to be OFF, should be unbound.
Any non-cognate binding is a crosstalk state:

\begin{align}
x_{2}^{A}= & \dfrac{CS}{e^{-E_{a}}+CS}\qquad(M_{A}-Q_{A}\text{ out of }M\text{ genes}).
\end{align}

A repressor-regulated gene that needs to be ON, should be unbound.
Any non-cognate binding is a crosstalk state:

\begin{align}
x_{1}^{R}= & \dfrac{CS}{e^{-E_{a}}+CS}\qquad(Q_{R}\text{ out of }M\text{ genes}).
\end{align}

Lastly, a repressor-regulated gene that needs to be OFF, should be bound
by the cognate repressor. The unbound state and any non-cognate binding
(non-cognate repressor or activator) are crosstalk states:

\begin{align}
x_{2}^{R}= & \dfrac{e^{-E_{a}}+CS}{\frac{C}{T}+e^{-E_{a}}+CS}\qquad(M_{R}-Q_{R}\text{ out of }M\text{ genes}).
\end{align}

As $x_{1}^{A}=x_{2}^{R}$ and $x_{2}^{A}=x_{1}^{R}$,
the overall crosstalk error reads

\be
\begin{split}
X_{\rm mixed, full}(Q_{A},Q_{R},M_{A},M_{R}) & =x_{1}^{A}\dfrac{Q_{A}}{M}+x_{2}^{A}\dfrac{M_{A}-Q_{A}}{M}+x_{1}^{R}\dfrac{Q_{R}}{M}+x_{2}^{R}\dfrac{M_{R}-Q_{R}}{M}\\
 & =x_{1}^{A}\dfrac{M_{R}+Q_{A}-Q_{R}}{M}+x_{2}^{A}\dfrac{M_{A}+Q_{R}-Q_{A}}{M}\\
 & =x_{1}^{A}\dfrac{T}{M}+x_{2}^{A}\dfrac{M-T}{M}\\
 & =X(Q_{\rm eff}=T,M_{\rm eff}=M).
\end{split}
\ee

Hence, given a set of $(Q_{A},Q_{R},M_{A},M_{R})$ of the mixed model,
crosstalk is same as that in an equivalent baseline activator model
with $Q_{\rm eff}=T=M_{R}+Q_{A}-Q_{R}$ and $M_{\rm eff}=M=M_{A}+M_{R}$.

For a given $M$, different ($M_A$,$M_R$) partitions are possible, which differ in the number of genes under activator or repressor control. This can be tuned on an evolutionary timescale.
Once $M_{A}$ is chosen, different selections of $Q$ genes that should be active potentially have different
numbers of genes under the control of activators $(Q_{A})$ and repressors
$(Q_{R}=Q-Q_{A})$. However, the optimal TF concentration $C^*$ and the minimal crosstalk $X^*$ only depend on the total number of TFs $T$.

\begin{figure}
\centering{\includegraphics[scale=0.6]{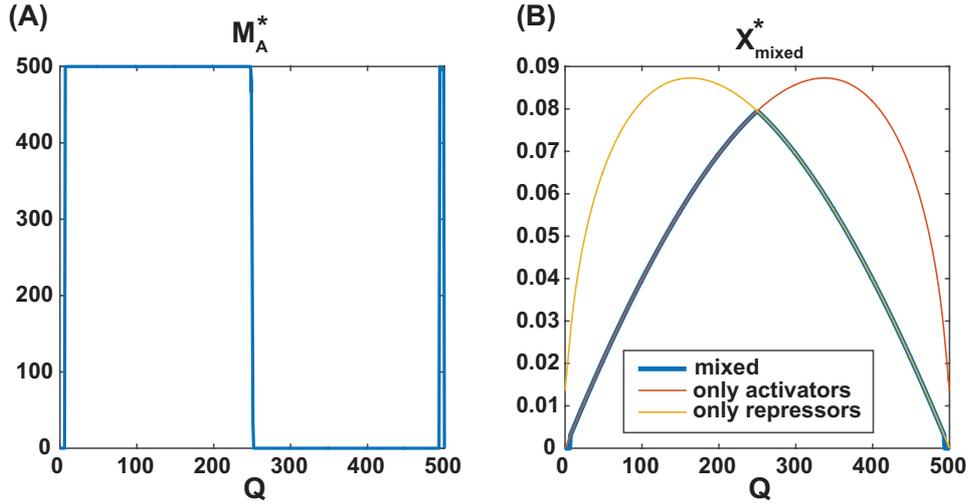}}

\caption{\textbf{Mixed model at best $M_{A}$}. On
the left, we plot the optimal number of activated genes $M_{A}^{*}$
for different $Q$ at $M=500$ and $\log(S)=-10.5$. For $Q<250$,
it is best to have all genes under activator control ($M_{A}^{*}=500$) and for
$Q\ge250$, it is best to have all genes under repressor control ($M_{A}^{*}=0$).
On the right, we plot the optimal mixed crosstalk, computed at $M_{A}^{*}$,
and averaged over different gene selections using $P_{Q_{A}}$. \label{fig:mixed}}
\end{figure}

For given $M,Q$, and $S$, we find the best possible $M_{A}$, which
minimizes the crosstalk. For a particular $M_{A}$, we define the optimal
crosstalk as the average optimal mixed crosstalk for all selections
of $Q$ genes out of $M$ (averaged over different choices of $Q_{A}$),

\begin{eqnarray}
X^{*}(M,Q,S,M_{A})={\displaystyle \sum_{Q_{A}}P_{Q_{A}}X_{\rm mixed, full}^{*}(Q_{A},M,Q,S,M_{A})},
\end{eqnarray}
where $P_{Q_{A}}$ is the fraction of $Q$ gene selections that have
$Q_{A}$ activated genes. We have

\begin{eqnarray}
P_{Q_{A}}=\dfrac{\binom{M_{A}}{Q_{A}}\binom{M-M_{A}}{Q-Q_{A}}}{\binom{M}{Q}},
\end{eqnarray}

\begin{eqnarray}
X_{\rm mixed}^{*}(M,Q,S)=\min\Big[X^{*}(M,Q,S,M_{A})\Big],
\end{eqnarray}

\begin{equation}
M_A^*=\argmin_{M_{A}}X^{*}(M,Q,S,M_{A}),
\end{equation}

where $M_{A}^{*}$ is the  $M_{A}$ value which minimizes crosstalk for a given $Q$.  In Fig.~\ref{fig:mixed},
we see that for $Q<M/2$, the best strategy is to use all activators
($M_{A}=M$), and for $Q\geq M/2$, the best strategy is to use all repressors; optimization of crosstalk in mixed models therefore always picks out one of the two ``pure'' regulatory strategies and does not yield an optimal mixed model.


To see if the pure strategies get chosen because the activation of all genes is symmetric in all environments,
 we studied a simple system in which different subsets
of genes are required to be activated with different probabilities.
So far, when $Q$ genes are required to be ON, each gene had the same
probability, $Q/M$, to be among the $Q$ out of $M$ required genes, i.e. $Q/M$ is the probability of each gene to be activated.

\begin{figure}
\centering{\includegraphics[scale=0.75]{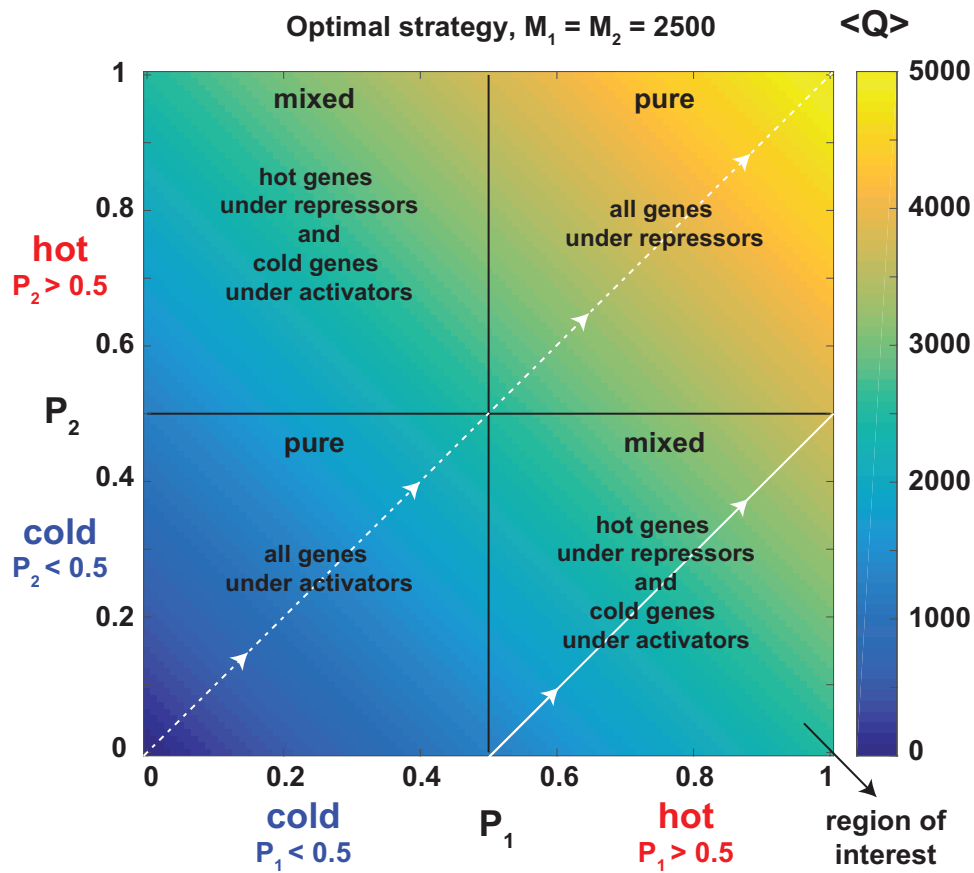}\caption{\textbf{When some genes are hot and other genes are cold, the optimal
mixed strategy puts hot genes under repressors and cold genes under
activators.} Here we show how the optimal strategy and $\langle Q\rangle$
vary as a function of $P_{1}$ and $P_{2}$ for a fixed choice of
$M_{1}=M_{2}=2500$.  $\langle Q\rangle$ increases in any
direction that increases $P_{1}$ or $P_{2}$. When $P_{1}<0.5$ and
$P_{2}<0.5$ (all genes are cold), the optimal strategy is a pure
one (all genes under activator control), while when $P_{1}>0.5$ and
$P_{2}>0.5$ (all genes are hot), the optimal strategy is to put all
genes under repressors, which is also a pure strategy. But when $P_{1}>0.5,P_{2}<0.5$
or $P_{1}<0.5,P_{2}>0.5$ (one class is hot, while the other is cold),
the optimal strategy is ``mixed'': hot genes are under repressor control and
  cold genes under activator control.
 \label{fig:mixed-phase}}
}
\end{figure}

 Here, we introduce two classes $(1\text{ and }2)$
of genes, with $M_{1}$ genes in the first class and $M_{2}=M-M_{1}$
genes in the second class. Genes in each of the two classes have different probabilities of
requiring activation across environments: $P_{1}$ for the first class and $P_{2}$ for the second class. If $P_{i}>0.5$,
then genes in class $i$ are called ``hot'' genes, and if $P_{i}<0.5$,
genes in class $i$ are called ``cold'' genes. Given certain $M_{1},M_{2},P_{1}$,
and $P_{2}$, different environments correspond to different choices of the
$Q$ genes that should be active, where $Q$ is no longer constant as before, but a random variable with mean
\begin{eqnarray*}
\langle Q\rangle=P_{1}M_{1}+P_{2}M_{2}.
\end{eqnarray*}
In a similar fashion as before, we compute the crosstalk (at optimal
$C^{*}$) for different choices of mixed models (how many class $i$
genes are under activators or repressors). Then, we obtain the optimal ($M_A$,$M_R$)
strategy among these mixed models that minimizes crosstalk. In Fig.~\ref{fig:mixed-phase}, we show how this optimal strategy varies,
along with $\langle Q\rangle$, as a function of $P_{1}$ and $P_{2}$
for a fixed choice of $M_{1}=M_{2}=2500$. First, we note that $\langle Q\rangle$
increases in any direction that increases $P_{1}$ or $P_{2}$. In
the symmetric mixed model setup, we essentially studied the system
along the diagonal from $(0,0)$ to $(1,1)$ on the $(P_{1},P_{2})$
plane (dashed white line), increasing $\langle Q\rangle$ from $0$
to $M$. The previously studied results yielded two ``pure'' strategies---all activators or all repressors, depending on whether $Q$ is bigger
or smaller than $M/2$---which is consistent with the following observations
in the asymmetric mixed models. When $P_{1}<0.5$ and $P_{2}<0.5$
(all genes are cold), the optimal strategy is a pure one, namely, to put all
genes under activators;  when $P_{1}>0.5$ and $P_{2}>0.5$ (all
genes are hot), the optimal strategy is to put all genes under repressors,
which is also a pure strategy. But when $P_{1}>0.5,P_{2}<0.5$ or
$P_{1}<0.5,P_{2}>0.5$ (one class is hot, while the other is cold),
the optimal strategy is ``mixed'': put hot genes under repressors
and cold genes under activators. Note that not all $\langle Q\rangle$
are possible with these optimal mixed strategies. From here onwards, we study
mixed models in the bottom right square of Fig.~\ref{fig:mixed-phase},
where $P_{1}>0.5$ and $P_{2}<0.5$, i.e., class $1$ is hot and
class $2$ is cold.

\begin{figure}
\centering{\includegraphics[scale=0.75]{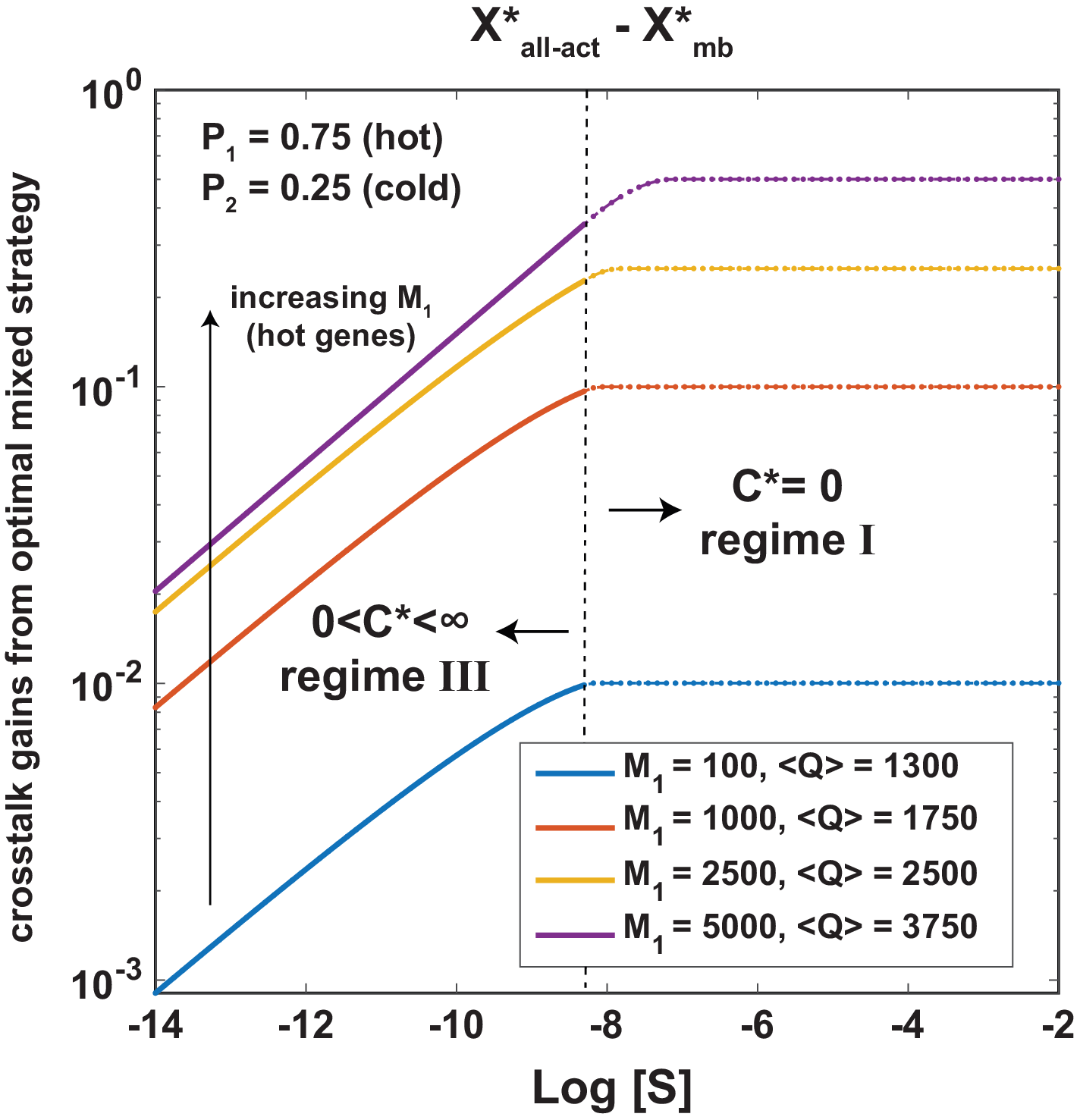}\caption{\textbf{Crosstalk gains from using the optimal mixed strategy instead
of all activators}. Plotted is the difference in optimal crosstalk (crosstalk gain), $X_{\rm all-act}^{*}-X_{\rm mb}^{*}$,
 between the pure strategy of using all activators
and the optimal mixed strategy of putting hot genes under repressors
and cold genes under activators, as a function of $S$, with fixed $P_{1}=0.75$ and $P_{2}=0.25$.
As $S$ increases, we cross from the regulatory regime III to regime
I in which $C^{*}=0$ . The optimal mixed strategy becomes increasingly
better (than the all activators pure strategy at reducing crosstalk)
as $S$ and $M_{1}$ increase. \label{fig:mixed-diffMhot} }
}
\end{figure}

At fixed $P_{1}$ and $P_{2}$, crosstalk gains from using the optimal
mixed strategy (instead of using all activators) increase with both
$S$ and the number of hot genes $M_{1}$, as shown in Fig.~\ref{fig:mixed-diffMhot}.

\begin{figure}
\centering{\includegraphics{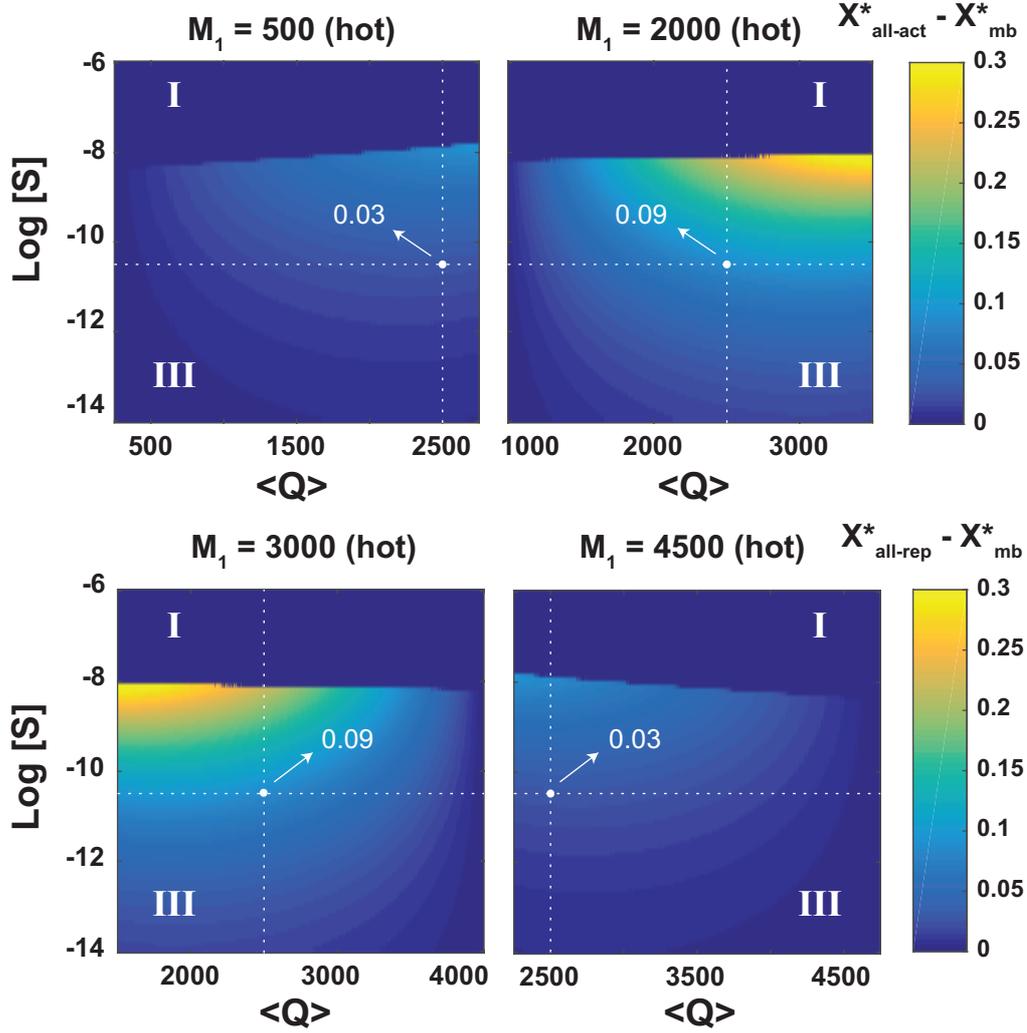}\caption{\textbf{Optimal mixed strategy is increasingly better than the optimal
pure strategy at intermediate $M_{1}$ and larger $S$, at the border
of the two regimes}. Here, we plot the crosstalk gains, ($X_{\rm all-act}^{*}-X_{\rm mb}^{*}$
in the top row, or $X_{\rm all-rep}^{*}-X_{\rm mb}^{*}$ in the bottom row)
from using the optimal mixed strategy instead of the optimal pure
strategy as a function of the average number of genes required, $\langle Q\rangle$,
and $S$, for different $M_{1}$. For $M_{1}<M/2=2500$, the optimal
pure strategy is to use all activators and for $M_{1}>M/2=2500$,
the optimal pure strategy is to use all repressors. Note that for
$M_{1}>M/2$, $X_{\rm all-rep}^{*}-X_{\rm mb}^{*}$ at $(\langle Q\rangle,S)$
is equal to $X_{\rm all-act}^{*}-X_{\rm mb}^{*}$ at $M'_{1}=M-M_{1}<M/2$
and $(M-\langle Q\rangle,S)$; they are laterally inverted mirror
images. In general, the optimal mixed strategy gives a lower crosstalk
than the optimal pure strategy for intermediate $M_{1}$. At the baseline
parameters of $\langle Q\rangle=2500,M=5000,\log(S)=-10.5$, for $M_{1}=500$
and $4500$ both, the crosstalk gain is $0.03$, while for $M_{1}=2000$
and $3000$, the crosstalk gain is $0.09$. For a particular $M_{1}$,
crosstalk gains are larger both at larger $S$ and larger (smaller)
$\langle Q\rangle$ for $M_{1}>M/2$ ($M_{1}<M/2$). We obtain different
$\langle Q\rangle$ on the x-axes as $\langle Q\rangle=P_{1}M_{1}+P_{2}M_{2}$
by varying $(P_{1},P_{2})$ along the solid white line of Fig. \ref{fig:mixed-phase}
from $(0.5,0)$ to $(1,0.5)$. \label{fig:mixed-varyingPhot}}
}
\end{figure}

In Fig.~\ref{fig:mixed-varyingPhot}, we show in detail the crosstalk
gains  from using the optimal mixed strategy instead of the optimal
pure strategy (either all activators or all repressors), for different
$\langle Q\rangle$ and $S$, for four different $M_{1}=500,2000,3000$
and $4500$.

\section{Alternative crosstalk definition}
\label{sec:Alt_xtalk}
In the basic setup presented in the main text, we considered ``activation out-of-context''---i.e., activation by the binding of a noncognate TF when the cognate TF is present (but not bound)---to be a crosstalk state. Our reasoning was motivated by viewing transcriptional regulation as a signal transmission apparatus. In this interpretation, gene activation by a noncognate TF amounts to generating a response (transcriptional activity) to a wrong input signal. Consequently, this should count as crosstalk, despite the fact that (by chance) the correct signal was simultaneously present in the cell. This is perhaps easiest to appreciate if one considers more realistic setups in which genes are not simply ``ON'' and ``OFF'', but can be quantitatively regulated by the level of their cognate TF.  In such a model, there might be two TFs present and varying in concentration as a function of time: one cognate for the gene of interest and one not. In this case it is clear that the correct response of the gene is to  track the changes in the cognate TF, and not  to simply be expressed in a constant ``ON'' state; consequently, tracking the noncognate TF due to crosstalk is  obviously an error, even if the cognate TF is present at the same time.

One could, however, argue that ``activation-out-of-context'' shouldn't be considered as an error state.
If the presence or absence of TF signals is a binary variable and if the binary response is defined solely by the state of transcriptional activity (activation/inactivation of gene), then when the presence of the signal matches the response state, the regulation outcome is correct, irrespective of the molecular details on the promoter. For example, for a gene whose cognate TF is present, activation by any means (either by cognate or noncognate binding) is the correct response. In this scenario, the "out-of-context activation" is actually what one might call beneficial crosstalk: here, noncognate TF can be seen as helping to activate the gene when the cognate TF is also present. For a gene whose cognate TF is absent, activation
is still an incorrect response, like before.

Hence, $x_{2}(i)$ retains the same expression,
but $x_{1}(i)$ changes to

\begin{eqnarray}
x_{1}(i)=\dfrac{e^{-E_{a}}}{C_{i}+e^{-E_{a}}+{\displaystyle \sum_{j\ne i}C_{j}e^{-\epsilon d_{ij}}}}.
\end{eqnarray}

As shown in Fig.~\ref{fig:alt_err}, optimizing $C$ results in three distinct regulatory regimes, like
in the default basic setup. For small $S$ in the regulation regime, the optimal
$C$ is given to the leading order by:
\begin{eqnarray}
C^{*} & \sim & \dfrac{e^{-E_{a}}}{\sqrt{S}}\dfrac{Q}{\sqrt{M-Q}}
\end{eqnarray}

The minimal crosstalk error at the optimal concentration $C^{*}$ is given by
\begin{eqnarray}
X^{*}=-SQ+2\dfrac{Q}{M}\sqrt{S(M-Q)(1+SQ)}
\end{eqnarray}

\begin{center}
\includegraphics[scale=0.65]{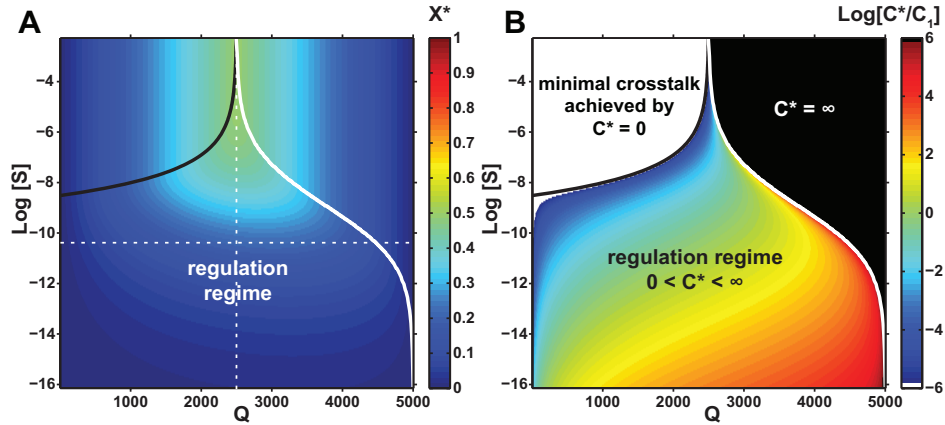}
\captionof{figure}{\textbf{Basic model with alternative crosstalk definition also exhibits
three distinct regulation regimes.} The alternative definition does not count ``activation out-of-context'' as an error state.
{\bf (A)}
Minimal crosstalk error, $X^{*}$, shown in color, as a function of
the number of coactivated genes $Q$, and binding site similarity
$S$. {\bf (B)} Optimal TF concentration $C^{*}$, that minimizes the crosstalk,
relative to $C_{0}$, the optimal concentration at the baseline parameters
(see main text).}\label{fig:alt_err}
\end{center}

\section{Estimating the binding site similarity, $S$}

\subsection{Optimal packing}

In real organisms, binding site sequences for different genes could depart from a random distribution (even after taking into account the statistical structure of the genomic background). For example, to achieve high specificity of regulation, we could hypothesize that binding site sequences evolved to minimize the overlap between any pair of consensus sequences.
To explore the crosstalk limit under such optimal use of sequence space and contrast it with the random choice of binding sites, we synthetically constructed  binding site sequences that are as distinct as possible. Specifically, our optimal codes are described by a parameter $d_{\min}$, which is the minimum required
number of basepair differences between any pair of binding site sequences. This is the Hamming distance, $HD$, between sequences. The problem of choosing $M$ sequences of length $L$ such that each
pair differs by at least $d_{\min}$ is not tractably solvable in general. We
construct numerical approximations to these optimal codes using the following algorithm:
\begin{enumerate}
\item Generate all possible sequences of length $L$ and store them in a list called $words$. Create an empty list, called $codewords$, which will store the binding site sequences.
\item Pick the first entry, $s$, from the list $words$, to be a binding site sequence, and append it to the list $codewords$.
\item Erase $s$ and all of its Hamming neighbours at distance strictly less than $d_{\min}$ from the list $words$.
\item If the list $words$ is not empty, repeat from step $2$. If the list $words$ is empty, stop.
\end{enumerate}

When the procedure terminates, the list $codewords$ will contain binding site sequences that are separated by at least $d_{\min}$ mismatches. The outcome of this procedure depends on the initial ordering of the list of all possible sequences. The procedure is not guaranteed to generate the maximal set of sequences satisfying the Hamming distance criteria. From
the list of generated binding site sequences, we obtain $P(d)$, the distribution of mismatch distances between
all pairs of binding sites, and hence obtain the value of $S$ as

\begin{eqnarray}
\tilde{S}(d_{\min})={\displaystyle \sum_{d\ge d_{\min}}P(d)e^{-\epsilon d}}.
\end{eqnarray}

\begin{center}
\includegraphics[width=0.45\textwidth]{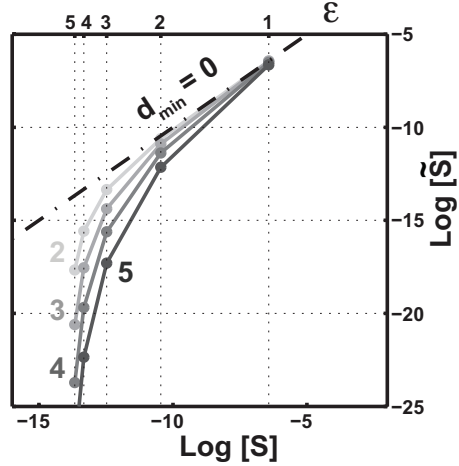}
\captionof{figure}{\textbf{Optimal packing.} This alternative model with optimal packing of binding sites in sequence space
leads to values for $\tilde{S}$ (y-axis) that can be remapped to the $S(\epsilon, L)$ (x-axis) for the random code with the
mismatch energy model, $E(d) = \epsilon d$ and $L = 10$ bp binding sites (corresponding scale for $\epsilon$ shown in the top
axis). Dashed lines denote equality. Optimally designed binding sites effectively decrease $S$. Here, their sequences are at
least $d_{\rm min}$ bp distant from each other (gray lines = different $d_{\rm min}$ as indicated).}\label{fig:2D}
\end{center}

$d_{\min}=0$ corresponds to the "random code'' and results in
$\tilde{S}(d_{\min}=0)=S=(\frac{1}{4}+\frac{3}{4}e^{-\epsilon})^{L}$.
Note that increasing $d_{\min}$ decreases the maximum possible $M$
as sequences move further apart in sequence
space whose size is fixed. A well-known upper bound on the number of sequences satisfying the Hamming distance criterion is the Singleton bound~\cite{_error_2004}:
$M(d_{\min},L)\le4^{L-d_{\min}+1}$. As shown in Fig.~\ref{fig:coding_bounds}, with $L=8$ and $d_{\min}=3$,
we already have $M\le4096$. With $L=10$ and $d_{\min}=4$, we have $M\le16384$. As $L$ becomes smaller, the possible
range of $M$ also decreases. This suggests that prokaryotes are
capable of having optimally packed binding site sequences, because
they typically have $L>10$  and $M<10^4$. On the other hand, eukaryotes
have smaller $L$ and larger $M$ and might not have enough sequence space to pack it optimally.

\subsection{Reverse complemented sequences}

We have also considered a different definition of distance between sequences that takes the double-stranded nature of DNA into account. This brings into picture the reverse complement of both  sequences in question. If $s_{i}$ and $s_{j}$ are two sequences with reverse complements $r_{i}$ and $r_{j}$ respectively, this new definition of Hamming distance is

\begin{eqnarray}
HD_{rc}(s_{i},s_{j})=\min\Big[HD(s_{i},s_{j}),HD(r_{i},s_{j}),HD(s_{i},r_{j}),HD(r_{i},r_{j})\Big]
\end{eqnarray}

where $HD(s_i,s_j)$ is the usual Hamming distance as considered previously. This restricts the sequence space much more than with the usual definition
and as such, as seen in Fig. \ref{fig:coding_bounds}, we can pack fewer binding
sites in the sequence space at a specific $d_{\min}$. Given that there are enough sequences under $HD_{rc}$ measure in the sequence space, we can also ask how $S$ changes in relative to the random code. Intuitively, $S$ should increase since each binding site sequence also contributes its reverse complement into the pool of sequences to which TFs can bind non-cognately. Indeed,  Fig.~\ref{fig:Src}, which maps $S$ from the reverse complement code to $S$ from a random code, shows that  $S$ increases by about a factor of $2$ due to the addition of reverse-complemented sites.

\begin{center}
\includegraphics[scale=0.7]{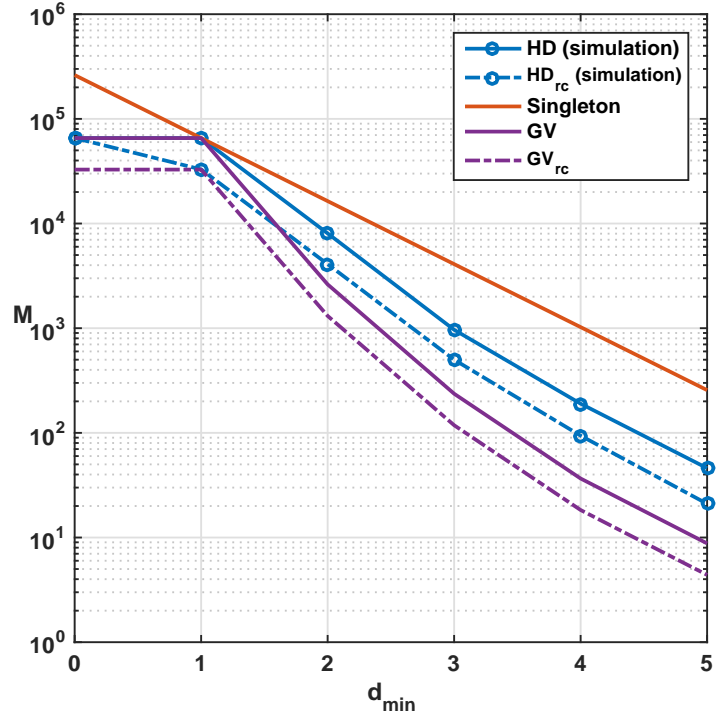}

\captionof{figure}{\textbf{Bounds on the maximal number of binding site sequences for different $d_{\min}$ with binding sites of length $L=8$.}
Two bounds from the coding theory (Singleton upper bound and Gilbert-Varshamov (GV) lower bound~\cite{_error_2004}) are shown together with the values of $M$ obtained by our numerical approximation procedure. These are shown both for the usual definition of distance between sequences as the Hamming distance, $HD$, as well as for a definition that considers the reverse complements of the sequences, $HD_{rc}$. For $d_{\rm min}=0$ there are $M=4^{8}\approx 65000$ possible sequences where all sequence pairs are at least $d_{\rm min}$ distant from each other, but the number quickly decreases with increasing $d_{\rm min}$. From the $HD$ to $HD_{rc}$, the Singleton bound doesn't change from the usual situation
but the Gilbert-Varshamov (GV) bound, which takes into account the ``volume
of restricted ball'' around each sequence, goes down. Because of stronger
constraints, the number of sequences that can be packed goes down
from the usual situation but only by a factor of $\approx2$.}\label{fig:coding_bounds}
\end{center}

\begin{center}
\includegraphics[width=0.45\textwidth]{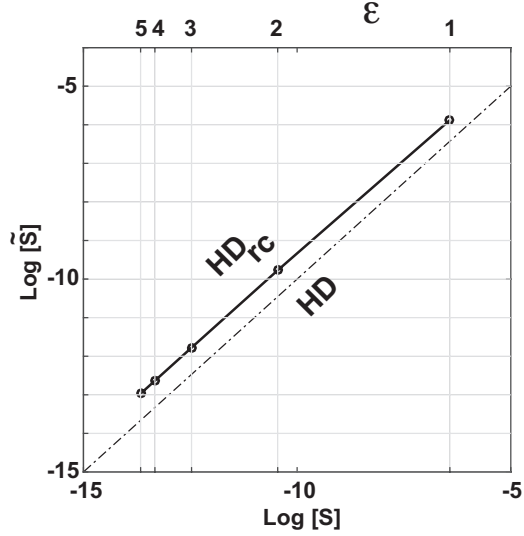}
\captionof{figure}{\textbf{Reverse complemented sequences.} Using an alternative definition of distance ($HD_{rc}$) between binding site sequences, which takes into account the double-stranded nature of DNA by considering the reverse complements as well of the sequences in question, leads to values for $\tilde{S}$ (y-axis) that can be remapped to the $S(\epsilon, L)$ (x-axis) for the random code with the usual Hamming distance definition, $HD$. Here, we have considered $L = 8$ bp binding sites (corresponding scale for $\epsilon$ shown in the top axis). Dashed lines denote equality. This alternative definition increases $S$ because more sequences are now found in the ``shells'' around the consensus to which the TF can bind on the reverse strand. $S$ increases by about a factor of $2$.}\label{fig:Src}
\end{center}

\subsection{Saturating model of TF-DNA binding energy}

It has been experimentally observed that the binding energy between
TF and DNA saturates to some nonspecific value after a certain number
of mismatches between the TF's cognate sequence and the DNA sequence
in question \cite{maerkl_systems_2007}. We consider such a saturating
energy model, characterized by a parameter $d_{0}$, the number of
mismatches after which binding energy saturates. The binding energy
is given by $E(d)=\epsilon\min(d,d_{0})$. We obtain $S$ as

\begin{eqnarray}
\tilde{S}(d_{0})={\displaystyle \sum_{d}P(d)e^{-E(d)}},
\end{eqnarray}

where $P(d)$ is the distribution of mismatch distances between all
pairs of binding sites picked at random from the sequence space. $d_{0}=L$
corresponds to a mismatch model with non-saturating energy. Decreasing $d_0$ limits the specificity of the TF towards binding site sequences far away from the consensus and thereby increases $\tilde{S}(d_{0})$.

\begin{center}
\includegraphics[width=0.45\textwidth]{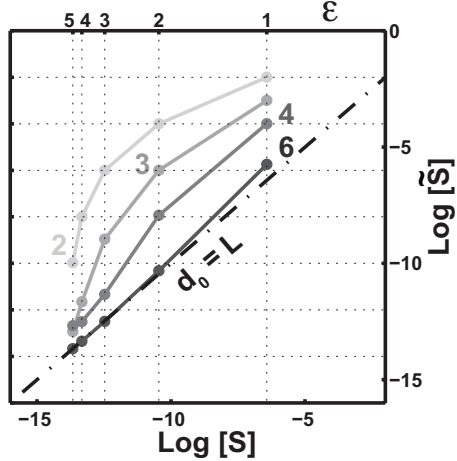}
\captionof{figure}{\textbf{Saturating energy model.} An improved affinity model where the mismatch energy saturates after
$d_0$ mismatches, $E(d) = \epsilon\min(d,d_0)$ (gray lines = different $d_0$ as indicated), effectively increases $S$.
$d_0 \sim 4$ has been reported experimentally \cite{maerkl_systems_2007}. This alternative model leads to values for
$\tilde{S}$ (y-axis) that can be remapped to the $S(\epsilon, L)$ (x-axis) for the random code with the mismatch energy model,
$E(d) = \epsilon d$ and $L = 10$ bp binding sites (corresponding scale for $\epsilon$ shown in the top axis).
Dashed lines denote equality.}\label{fig:2C}
\end{center}

\subsection{Empirical values}

We obtain organism-specific estimates of $S$ from known databases \cite{gama-castro_regulondb_2011, mathelier_jaspar_2013, spivak_scertf:_2012}
of the binding site sequences of different TFs. In the main text,
for a particular genome, we defined $S$ for a collection of TFs with
the same mismatch penalty $\epsilon$ and binding sites of a specific constant length
$L$. In real organisms, different TFs have different $\epsilon$
and $L$, making it difficult to directly calculate $S$ for a genome.
Instead we obtain a value of $S$ for each TF by defining it as the
value of $S$ of a hypothetical genome in which all TFs have the same
binding site properties $(\epsilon,L)$ as our TF. Hence, for each
organism, we obtain a set of $S$ values.

Many databases document the binding site sequences of TFs in Position
Count Matrices (PCMs). The PCM of a TF with a binding site of length
$L$ is a $4\times L$ matrix $B$ with $b_{ij}$ denoting the number
of known TF binding site sequences that have nucleotide $i$ in position
$j$. One can obtain estimates of $\epsilon$ and $L$ from $B$,
and use them to calculate $S$. There are two broad ways to estimate
$\epsilon$ and $L$ (and hence, $S$) of a TF: (a) Information method,
(b) Pseudo-count method. In (a), we calculate the information contained
in the whole binding site motif and obtain an $\epsilon$ that distributes
this information uniformly among all sites in an equivalent "effective" motif that has the same length as the original, but only has $0$ or $\epsilon$ mismatch energy values. In (b), we
obtain $\epsilon$ for all entries of the PCM and calculate an average
$\epsilon$ from these entries. To handle zeros in the PCM which lead
to undefined $\epsilon$, (b) uses an arbitrary pseudo-count. Method (a) can, in contrast, avoid the use of pseudo-counts and, additionally, reproduces by construction the information content of each known motif, which is the key statistical property of TF specificity \cite{wunderlich_different_2009, schneider_information_1986}. Hence,
we used (a) to infer $S$ values. In both the methods, we used PCMs
that have that have been constructed from at least $10$ distinct binding site sequences.

\subsubsection{Information method }

In this method, we first obtain the binding site length $L$ and also the total information $I$, contained
in the binding site sequences of the TF.

\begin{eqnarray}
I & ={\displaystyle \sum_{j}I_{j}}={\displaystyle \sum_{j}\sum_{i}p_{ij}\log_{2}\dfrac{p_{ij}}{q_{ij}}},
\end{eqnarray}

where $I_{j}$ is the information contained in position $j$, $p_{ij}$
is the frequency of nucleotide $i$ in position $j$, obtained in a straightforward
way from $B$, and $q_{ij}$ is the expected background frequency.
To get rid of non-specific positions, we neglect all positions that
contain information less than a certain threshold ($I_{j}>0.2$ bits for position $j$ to be considered part of the binding site). For a random genome,
$q_{ij}=0.25$ $\forall$ $i,j$, resulting in

\begin{eqnarray}
I=2L+\sum_{i,j}p_{ij}\log_{2}p_{ij}
\end{eqnarray}
The maximum information in the motif is $2L$ bits (when $\epsilon \rightarrow \infty$) with each position
contributing a maximum of $2$ bits, which for finite $\epsilon$, is reduced by an entropy term. Obtaining information per position
$I_{pos}=I/L$, we infer an $\epsilon$ that uniformly distributes
the information in the motif among individual positions. At a specific
position $j^{*}$, without loss of generality, assume that $i=4$
has the best binding energy $(=0)$. The probability of observing
$i=4$ at $j^{*}$ is given by $p_{4}=1/Z$ while the probability
of observing any of the three other possible nucleotides is given
by $p_{1,2,3}=e^{-\epsilon}/Z$, with $Z=1+3e^{-\epsilon}$ \cite{berg_selection_1987}. Hence,

\begin{eqnarray}
I_{pos}= & 2+{\displaystyle \sum_{i}}p_{i}\log_{2}p_{i}\\
= & 2-\dfrac{1}{Z}\log_{2}Z+3\dfrac{1}{Z\:\ln2}\epsilon e^{-\epsilon}-3\dfrac{e^{-\epsilon}}{Z}\log_{2}Z\\
= & 2-\log_{2}Z+3\dfrac{1}{Z\:\ln2}\epsilon e^{-\epsilon}
\end{eqnarray}
The mismatch energy $\epsilon$ can be obtained from the above expression, and from $\epsilon$
and $L$, we obtain $S(\epsilon,L)=(\frac{1}{4}+\frac{3}{4}e^{-\epsilon})^L$.

\subsubsection{Pseudo-count method}

In this method, we infer $\epsilon$ for all three non-cognate nucleotides
in each position, and obtain $\epsilon$ for the TF as an average
of these $3L$ values. For an arbitrary position $j$, as before, assume that
$i=4$ has the maximum counts $(b_{4j}>b_{ij}\:,i=1,2,3)$. We obtain
$\epsilon_{ij}=\log\frac{b_{4j}}{b_{ij}}$ and mismatch penalty for
position $j$ as $\epsilon_{j}=\frac{1}{3}(\epsilon_{1j}+\epsilon_{2j}+\epsilon_{3j})$.
If some entry $b_{kj}=0$, $\epsilon_{kj}$ is undefined. To take
care of this, we first add a pseudocount $\delta$ to all entries
of $B$ and obtain a modified PCM $B_{\delta}$ to infer $\epsilon$.
The value of $\delta$ chosen is arbitrary and it is common practice
to use $\delta=0.5$ or $\delta=1$. As before, to get rid of non-specific
positions, we consider positions that have $\epsilon_{j}\ge1$.
From the remaining, we take a mean to obtain $\epsilon=\frac{1}{L}{\displaystyle \sum_{j}\epsilon_{j}}$,
and finally obtain $S(\epsilon,L)=(\frac{1}{4}+\frac{3}{4}e^{-\epsilon})^L$.

\begin{center}
\includegraphics[scale=0.6]{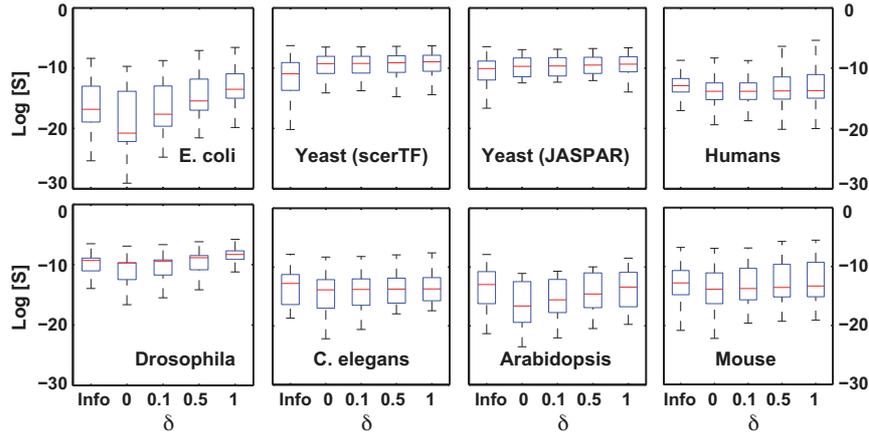}

\captionof{figure}{\textbf{Boxplots of $S$ for TFs from different databases.} In each
panel, organism-specific (from a single database) boxplots of $S$
are shown. The first boxplot in each panel corresponds to $S$ values
obtained from information estimates, and the remaining four correspond
to $S$ values obtained using the psuedo-count method with $\delta=0,0.1,0.5,1$
from left to right.\textit{ E. coli} TFs were obtained from RegulonDB
\cite{gama-castro_regulondb_2011} and yeast (\emph{S. cerevisiae}) from two different databases - scerTF \cite{spivak_scertf:_2012} and JASPAR \cite{mathelier_jaspar_2013}. All the other organism specific TFs were obtained from JASPAR. Notice that in the pseudo-count method, $\delta$ has
the biggest influence on the estimates in \textit{E. coli. } Importantly, for all other organisms, the estimates are invariant to $\delta$ and agree well with the information estimate.}\label{fig:boxplots_A}
\end{center}


\section{Cooperative regulation}
So far, we assumed a single binding site for every gene. Yet, some genes employ combinatorial regulation, with several binding sites regulated by a number of transcription factors. As a next step in extending our model we consider cooperative regulation, where every gene has two binding sites that are bound by two copies of the same type of transcription factor.

We assume 2 binding sites per gene, with energy gap $E_a$ between cognate-bound and unbound states. An additional energy contribution $\Delta$ is obtained if both sites are bound by cognate factors, which then interact with each other. We consider also the configuration that two noncognate factors \emph{of the same type} bind to the double binding sites and interact with each other as well.
In the limit that $\Delta\gg E_a$ once one of the sites is bound, the binding of the other becomes energetically favorable. This cooperative binding energy only applies for two molecules of the same type. Thus, if one site is bound by the cognate and the other by a noncognate molecule, cooperative interaction doesn't apply.
We assume that only binding of one of the two sites induces transcription. The reasoning for this assumption is that for many bacterial and yeast genes activators are thought to work by recruiting the transcriptional machinery to the DNA~\cite{ptashne_transcriptional_1997}. Following this rationale, only one of the two sites is in the correct physical location (in bacteria, the proximal one) to do so successfully.  Technically, if we assume that only one of the two sites determines transcription, for $\Delta=0$, the cooperativity case reduces back to the basic model (Section \ref{sec:BasicModel}). We list the possible binding configurations of the two sites, their energies and statistical weight in Table \ref{Tab:CooperativeConfig}.

\begin{table}[h!]
\begin{tabular}{|c|c|c|c|c|c|c|c|}
  \hline
  & configuration & activity &  crosstalk  & crosstalk  & strong         & Energy             & Weight \\
  &               &          &   if ON     &  if OFF    &  cooperativity &                    &         \\&&&&&&&\\
  \hline
  1 & CC          & ON       & -           &            & + & 0                            & $(C/Q)^2$
  \\&&&&&&&\\
  2 & UC          & ON       & -           &            &   & $E_a + \Delta$               & $C/Q e^{-E_a-\Delta}$ \\&&&&&&&\\
  3 & NC          & ON       & -           &            &   & $\Delta + \epsilon d$        & $C^2/Q S e^{-\Delta}$ \\&&&&&&&\\
  4 & UU          & OFF      & +           &  -         & + & $2E_a + \Delta$              & $e^{-2E_a-\Delta}$     \\&&&&&&&\\
  5 & CU          & OFF      & +           &  -         &   & $E_a + \Delta$               & $C/Q e^{-E_a-\Delta}$ \\&&&&&&&\\
  6 & NU          & OFF      & +           &  -         &   & $E_a + \Delta + \epsilon d$  & $CS e^{-E_a-\Delta}$ \\&&&&&&&\\
  7 & UN          & *        & +           &  +         &   & $E_a + \Delta + \epsilon d$  & $CS e^{-E_a-\Delta}$ \\&&&&&&&\\
  8 & CN          & *        & +           &            &   & $\Delta + \epsilon d$        & $C^2/Q S e^{-\Delta}$ \\&&&&&&&\\
  9 & $N_xN_y$    & *        & +           &  +         &   & $\Delta + \epsilon(d_1+d_2)$ & $C^2S^2 e^{-\Delta}$  \\&&&&&&&\\
  10& $N_xN_x$    & *        & +           &  +         & + & $2\epsilon d$                & $\frac{C^2}{Q}S(2\epsilon,L)$ \\&&&&&&&\\
  \hline
\end{tabular}
\caption {
\label{Tab:CooperativeConfig}
All possible binding configurations and the corresponding energies for a two-binding site model with cooperative interaction. 'C' denotes binding by cognate factor, 'N' - binding by noncognate and 'U' - means that the site is unbound. We distinguish between binding of noncognate molecules of the same type ($N_xN_x$) and different types ($N_xN_y$), where in the former there is also cooperative interaction between the molecules. We define the reference energetic level $E=0$ as the state 'CC' when both sites are bound by cognate factors with cooperative interaction, such that all other energies are positive. We assume that the left binding site is the auxiliary and only the right one determines the state of activity. Note that the statistical weight of the last binding configuration $N_xN_x$ uses $S(2\epsilon,L)$ instead of the otherwise $S(\epsilon,L)$.
The column 'activity' denotes whether in the given configuration the gene is either ON, OFF or * - could be either active or inactive (possibly active in response to noncognate signal). Blank space denotes a non-existing configuration (or one which is not accounted for): these are the configurations including a cognate factor bound in the situation that it is absent because the gene should be silent. The next two columns denote whether this configuration was counted as crosstalk (+) or not (-) if the cognate transcription factor is present and the gene should be activated or if it is absent (and the gene should be silenced). The 'Strong Cooperativity' column denotes the configuration included under strong cooperativity approximation.  }
\end{table}

The general case of this model, incorporating all possible binding configurations yields a 6th order equation in the TF concentration $C$, which we only handle numerically. The following limiting cases are however analytically solvable:

\begin{enumerate}
\item
Limit of strong cooperativity: Assume that the cooperative interaction is strong compared to the individual protein-DNA binding energies $\Delta \gg E_a$. We can then neglect binding configurations in which only one of the sites is bound and the other is vacant, and the ones in which both are bound, but by molecules that do not interact cooperatively. That leaves us with only 3 possible binding configurations: both sites unbound, both bound by cognate TF or both bound by noncognate TF molecules of the same type with cooperative interaction (configurations 1,4 and 10 in Table \ref{Tab:CooperativeConfig}). By proper change of variables this case can be reduced back to the basic single-binding-site model.
The minimal crosstalk then reads:
\be
X_{\mbox{coop}}^\ast = \frac{-Q \left(\tilde{S} (M - Q)+2 \sqrt{\tilde{S} (M-Q)}\right)}{M},
\ee
where $\tilde{S}=S(2\epsilon, L)$. This error is achievable with TF concentration
\be
C_{\mbox{coop}}^\ast = Q \sqrt{-\frac{e^{-\Delta -2 E_a} \left(\tilde{S} (M-Q)-1\right)}{\left(\tilde{S} \left(\tilde{S} Q (M-Q)+M-2 Q\right)+\sqrt{\tilde{S} (M-Q)}\right)}}.
\ee
Since the cooperative binding model allows for a binding site which is twice as long and higher total binding energy the parameters need to be correctly transformed to compare to the 1-site model. If we transform: $\tilde{S} \rightarrow S$ we obtain exactly the same minimal error as in the single-site model. By proper transformation of the energy of the unbound state $\tilde{E_a}=\Delta + 2E_a$ the TF concentration that minimizes the error is a square root of the one we had in the single-site model \eqref{eq:Copt}.
In similarity with the basic single-site model, here too we obtain different parameter regimes, whereas For $\tilde{S} = S(2\epsilon,L)>\frac{1}{M-Q}$ the minimal error is obtained by taking $C=0$, namely regulation is not advantageous.
While seemingly the cooperative binding is equivalent to a 1-site model which has twice as long binding site, this is not accurate. The reason is that cooperative interaction occurs only between two specific molecules, which limits the possible sequence space.
\item
Limit of weak cooperativity: If $\Delta=0$, the problem reduces to the basic single-site model.
\end{enumerate}

\subsection{Cooperativity with interactions between noncognate pairs}
\begin{figure}[h!]
\centering
\subfigure[]{
\includegraphics[width=0.45\textwidth]{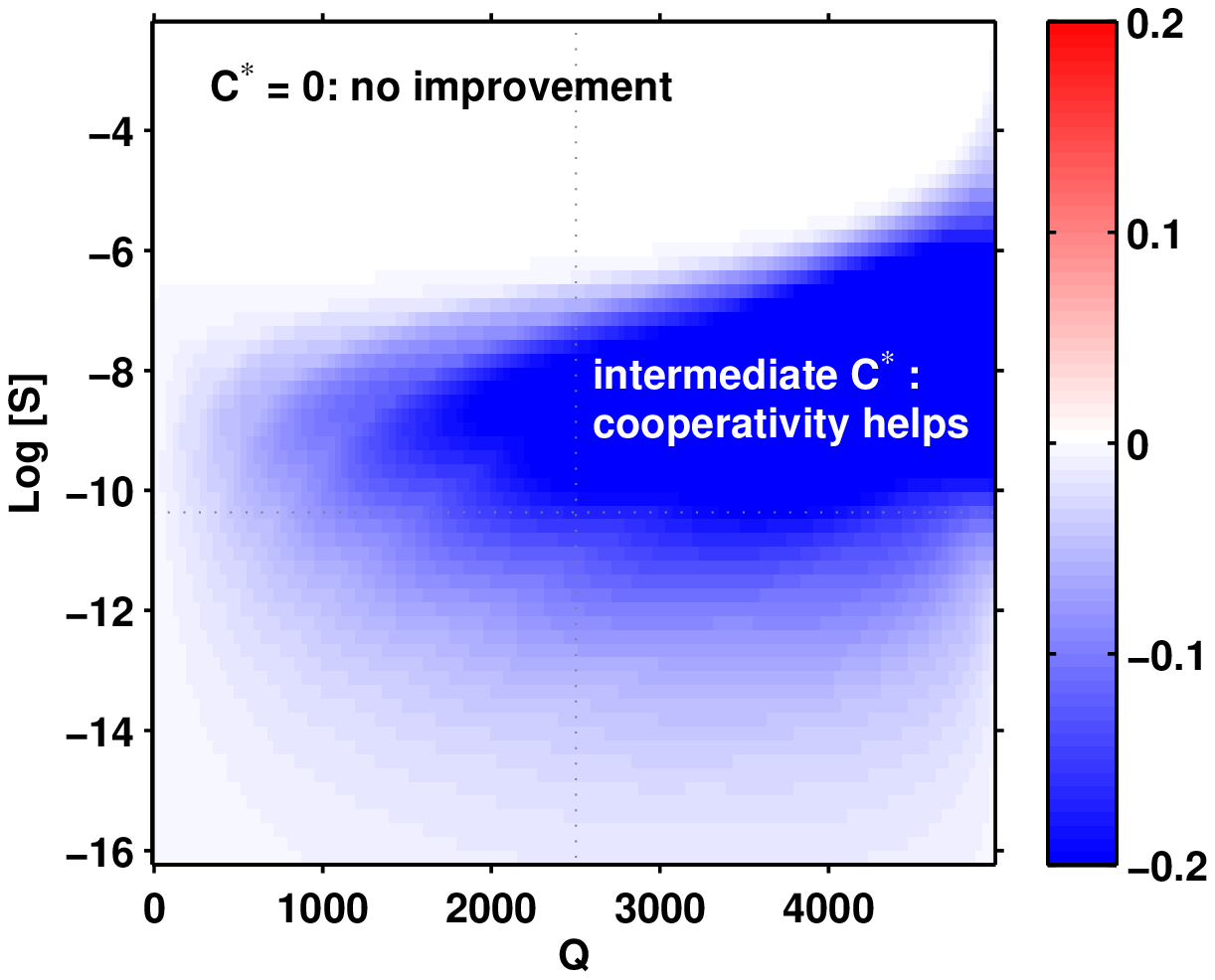}
\label{fig:5B_noncog} }
\subfigure[]{
\includegraphics[width=0.45\textwidth]{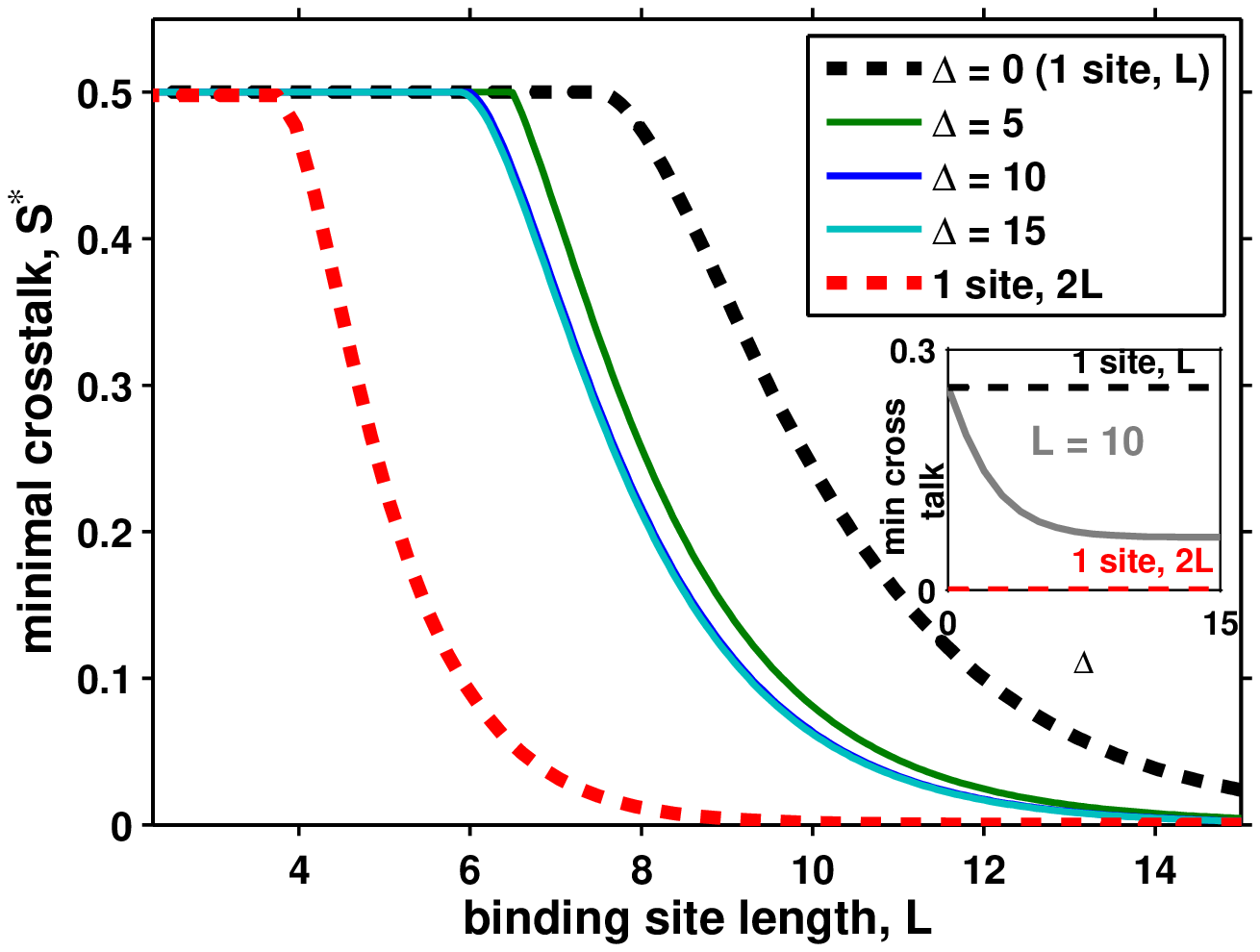}
\label{fig:5C_noncog} }
\subfigure[]{
\includegraphics[width=0.45\textwidth]{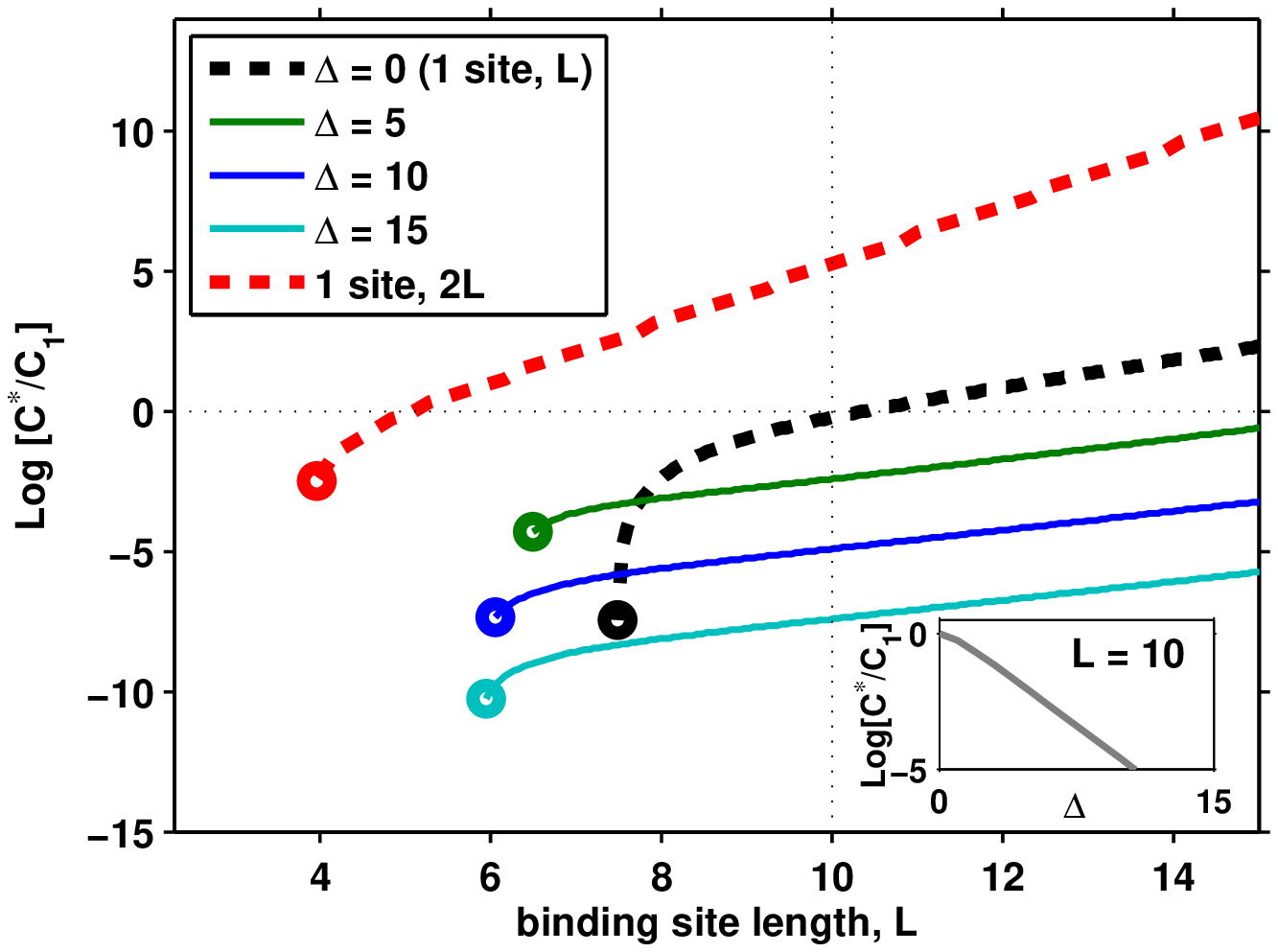}
\label{fig:5D_noncog} }
  \caption[]
 { \label{fig:noncog_coop} {\bf Crosstalk when any pair of the same type TFs interacts cooperatively, even if bound to noncognate site.} Here we repeat the calculation of Fig. 4 of the main text where we also account for cooperative interaction between the noncognate binders. This significantly decreases the benefit of cooperative interaction, although it still shows some improvement compared to the single-site basic model.
  (a): Difference in crosstalk compared to the basic model with single site, $X^*_{\mbox{coop}}-X^*$, where the strength of the cooperative interaction is $\Delta=10$. One outcome of this is that the $C^*=0$ (no regulation regime) becomes significantly larger (compare to Fig. 4B). (b): Minimal crosstalk obtained for different intensities of cooperative interaction. In contrast to the case shown in the main text Fig. 4C, where increased cooperativity always reduces crosstalk, here the improvement is limited. For example, increasing cooperativity from $\Delta=5$  to $\Delta=10$ brings about only a minor improvement.
  (c): Optimal TF concentration decreases with increased cooperativity, as in Fig. 4D. Circles denote transition to $C^*=0$ - no regulation regime.   }
  \end{figure}

In Fig. 4 of the main text  we neglected the possibility of cooperative interaction between pairs of noncognate molecules at the binding site of interest. This situation is plausible if the interaction between the molecules is facilitated by the specific binding sites.
However, the molecules can also cooperatively interact in solution before binding and then bind a noncognate site as a complex.
This possibility was not taken into account in Fig. 4 (main text).
In the following we repeat the calculation including this interaction too (state no. 10 in Table~\ref{Tab:CooperativeConfig}).
The results are illustrated in \figref{fig:noncog_coop}. Evidently, the improvement in crosstalk owing to cooperativity is now significantly smaller.

\section{Combinatorial regulation (AND gate)}
\label{sec:combReg}
So far, we have been dealing with models in which each gene is regulated
by a single type of TF, be it by a single activator, a single repressor,
or multiple TFs of the same type using cooperative interactions. Here, we will consider a simple model
of combinatorial regulation by a combination of two activators of different types, and compute optimal crosstalk for this
setup as a function of parameters of interest.

As before, we have $M$ genes in total, with each gene having two
binding sites, corresponding to two different (cognate) TF types. For a particular gene to be ON, we need the presence of both cognate TF types, which need to occupy both  binding sites. This  regulatory architecture corresponds to an AND gate. We don't specify how this AND gate is implemented  on the molecular level. Unlike in cooperative regulation, no additional energy gain is assumed here due to the interaction between the two TFs when bound to the DNA.

Each TF can pair  with various other TFs in regulating a particular gene. In the basic activation setup, the total number of TFs, $M$, was equal to the total number of genes. In the combinatorial regulation setup, which is an extension of the basic activation setup, the total number of genes $M$ will be equal to the total number of different TF-TF combinations that can exist. This will depend on the extent of combinatorial regulation, which we quantify using $f$, the fraction of TF-TF combinations each TF type realizes out of the theoretically maximal number of pairwise combinations  it could have.


If there are $T$ TFs in total, each TF can potentially pair with
$N_{\rm int}=f(T-1)$ other TF types, where $f$ is the fraction of pairs each TF type realizes. This gives us $M=TN_{\rm int}/2$, and thus $T\approx\sqrt{2M/f}$ and $N_{\rm int}\approx\sqrt{2Mf}$.
But each TF should pair with at least one other TF, so we require
$N_{\rm int}\ge$1. Taking both of these limits into account, we have, for $N_{\rm int}$,
the number of TFs each TF pairs with, and the number of total
TFs $T$,

\begin{eqnarray}
N_{\rm int}=\max(1,\sqrt{2Mf})\\
T=\dfrac{2M}{N_{\rm int}}.
\end{eqnarray}
If each TF pairs with all other TFs, we have $f=1$ and $N_{\rm int}=T-1$,
which gives us $T\approx\sqrt{2M}$. We call this ``perfect combinatorial
regulation'' because it minimizes the number of TFs needed to regulate
a certain number of genes.

If each TF realizes only a fraction $1/2M<f<1$ of its combinations,
we have $N_{\rm int}>1$ pairs for each TF, which gives us $T\approx\sqrt{2M/f}$.
We call this ``imperfect combinatorial regulation''.

If $f\le1/2M$, we have $N_{\rm int}=1$, which gives us $T=2M$. We call this ``worst combinatorial regulation''.

As before, we will compute the optimal crosstalk when $Q$ genes are
required to be ON. Here, we compute the ``typical'' number of TFs present at any one time, $t$, by following a similar recipe as before. We have $Q=tn_{\rm int}/2$,
where $n_{\rm int}$ is the number of pairs per TF present at any one time. This will
be smaller as there are fewer TFs present at any given time relative to the total number of TF types, i.e., $t\le T$. As before, we have

\begin{eqnarray}
n_{\rm int}=\max(1,\sqrt{2Qf})\\
t=\dfrac{2Q}{n_{\rm int}}.
\end{eqnarray}

When $f>1/2Q$, we have $t=\sqrt{2Q/f}$ and when $f\le1/2Q$, we
have $t=2Q$.

\begin{table}
\begin{tabular}[b]{|c|c|c|>{\centering}m{1cm}|>{\centering}m{1cm}|>{\centering}m{1cm}|>{\centering}m{1cm}|c|c|}
\hline
 &  &  & \multicolumn{4}{c|}{crosstalk if gene needs to be} &  & \tabularnewline
 & configuration & activity & \multirow{2}{1cm}{ON} & \multicolumn{3}{c|}{OFF, C can be} & Energy & Weight\tabularnewline
\cline{5-7}
 & (XY) &  &  & X & Y & none &  & \tabularnewline
\hline
 &  &  &  &  &  &  &  & \tabularnewline
1 & CC & ON & - &  &  &  & 0 & $(C/t)^{2}$\tabularnewline
 &  &  &  &  &  &  &  & \tabularnewline
2 & UC & OFF & + &  & - &  & $E_{a}$ & $e^{-E_{a}}(C/t)$\tabularnewline
 &  &  &  &  &  &  &  & \tabularnewline
3 & NC & ON & + &  & + &  & $\epsilon d$ & $(C/t)CS$\tabularnewline
 &  &  &  &  &  &  &  & \tabularnewline
4 & CU & OFF & + & - &  &  & $E_{a}$ & $e^{-E_{a}}(C/t)$\tabularnewline
 &  &  &  &  &  &  &  & \tabularnewline
5 & CN & ON & + & + &  &  & $\epsilon d$ & $(C/t)CS$\tabularnewline
 &  &  &  &  &  &  &  & \tabularnewline
6 & UU & OFF & + & - & - & - & $2E_{a}$ & $e^{-2E_{a}}$\tabularnewline
 &  &  &  &  &  &  &  & \tabularnewline
7 & UN & OFF & + & - & - & - & $E_{a}+\epsilon d$ & $e^{-E_{a}}CS$\tabularnewline
 &  &  &  &  &  &  &  & \tabularnewline
8 & NU & OFF & + & - & - & - & $E_{a}+\epsilon d$ & $e^{-E_{a}}CS$\tabularnewline
 &  &  &  &  &  &  &  & \tabularnewline
9 & $N_{x}N_{y}$ & ON & + & + & + & + & $\epsilon(d_{1}+d_{2})$ & $(CS)^{2}$\tabularnewline
 &  &  &  &  &  &  &  & \tabularnewline
10 & $N_{x}N_{x}$ & ON & + & + & + & + & $2\epsilon d$ & $(C/t)CS(2\epsilon,L)$\tabularnewline
 &  &  &  &  &  &  &  & \tabularnewline
\hline
\end{tabular}

\caption{All possible binding configurations and the corresponding energies
for a combinatorial regulation setup implementing an AND gate. Each
gene has two binding sites which bind two different cognate TF types.
The ``configuration'' column lists all the configurations of the
two binding sites of a gene. 'C' denotes binding by cognate factor,
'N' - binding by noncognate and 'U' - means that the site is unbound.
We distinguish between binding of noncognate molecules of the same
type ($N_{x}N_{x}$) and different types ($N_{x}N_{y}$). The ``activity''
column denotes whether in the given configuration the gene is either
ON or OFF. To implement the AND gate, we assume that transcription
occurs (ON) only when both the binding sites are bound. The next four
columns denote whether this configuration is counted as crosstalk
(+) or not (-). In the leftmost column ``ON'', both the cognate
transcription factors are present (and the gene should be ON). In
the next three ``OFF'' columns, at least one of the cognate TFs
is absent (and the gene should be OFF). In ``C can be X'' column,
the cognate TF of only the left binding site (X) is present, in ``C
can be Y'', the cognate TF of only the right binding site is present,
and in ``C can be none'' column, both the cognate TFs are absent.
Blank space denotes a non-existing configuration: these are the configurations
including a cognate factor bound in the situation that it is absent.
The column ``Energy'' specifies the energy of these configurations.
We define the reference energetic level $E=0$ as the state 'CC' when
both sites are bound by their cognate factors, such that all other
energies are positive. The column ``Weight'' denotes the statistical
weight of the configurations, taking into account the concentrations
of the relevant TFs and the energy of the configurations. Note that
the statistical weight of the last binding configuration $N_{x}N_{x}$
uses $S(2\epsilon,L)$ instead of the usual $S(\epsilon,L)$.}
\label{Tab:combreg}
\end{table}

Unlike in the basic activation setup, $Q$ genes that are required to be ON have two cognate TFs present, but genes that are required to be OFF have either none of the cognate types present, or one (but not both) of TF types present. As calculated above,
we have $t$ TFs and each TF has $n_{\rm int}$ combinations, while the
total number of combinations it can have are $N_{\rm int}$; each TF
that is present therefore has $N_{\rm int}-n_{\rm int}$ missing combinations. The
number of genes (that should be OFF)  which have only one TF present can be obtained as

\begin{eqnarray}
Q_{1}=\dfrac{t(N_{\rm int}-n_{\rm int})}{2}.
\end{eqnarray}

The number of genes with no cognate TFs present is $Q_{0}=M-Q-Q_{1}$. In Table~\ref{Tab:combreg}, we have listed all possible configurations for the two binding sites of a gene, along with details of crosstalk states and statistical weights. From this, we get the per-gene crosstalk for different types of genes. For genes that have both  cognate TFs present ($Q$ out of $M$), the per-gene crosstalk error is

\begin{eqnarray}
x_{\rm both}=1-\dfrac{(C/t)^2}{(C/t)^2 + 2e^{-E_a}(C/t) + 2(C/t)CS + 2e^{-E_a}CS + (CS)^2 + (C/t)CS(2\epsilon,L) + e^{-2E_a}}.
\end{eqnarray}

For genes that have only one of the two cognate TFs present ($Q_1$ out of $M$ genes), the per-gene crosstalk error is

\begin{eqnarray}
x_{\rm one}=\dfrac{(C/t)CS + (CS)^2 + (C/t)CS(2\epsilon,L)}{e^{-E_a}(C/t) + (C/t)CS + 2e^{-E_a}CS + (CS)^2 + (C/t)CS(2\epsilon,L) + e^{-2E_a}}.
\end{eqnarray}

For genes that don't have any of their two cognate TFs present ($M-Q-Q_1$ out of $M$ genes), the per-gene crosstalk error is

\begin{eqnarray}
x_{\rm none}=\dfrac{(CS)^2 + (C/t)CS(2\epsilon,L)}{2e^{-E_a}CS + (CS)^2 + (C/t)CS(2\epsilon,L) + e^{-2E_a}}.
\end{eqnarray}

The total crosstalk is:

\begin{eqnarray}
X=\dfrac{Q}{M}x_{\rm both} + \dfrac{Q_1}{M}x_{\rm one} + \left(1-\dfrac{Q+Q_1}{M}\right)x_{\rm none}.
\end{eqnarray}

For a given $M$ and $f$ and for each $(Q,S)$ pair, we compute the optimal concentration $C^*$ numerically, and obtain the minimal crosstalk $X_{\rm comb}^*$.

\begin{figure}
\centering{\includegraphics[scale=0.4]{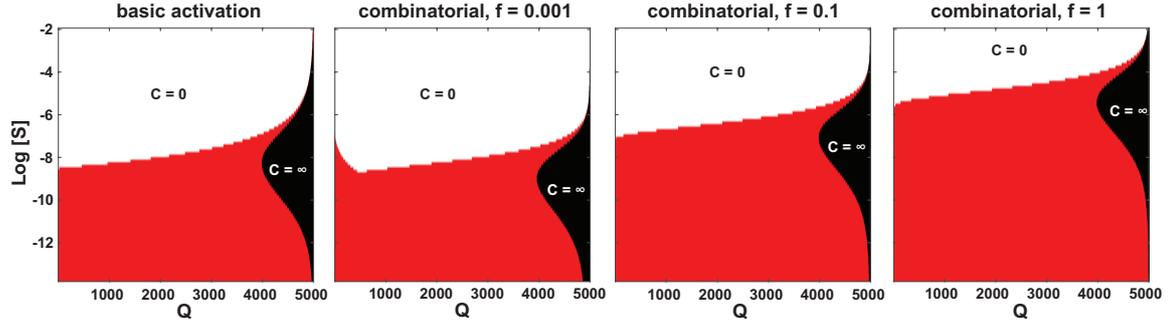}}
\caption{\textbf{Different regimes in the $(Q,S)$ plane for the basic and combinatorial setup.} Shifts in the regime boundaries in the basic activation setup vs. the combinatorial regulation setup. In the leftmost panel, we show the regimes for the basic activation setup. In  the other panels, we show the regimes for the combinatorial setup for $f=0.001, 0.1$, and $1$, respectively, from left to right. For $f=0.001$, the "regulation regime" is slightly smaller than in the basic activation setup. As $f$ increases, the "regulation regime" increases in size (and is bigger than in the basic activation setup) and the boundary with $C=0$ is pushed higher towards larger $S$.  \label{fig:comb_regimesf1}}
\end{figure}

\begin{figure}
\centering{\includegraphics[scale=0.4]{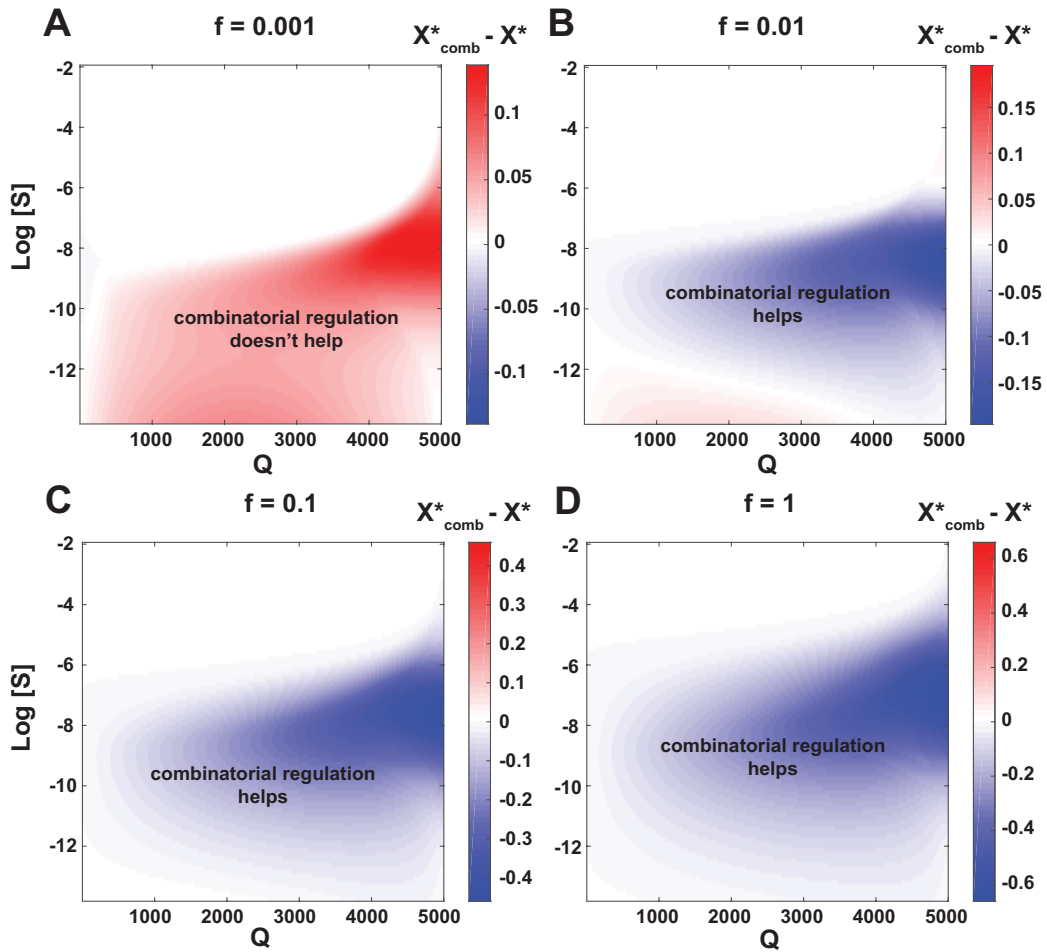}}
\caption{\textbf{Difference in minimal crosstalk between combinatorial setup and the basic activation setup for different $f$.}  Panel (a) shows $f=0.001$, where combinatorial regulation underperforms the basic regulation setup. (b,c,d) Increasing values of $f$ ($f=0.01, 0.1, 1$, respectively) can lower the crosstalk relative to the basic setup. At baseline parameters ($Q=2500, M=5000$ and $\log{(S)}=-10.5$), minimal crosstalk for the combinatorial setups reads $X_{\rm comb}^* = 0.28, 0.18, 0.11$ and $0.07$ for $f=0.001, 0.01, 0.1$ and $1$ respectively, compared to $X^*=0.23$ for the basic activation setup. \label{fig:comb_Xdiff}}
\end{figure}

As plotted in Fig.~\ref{fig:comb_regimesf1}, the boundaries between different regimes shift in the combinatorial setup. In particular, while at small $f$ the "regulation regime" shrinks in the $(Q,S)$ plane, as $f$ increases, it expands. As $f$ increases towards $1$, the boundary between the "regulation regime" and "$C=0$" regime moves towards larger $S$.  In Fig.~\ref{fig:comb_Xdiff}, we have plotted the difference in optimal crosstalk between combinatorial regulation and the basic activation setup. For $f=0.001$, combinatorial regulation doesn't improve from the basic activation setup in terms of optimal crosstalk. But for $f=0.01, 0.1,$ and $1$, combinatorial regulation gives a lower optimal crosstalk than the basic activation setup. So, there exists a threshold in $f$ such that for combinatorial regulation below that threshold, the "regulation regime" shrinks in comparison to the basic activation setup and performs worse. Above the threshold, the "regulation regime" expands towards larger $S$ and gives a lower optimal crosstalk than the basic activation setup. At the baseline parameters of $Q=2500, M=5000$ and $\log{(S)}=-10.5$, optimal crosstalk for the combinatorial setups reads as $X_{\rm comb}^* = 0.28, 0.18, 0.11$ and $0.07$ for $f=0.001, 0.01, 0.1$ and $1$ respectively, compared to $X^*=0.23$ for the basic activation setup.

\begin{figure}
\centering{\includegraphics[scale=0.6]{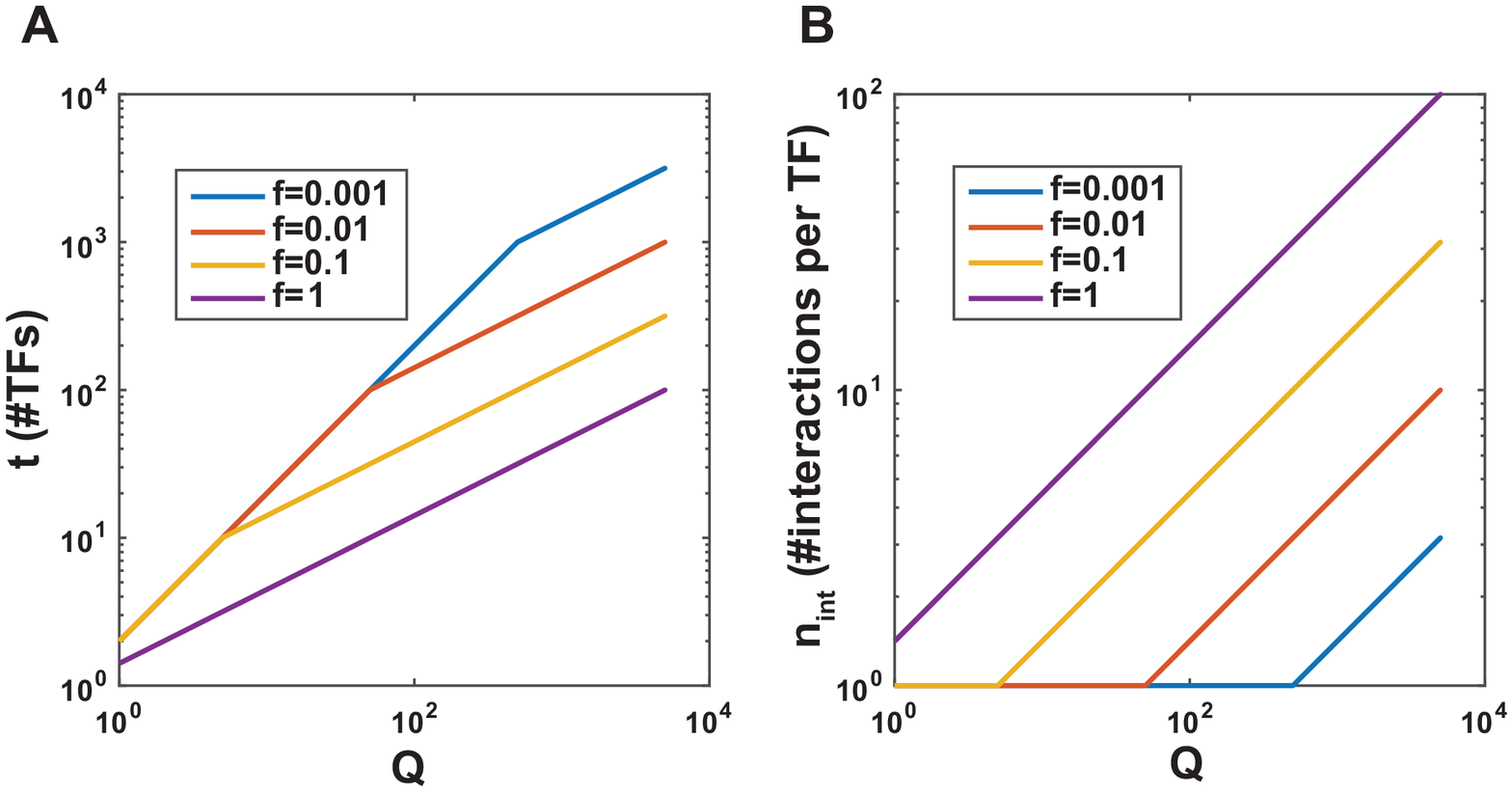}}
\caption{\textbf{Scaling of the typical number of TFs present ($t$) and number of interactions per TF ($n_{\rm int}$) as a function of $Q$ for different $f$.}  For each $f$, for $Q$ smaller than some threshold value which depends on $f$, the number of TFs $t$ varies as $Q=2t$ and the number of interactions per TF $n$ is constant at $1$. For all $Q$ greater than this threshold value, $\log n$ increases linearly with $\log Q$ ($n$ changes with $Q$ in a power-law fashion). \label{fig:interactions_f}}
\end{figure}

This decrease in crosstalk is consistent  with the reduction in the number of regulatory components ($T$ and $t$, the number of TFs, see Fig. \ref{fig:interactions_f}), as discussed in SI Section 1.5. In the case of perfect combinatorial regulation ($f=1$), we have roughly $\sqrt{2M}$ instead of $M$ TF species in the basic activation setup, which is a significant reduction in the number of regulatory components. Hence, each TF now effectively controls $\Theta = M/\sqrt{2M} = \sqrt{M/2}$ genes, and so the decrease in crosstalk is expected to be roughly $\sqrt{\Theta}$ compared to the basic activation setup. For $M=5000$ genes, this would suggest that perfect combinatorial regulation could decrease the crosstalk by $\sim 7$-fold over the basic model. The actual reduction in crosstalk (from 0.23 to 0.07) isn't as large because of certain differences between the combinatorial setup and $\Theta$-genes setup of SI Section~\ref{sec:thetagenes}. One major difference is that in the $\Theta$-genes setup, the cell can only activate sets of genes of size $\Theta$, while in the combinatorial setup, the cell has the power to activate single genes at will, albeit at the cost of partially activating genes that aren't needed (since a considerable fraction of genes that should be OFF must have one of the two activators present) and allowing new non-cognate configurations. Fundamentally, therefore, crosstalk reduction comes from the decrease in the number of regulatory components (TF species) needed in the system, which again points to the explosion in the number of possible noncognate interactions as the crucial origin of the crosstalk. In other words, what qualitatively seems to matter is $\Theta$, the number of regulated genes per TF, while the detailed manner in which these TFs regulate is less important for the actual numerical value of crosstalk (but \emph{is} important for the functioning of the cell; e.g., in combinatorial regulation genes can be addressed individually, while in the model of SI Section~\ref{sec:thetagenes} they cannot be).

We also note that while near-ideal combinatorial regulation appears to be a useful strategy to reduce the crosstalk, studies of scaling laws in gene regulatory networks do not appear to be consistent with the use of such a pure combinatorial strategy. In particular, the number of TFs scales at least linearly (quadratically, in prokaryotes) with the total number of genes~\cite{nimwegen_scaling_2003} across different organisms, while an efficient combinatorial strategy would suggest sub-linear (e.g., square-root) scaling. This clearly does not preclude the use of combinatorial regulation in some regulatory elements, but does show that even with the possible utilization of the combinatorial strategy the observed growth in the number of distinct TF species (which seems to be an important crosstalk parameter) is extensive.

\section{Weak global repressor}

So far we only considered gene regulation by activators. Cells however also have repression mechanisms as an additional means of regulation.
As a first step to account for that we incorporate in the model one type of an abundant weak global repressor that interacts with all binding sites with sequence-independent low affinity. Non-specific repression mechanisms such as the nuclear envelope, histones and DNA methylation are thought to mitigate spurious transcription~\cite{bird_gene_1995}. It was hypothesized that their emergence enabled the genome expansion in the transitions between prokaryotes to eukaryotes and from invertebrates to vertebrates \cite{bird_gene_1995}.
We include an additional molecule in the model, which is found in concentration $C_r$ and can bind all binding sites equally well with energy $0<E_r<E_a$, namely it is more favorable than the unbound state, but not as favorable as the specific cognate activator of each site.
Hence, our intuition was that such a global repressor cannot compete equally with specific binding, but it can reduce non-specific binding.
The crosstalk expressions now read:
\be
x_1^r = \frac{S C + C_r e^{-E_r}+e^{-E_a}}{S C+\frac{C}{Q}+C_r e^{-E_r}+e^{-E_a}}
\ee

\be
x_2^r = \frac{S C}{S C + C_r e^{-E_r} + e^{-E_a}}.
\ee

As before, we minimize the crosstalk with respect to the TF concentration. The optimal concentration is now:
\be
C^\ast_{GR} = -\frac{Q \left(C_r e^{-E_r} + e^{-E_a}\right)  \left(\sqrt{S (M-Q)}-S (S M Q-Q (S Q+2)+M)\right) }{S \left(-M (S Q+1)^2+S Q^2 (S Q+3)+Q\right)}.
\label{eq:Cr_opt}
\ee

This is the same optimal concentration $C^\ast$ as in \eqref{eq:Copt} only scaled by a factor $C_r e^{-E_r} + e^{-E_a}$, instead of $e^{-E_a}$ there. We conclude that the mere effect of a global repressor is to scale down the concentration of the specific activator. This is simply compensated for by a larger concentration of the activator. Hence, regardless of the global repressor affinity $E_r$ and concentration $C_r$ this additional regulatory mechanism cannot lower the crosstalk beyond what is possible with specific activators only. As before, the minimal crosstalk is:
\be
X^\ast_{GR}= \frac{Q}{M}\left(-S (M - Q) + 2 \sqrt{S(M-Q)}\right).
\label{eq:minX_globalR}
\ee

\section{Regulation by a combination of specific activators and specific repressors}

As the global repressor examined in the previous section did not show any additional improvement in crosstalk, we elaborate the model further to account for specific repressors, in similarity to the specific activators.
We extended the basic model (Section \ref{sec:BasicModel}) in which a gene had a single regulatory site and was regulated by an activator alone, to a more general model in which each gene has two regulatory sites: one compatible with a specific activator binding and the other with a specific repressor.
We assume that each gene has a unique  activator and unique repressor.
In the basic model (Section \ref{sec:BasicModel}), for a gene to be silent its binding site should be vacant. The only way to achieve this was to lower the activator concentration. On the other hand, to improve activation reliability, the activator concentration, should be increased! Thus, in the simple model there seemed to be a trade-off between reliable activation and elimination of undesirable activation.
The existence of a specific molecule that blocks the site from binding of other (potentially activating) molecules is thought to be a more reliable way to prevent undesired gene activation, not at the expense of the activation of other genes \cite{shinar_rules_2006}.

To be consistent with the basic model, we assume that the total concentration of all TFs (activators and repressors together) is constant $C$. As before, $Q$ genes need to be activated for which $Q$ specific activators are present. The other $M-Q$ genes need to be silent for which we now add their $M-Q$ specific repressors. All activators are found in equal concentrations $C_A/Q = \alpha * C / Q$ each. All repressors are in equal concentrations $C_R/(M-Q) = (1-\alpha)*C/(M-Q)$ each. We allow for different binding energies for the two binding sites $E_a$ and $E_r$. 
We assume that activation can only occur by binding of an activator molecule to the 'A' site. Repression is asymmetric in the sense that binding of any molecule to the repressor site prevents binding regardless of what is bound to the activator site. Thus a gene can only be active if the repressor site is empty and the activator site is bound by an activator. See the list of all possible states of the two binding sites in Tables \ref{Tab:ActivatorRepressor_A} and \ref{Tab:ActivatorRepressor_R} below.

\begin{table}[h!]
\begin{tabular}{|c|c|c|c|c|c|}
  \hline
   & configuration          & activity &  crosstalk  &  Energy             & Weight \\
   &(R-site,A-site)         &          &   if ON     &                     &         \\&&&&&\\
  \hline
  1 &   U, U                & OFF      & +           &$E_a+E_r$  & $e^{-(E_a+E_r)}$        \\&&&&\\
  2 &   U, $C_A$            &  ON      & -           & $E_r$                        &  $\frac{C}{Q}\alpha e^{-E_r}$ \\&&&&&\\
  3 &   U, $N_A$            & *        & +           & $E_r + \epsilon d$           & $C \alpha S e^{-E_r}$ \\&&&&&\\
  4 &   U, $N_R$            & OFF      & +           & $E_r + \epsilon d$           & $C (1-\alpha) S e^{-E_r}$ \\&&&&&\\
  5 &   $C_A$, U            & OFF      & +           & $E_a + \epsilon d$           & $\frac{C}{Q}\alpha S e^{-E_a}$ \\&&&&&\\
  6 & $N_A$, U              & OFF      & +           & $E_a + \epsilon d$           & $C\frac{Q-1}{Q}\alpha S e^{-E_a}$ \\&&&&&\\
  7 & $N_R$, U              & OFF      & +           & $E_a + \epsilon d$           & $C (1-\alpha) S e^{-E_a}$ \\&&&&&\\
  \hline\hline
  8 & ($N_A$, $C_A$),$C_A$  & OFF      & +           & $\epsilon d$           &  $\frac{(C\alpha)^2}{Q}S$  \\&&&&&\\
  9 & $C_A$,$N_A$           & OFF      & +           & $\epsilon (d_1+d_2)$   &  $\frac{(C\alpha)^2}{Q}S^2\frac{Q-1}{Q}$ \\&&&&&\\
  10 &$N_R$, $C_A$          & OFF      & +           & $\epsilon d$           &  $\frac{C^2}{Q}S \alpha(1-\alpha)$ \\&&&&&\\
  11 &($N_A$, $N_R$),$N_A$  & OFF      & +           & $\epsilon (d_1+d_2)$   &   $C^2S^2\alpha\frac{Q-1}{Q}\frac{Q-\alpha}{Q}$ \\&&&&&\\
  12 & ($N_R$, $N_A$, $C_A$),$N_R$& OFF& +           & $\epsilon (d_1+d_2)$ &   $C^2S^2(1-\alpha)$         \\&&&&&\\
  \hline
\end{tabular}
\caption {
\label{Tab:ActivatorRepressor_A}
All possible binding configurations, corresponding energies and statistical weights for a two-binding site (A,R)-model: a gene that needs to be activated (hence its cognate activator is present and its cognate repressor is absent). The subscripts 'A' and 'R' refer to activator and repressor. We assume that the site to which the molecule binds determines the activity state, where binding to A-site can activate the gene and binding to the R-site (even if it is an activator!) hinders activation. 'C' denotes binding by cognate factor, N - binding by noncognate and U - site is unbound.  $E_a$ and $E_r$ are the energy gaps between unbound and cognate-bound states of the corresponding binding sites.
In the upper part of the table (above the double line) we enumerate only states possible when both sites cannot be bound simultaneously (simplified model). If the two sites can be bound simultaneously, there are additional binding configurations, which are detailed below the line. The column 'crosstalk if ON' lists all binding configurations that were accounted for as crosstalk in $x_1$ calculation - in this case all except for no. 2 ( U, $C_A$). }
\end{table}

\begin{table}[h!]
\begin{tabular}{|c|c|c|c|c|c|}
  \hline
   & configuration                 & activity &  crosstalk   &  Energy             & Weight \\
   &(R-site,A-site)                &          &   if OFF     &                     &         \\&&&&&\\
  \hline
  1 &   U, U                       & OFF       &    -        & $E_a+E_r$                           & $e^{-(E_a+E_r)}$        \\&&&&&\\
  2 &   $C_R$, U                   & OFF       &    -        &  $E_a$                               &  $\frac{C(1-\alpha)}{M-Q}e^{-E_a}$ \\&&&&&\\
  3 &   $N_A$, U                   & OFF       &    -        &  $E_r + \epsilon d$                  & $C S\alpha e^{-E_a}$ \\&&&&&\\
  4 &   $N_R$, U                   & OFF       &    -        &  $E_r + \epsilon d$                  & $C S (1-\alpha) e^{-E_a}$ \\&&&&&\\
  5 &   U, $N_A$                   & *         &    +        &  $E_a + \epsilon d$                  & $C S \alpha e^{-E_r}$ \\&&&&&\\
  6 &   U, ($C_R$, $N_R$)          &   OFF     &    -    & $E_a + \epsilon d$       & $C S (1-\alpha) e^{-E_r}$ \\&&&&&\\
  \hline\hline
  7 &   $C_R$, ($C_R$ $N_R$, $N_A$)& OFF       &    -        &  $E_a + \epsilon d$   & $\frac{C (1-\alpha)}{M-Q} CS$ \\&&&&&\\
  8 & $N_R$, ($C_R$ $N_R$, $N_A$)  & OFF       &    -        &   $\epsilon d$           & $C^2 S^2 (1-\alpha^2)$  \\&&&&&\\
  9 & $N_A$, ($C_R$ $N_R$)         & OFF       &    -        &   $\epsilon (d_1+d_2)$ & $C^2 S^2 (1-\alpha^2)$ \\&&&&&\\
  10 &$N_A$, $N_A$                 & OFF       &    -        &  $\epsilon d$                  & $C^2 S^2 \alpha^2$ \\&&&&&\\
  \hline
\end{tabular}
\caption {
\label{Tab:ActivatorRepressor_R}
All possible binding configurations, corresponding energies and statistical weights for a two-binding site (A,R)-model: a gene that needs to be silent (hence its cognate repressor is present and its cognate activator is absent). All notation is the same as in Table~\ref{Tab:ActivatorRepressor_A}.  The column 'crosstalk if OFF' lists binding configurations that were accounted for as crosstalk in $x_2$ calculation - in this case only no. 5.  }
\end{table}

\subsection{Overlapping activator and repressor binding sites}
\begin{figure}[h!]
\centering
\includegraphics[width=0.45\textwidth]{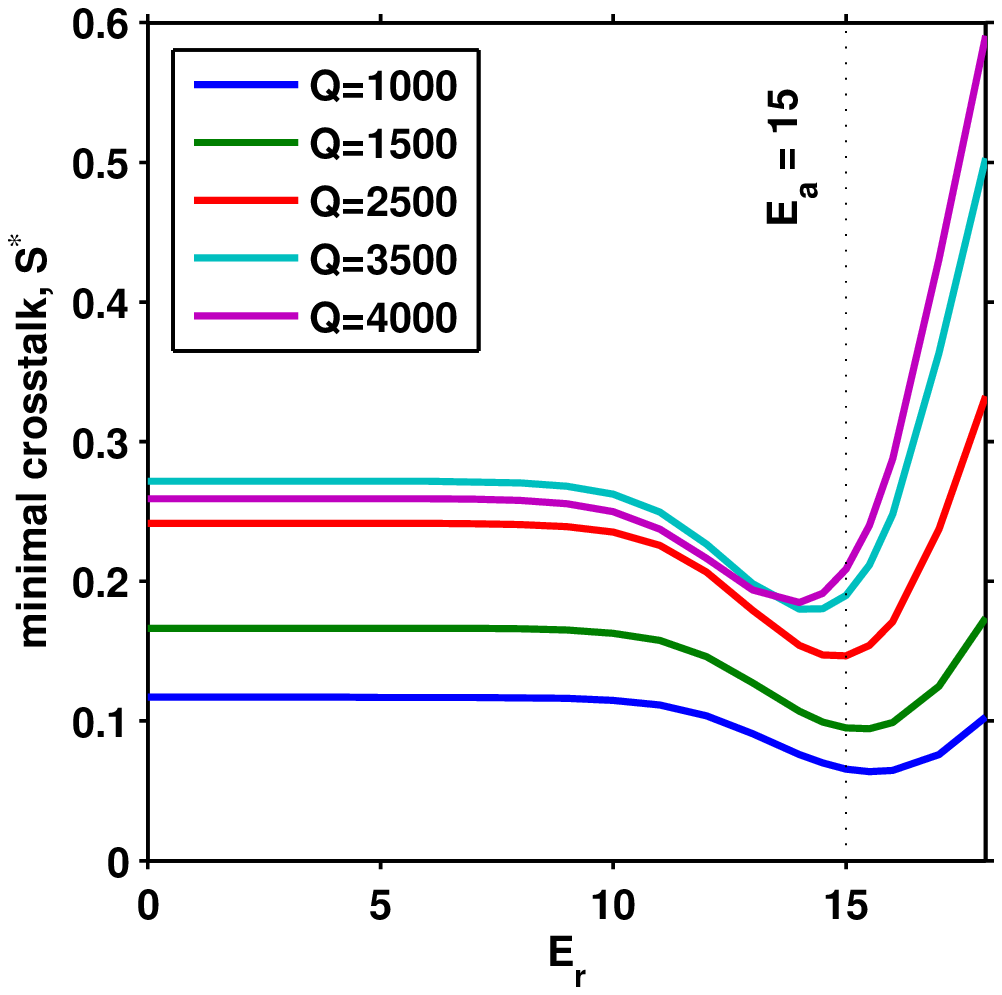}
  \caption[]
 { \label{fig:xtalk_vs_Er_diff_Q_AR_1_mol} \textbf{Activator-repressor overlapping binding sites, different $Q$ values}. $E_r^\ast$ - the energy gap between unbound and repressor-bound states - that minimizes crosstalk depends on the number of co-activated genes $Q$. Here we show numerical results for the minimal crosstalk $X^\ast$ as a function of the repressor binding affinity $E_r$ (with constant activator affinity $E_a=15$) for different numbers of co-activated genes $Q$, in the model where activator and repressor binding sites overlap. We find that when the number of co-activated genes decreases (so that more genes need to be repressed) the optimal repressor affinity $E_r^*$ increases, so that repressors more effectively bind their cognate binding sites and eliminate spurious transcription. When the number of genes that need to be activated equals the numbers of genes that need to be repressed $Q=M-Q$, we obtain that full symmetry between activator and repressor $E_r^\ast = E_a$ provides minimal crosstalk - this case is shown in the main text, Fig. 5. Parameters: $M=5000$, $S=10^{-4.5}$. }
  \end{figure}

 For some genes, the regulatory sites of the activator and repressor partially overlap. Another possibility is "negative cooperativity" - when one molecule repels the other. The outcome of either option is that either an activator or a repressor could be bound at any given time, but not both of them simultaneously. In Tables \ref{Tab:ActivatorRepressor_A}-\ref{Tab:ActivatorRepressor_R} all the states above the double horizontal line are such that only one site can be bound at any given time ('overlapping sites'). The additional states below the line are only possible if both sites can be bound simultaneously ('non-overlapping sites').
\figref{fig:xtalk_vs_Er_diff_Q_AR_1_mol} illustrates the dependence of crosstalk on the energy $E_r$ (energy gap between unbound and repressor-bound states) for different values of co-activated genes $Q$. Crosstalk is minimized for $E_r=E_a$ exactly when $Q=M-Q$, meaning equal number of activated and repressed genes. However, for other values of $Q \neq M-Q$, $E_r$ is also not significantly different from $E_a$.

\clearpage
\bibliographystyle{unsrt}
\bibliography{ZotOutput_corrected_290216}


\end{document}